\documentclass[USenglish,12pt,fleqn]{article} % Set [final] to switch off todonotes, Set [draft] for faster minted processing
\usepackage[USenglish=usenglishmax]{hyphsubst}
\usepackage{babel}
\usepackage{ragged2e}
\usepackage{amsmath,amssymb,amsfonts}
 
\usepackage{verbatim}
\usepackage[final]{graphicx}

\usepackage[epsilon,tstt]{backnaur}

\usepackage[fixpdftex,xcdraw]{xcolor}
\usepackage[margin=1in,footskip=0.25in]{geometry}
\usepackage[numbers,sort&compress]{natbib}
\usepackage[noend]{algpseudocode}
\usepackage{array,booktabs,threeparttable}
\newcolumntype{L}[1]{>{\RaggedRight\let\newline\\\arraybackslash\hspace{0pt}}p{#1}}
\newcolumntype{C}[1]{>{\Centering\let\newline\\\arraybackslash\hspace{0pt}}p{#1}}
\newcolumntype{R}[1]{>{\RaggedLeft\let\newline\\\arraybackslash\hspace{0pt}}p{#1}}
\usepackage[margin=20pt,font=small,labelfont=bf]{caption}
\usepackage[marginal]{footmisc}
\usepackage{pdflscape}
\usepackage{rotating}
\usepackage{tikz}
\usepackage[colorlinks=true]{hyperref}
\usepackage{url}
\usepackage[draft]{minted}

\makeatletter
\def\PYGdefault@reset{\let\PYGdefault@it=\relax \let\PYGdefault@bf=\relax%
    \let\PYGdefault@ul=\relax \let\PYGdefault@tc=\relax%
    \let\PYGdefault@bc=\relax \let\PYGdefault@ff=\relax}
\def\PYGdefault@tok#1{\csname PYGdefault@tok@#1\endcsname}
\def\PYGdefault@toks#1+{\ifx\relax#1\empty\else%
    \PYGdefault@tok{#1}\expandafter\PYGdefault@toks\fi}
\def\PYGdefault@do#1{\PYGdefault@bc{\PYGdefault@tc{\PYGdefault@ul{%
    \PYGdefault@it{\PYGdefault@bf{\PYGdefault@ff{#1}}}}}}}
\def\PYGdefault#1#2{\PYGdefault@reset\PYGdefault@toks#1+\relax+\PYGdefault@do{#2}}

\expandafter\def\csname PYGdefault@tok@w\endcsname{\def\PYGdefault@tc##1{\textcolor[rgb]{0.73,0.73,0.73}{##1}}}
\expandafter\def\csname PYGdefault@tok@c\endcsname{\let\PYGdefault@it=\textit\def\PYGdefault@tc##1{\textcolor[rgb]{0.25,0.50,0.50}{##1}}}
\expandafter\def\csname PYGdefault@tok@cp\endcsname{\def\PYGdefault@tc##1{\textcolor[rgb]{0.74,0.48,0.00}{##1}}}
\expandafter\def\csname PYGdefault@tok@k\endcsname{\let\PYGdefault@bf=\textbf\def\PYGdefault@tc##1{\textcolor[rgb]{0.00,0.50,0.00}{##1}}}
\expandafter\def\csname PYGdefault@tok@kp\endcsname{\def\PYGdefault@tc##1{\textcolor[rgb]{0.00,0.50,0.00}{##1}}}
\expandafter\def\csname PYGdefault@tok@kt\endcsname{\def\PYGdefault@tc##1{\textcolor[rgb]{0.69,0.00,0.25}{##1}}}
\expandafter\def\csname PYGdefault@tok@o\endcsname{\def\PYGdefault@tc##1{\textcolor[rgb]{0.40,0.40,0.40}{##1}}}
\expandafter\def\csname PYGdefault@tok@ow\endcsname{\let\PYGdefault@bf=\textbf\def\PYGdefault@tc##1{\textcolor[rgb]{0.67,0.13,1.00}{##1}}}
\expandafter\def\csname PYGdefault@tok@nb\endcsname{\def\PYGdefault@tc##1{\textcolor[rgb]{0.00,0.50,0.00}{##1}}}
\expandafter\def\csname PYGdefault@tok@nf\endcsname{\def\PYGdefault@tc##1{\textcolor[rgb]{0.00,0.00,1.00}{##1}}}
\expandafter\def\csname PYGdefault@tok@nc\endcsname{\let\PYGdefault@bf=\textbf\def\PYGdefault@tc##1{\textcolor[rgb]{0.00,0.00,1.00}{##1}}}
\expandafter\def\csname PYGdefault@tok@nn\endcsname{\let\PYGdefault@bf=\textbf\def\PYGdefault@tc##1{\textcolor[rgb]{0.00,0.00,1.00}{##1}}}
\expandafter\def\csname PYGdefault@tok@ne\endcsname{\let\PYGdefault@bf=\textbf\def\PYGdefault@tc##1{\textcolor[rgb]{0.82,0.25,0.23}{##1}}}
\expandafter\def\csname PYGdefault@tok@nv\endcsname{\def\PYGdefault@tc##1{\textcolor[rgb]{0.10,0.09,0.49}{##1}}}
\expandafter\def\csname PYGdefault@tok@no\endcsname{\def\PYGdefault@tc##1{\textcolor[rgb]{0.53,0.00,0.00}{##1}}}
\expandafter\def\csname PYGdefault@tok@nl\endcsname{\def\PYGdefault@tc##1{\textcolor[rgb]{0.63,0.63,0.00}{##1}}}
\expandafter\def\csname PYGdefault@tok@ni\endcsname{\let\PYGdefault@bf=\textbf\def\PYGdefault@tc##1{\textcolor[rgb]{0.60,0.60,0.60}{##1}}}
\expandafter\def\csname PYGdefault@tok@na\endcsname{\def\PYGdefault@tc##1{\textcolor[rgb]{0.49,0.56,0.16}{##1}}}
\expandafter\def\csname PYGdefault@tok@nt\endcsname{\let\PYGdefault@bf=\textbf\def\PYGdefault@tc##1{\textcolor[rgb]{0.00,0.50,0.00}{##1}}}
\expandafter\def\csname PYGdefault@tok@nd\endcsname{\def\PYGdefault@tc##1{\textcolor[rgb]{0.67,0.13,1.00}{##1}}}
\expandafter\def\csname PYGdefault@tok@s\endcsname{\def\PYGdefault@tc##1{\textcolor[rgb]{0.73,0.13,0.13}{##1}}}
\expandafter\def\csname PYGdefault@tok@sd\endcsname{\let\PYGdefault@it=\textit\def\PYGdefault@tc##1{\textcolor[rgb]{0.73,0.13,0.13}{##1}}}
\expandafter\def\csname PYGdefault@tok@si\endcsname{\let\PYGdefault@bf=\textbf\def\PYGdefault@tc##1{\textcolor[rgb]{0.73,0.40,0.53}{##1}}}
\expandafter\def\csname PYGdefault@tok@se\endcsname{\let\PYGdefault@bf=\textbf\def\PYGdefault@tc##1{\textcolor[rgb]{0.73,0.40,0.13}{##1}}}
\expandafter\def\csname PYGdefault@tok@sr\endcsname{\def\PYGdefault@tc##1{\textcolor[rgb]{0.73,0.40,0.53}{##1}}}
\expandafter\def\csname PYGdefault@tok@ss\endcsname{\def\PYGdefault@tc##1{\textcolor[rgb]{0.10,0.09,0.49}{##1}}}
\expandafter\def\csname PYGdefault@tok@sx\endcsname{\def\PYGdefault@tc##1{\textcolor[rgb]{0.00,0.50,0.00}{##1}}}
\expandafter\def\csname PYGdefault@tok@m\endcsname{\def\PYGdefault@tc##1{\textcolor[rgb]{0.40,0.40,0.40}{##1}}}
\expandafter\def\csname PYGdefault@tok@gh\endcsname{\let\PYGdefault@bf=\textbf\def\PYGdefault@tc##1{\textcolor[rgb]{0.00,0.00,0.50}{##1}}}
\expandafter\def\csname PYGdefault@tok@gu\endcsname{\let\PYGdefault@bf=\textbf\def\PYGdefault@tc##1{\textcolor[rgb]{0.50,0.00,0.50}{##1}}}
\expandafter\def\csname PYGdefault@tok@gd\endcsname{\def\PYGdefault@tc##1{\textcolor[rgb]{0.63,0.00,0.00}{##1}}}
\expandafter\def\csname PYGdefault@tok@gi\endcsname{\def\PYGdefault@tc##1{\textcolor[rgb]{0.00,0.63,0.00}{##1}}}
\expandafter\def\csname PYGdefault@tok@gr\endcsname{\def\PYGdefault@tc##1{\textcolor[rgb]{1.00,0.00,0.00}{##1}}}
\expandafter\def\csname PYGdefault@tok@ge\endcsname{\let\PYGdefault@it=\textit}
\expandafter\def\csname PYGdefault@tok@gs\endcsname{\let\PYGdefault@bf=\textbf}
\expandafter\def\csname PYGdefault@tok@gp\endcsname{\let\PYGdefault@bf=\textbf\def\PYGdefault@tc##1{\textcolor[rgb]{0.00,0.00,0.50}{##1}}}
\expandafter\def\csname PYGdefault@tok@go\endcsname{\def\PYGdefault@tc##1{\textcolor[rgb]{0.53,0.53,0.53}{##1}}}
\expandafter\def\csname PYGdefault@tok@gt\endcsname{\def\PYGdefault@tc##1{\textcolor[rgb]{0.00,0.27,0.87}{##1}}}
\expandafter\def\csname PYGdefault@tok@err\endcsname{\def\PYGdefault@bc##1{\setlength{\fboxsep}{0pt}\fcolorbox[rgb]{1.00,0.00,0.00}{1,1,1}{\strut ##1}}}
\expandafter\def\csname PYGdefault@tok@kc\endcsname{\let\PYGdefault@bf=\textbf\def\PYGdefault@tc##1{\textcolor[rgb]{0.00,0.50,0.00}{##1}}}
\expandafter\def\csname PYGdefault@tok@kd\endcsname{\let\PYGdefault@bf=\textbf\def\PYGdefault@tc##1{\textcolor[rgb]{0.00,0.50,0.00}{##1}}}
\expandafter\def\csname PYGdefault@tok@kn\endcsname{\let\PYGdefault@bf=\textbf\def\PYGdefault@tc##1{\textcolor[rgb]{0.00,0.50,0.00}{##1}}}
\expandafter\def\csname PYGdefault@tok@kr\endcsname{\let\PYGdefault@bf=\textbf\def\PYGdefault@tc##1{\textcolor[rgb]{0.00,0.50,0.00}{##1}}}
\expandafter\def\csname PYGdefault@tok@bp\endcsname{\def\PYGdefault@tc##1{\textcolor[rgb]{0.00,0.50,0.00}{##1}}}
\expandafter\def\csname PYGdefault@tok@vc\endcsname{\def\PYGdefault@tc##1{\textcolor[rgb]{0.10,0.09,0.49}{##1}}}
\expandafter\def\csname PYGdefault@tok@vg\endcsname{\def\PYGdefault@tc##1{\textcolor[rgb]{0.10,0.09,0.49}{##1}}}
\expandafter\def\csname PYGdefault@tok@vi\endcsname{\def\PYGdefault@tc##1{\textcolor[rgb]{0.10,0.09,0.49}{##1}}}
\expandafter\def\csname PYGdefault@tok@sb\endcsname{\def\PYGdefault@tc##1{\textcolor[rgb]{0.73,0.13,0.13}{##1}}}
\expandafter\def\csname PYGdefault@tok@sc\endcsname{\def\PYGdefault@tc##1{\textcolor[rgb]{0.73,0.13,0.13}{##1}}}
\expandafter\def\csname PYGdefault@tok@s2\endcsname{\def\PYGdefault@tc##1{\textcolor[rgb]{0.73,0.13,0.13}{##1}}}
\expandafter\def\csname PYGdefault@tok@sh\endcsname{\def\PYGdefault@tc##1{\textcolor[rgb]{0.73,0.13,0.13}{##1}}}
\expandafter\def\csname PYGdefault@tok@s1\endcsname{\def\PYGdefault@tc##1{\textcolor[rgb]{0.73,0.13,0.13}{##1}}}
\expandafter\def\csname PYGdefault@tok@mb\endcsname{\def\PYGdefault@tc##1{\textcolor[rgb]{0.40,0.40,0.40}{##1}}}
\expandafter\def\csname PYGdefault@tok@mf\endcsname{\def\PYGdefault@tc##1{\textcolor[rgb]{0.40,0.40,0.40}{##1}}}
\expandafter\def\csname PYGdefault@tok@mh\endcsname{\def\PYGdefault@tc##1{\textcolor[rgb]{0.40,0.40,0.40}{##1}}}
\expandafter\def\csname PYGdefault@tok@mi\endcsname{\def\PYGdefault@tc##1{\textcolor[rgb]{0.40,0.40,0.40}{##1}}}
\expandafter\def\csname PYGdefault@tok@il\endcsname{\def\PYGdefault@tc##1{\textcolor[rgb]{0.40,0.40,0.40}{##1}}}
\expandafter\def\csname PYGdefault@tok@mo\endcsname{\def\PYGdefault@tc##1{\textcolor[rgb]{0.40,0.40,0.40}{##1}}}
\expandafter\def\csname PYGdefault@tok@ch\endcsname{\let\PYGdefault@it=\textit\def\PYGdefault@tc##1{\textcolor[rgb]{0.25,0.50,0.50}{##1}}}
\expandafter\def\csname PYGdefault@tok@cm\endcsname{\let\PYGdefault@it=\textit\def\PYGdefault@tc##1{\textcolor[rgb]{0.25,0.50,0.50}{##1}}}
\expandafter\def\csname PYGdefault@tok@cpf\endcsname{\let\PYGdefault@it=\textit\def\PYGdefault@tc##1{\textcolor[rgb]{0.25,0.50,0.50}{##1}}}
\expandafter\def\csname PYGdefault@tok@c1\endcsname{\let\PYGdefault@it=\textit\def\PYGdefault@tc##1{\textcolor[rgb]{0.25,0.50,0.50}{##1}}}
\expandafter\def\csname PYGdefault@tok@cs\endcsname{\let\PYGdefault@it=\textit\def\PYGdefault@tc##1{\textcolor[rgb]{0.25,0.50,0.50}{##1}}}

\makeatother

\makeatletter
\def\PYG@reset{\let\PYG@it=\relax \let\PYG@bf=\relax%
    \let\PYG@ul=\relax \let\PYG@tc=\relax%
    \let\PYG@bc=\relax \let\PYG@ff=\relax}
\def\PYG@tok#1{\csname PYG@tok@#1\endcsname}
\def\PYG@toks#1+{\ifx\relax#1\empty\else%
    \PYG@tok{#1}\expandafter\PYG@toks\fi}
\def\PYG@do#1{\PYG@bc{\PYG@tc{\PYG@ul{%
    \PYG@it{\PYG@bf{\PYG@ff{#1}}}}}}}
\def\PYG#1#2{\PYG@reset\PYG@toks#1+\relax+\PYG@do{#2}}

\expandafter\def\csname PYG@tok@w\endcsname{\def\PYG@tc##1{\textcolor[rgb]{0.73,0.73,0.73}{##1}}}
\expandafter\def\csname PYG@tok@c\endcsname{\let\PYG@it=\textit\def\PYG@tc##1{\textcolor[rgb]{0.25,0.50,0.50}{##1}}}
\expandafter\def\csname PYG@tok@cp\endcsname{\def\PYG@tc##1{\textcolor[rgb]{0.74,0.48,0.00}{##1}}}
\expandafter\def\csname PYG@tok@k\endcsname{\let\PYG@bf=\textbf\def\PYG@tc##1{\textcolor[rgb]{0.00,0.50,0.00}{##1}}}
\expandafter\def\csname PYG@tok@kp\endcsname{\def\PYG@tc##1{\textcolor[rgb]{0.00,0.50,0.00}{##1}}}
\expandafter\def\csname PYG@tok@kt\endcsname{\def\PYG@tc##1{\textcolor[rgb]{0.69,0.00,0.25}{##1}}}
\expandafter\def\csname PYG@tok@o\endcsname{\def\PYG@tc##1{\textcolor[rgb]{0.40,0.40,0.40}{##1}}}
\expandafter\def\csname PYG@tok@ow\endcsname{\let\PYG@bf=\textbf\def\PYG@tc##1{\textcolor[rgb]{0.67,0.13,1.00}{##1}}}
\expandafter\def\csname PYG@tok@nb\endcsname{\def\PYG@tc##1{\textcolor[rgb]{0.00,0.50,0.00}{##1}}}
\expandafter\def\csname PYG@tok@nf\endcsname{\def\PYG@tc##1{\textcolor[rgb]{0.00,0.00,1.00}{##1}}}
\expandafter\def\csname PYG@tok@nc\endcsname{\let\PYG@bf=\textbf\def\PYG@tc##1{\textcolor[rgb]{0.00,0.00,1.00}{##1}}}
\expandafter\def\csname PYG@tok@nn\endcsname{\let\PYG@bf=\textbf\def\PYG@tc##1{\textcolor[rgb]{0.00,0.00,1.00}{##1}}}
\expandafter\def\csname PYG@tok@ne\endcsname{\let\PYG@bf=\textbf\def\PYG@tc##1{\textcolor[rgb]{0.82,0.25,0.23}{##1}}}
\expandafter\def\csname PYG@tok@nv\endcsname{\def\PYG@tc##1{\textcolor[rgb]{0.10,0.09,0.49}{##1}}}
\expandafter\def\csname PYG@tok@no\endcsname{\def\PYG@tc##1{\textcolor[rgb]{0.53,0.00,0.00}{##1}}}
\expandafter\def\csname PYG@tok@nl\endcsname{\def\PYG@tc##1{\textcolor[rgb]{0.63,0.63,0.00}{##1}}}
\expandafter\def\csname PYG@tok@ni\endcsname{\let\PYG@bf=\textbf\def\PYG@tc##1{\textcolor[rgb]{0.60,0.60,0.60}{##1}}}
\expandafter\def\csname PYG@tok@na\endcsname{\def\PYG@tc##1{\textcolor[rgb]{0.49,0.56,0.16}{##1}}}
\expandafter\def\csname PYG@tok@nt\endcsname{\let\PYG@bf=\textbf\def\PYG@tc##1{\textcolor[rgb]{0.00,0.50,0.00}{##1}}}
\expandafter\def\csname PYG@tok@nd\endcsname{\def\PYG@tc##1{\textcolor[rgb]{0.67,0.13,1.00}{##1}}}
\expandafter\def\csname PYG@tok@s\endcsname{\def\PYG@tc##1{\textcolor[rgb]{0.73,0.13,0.13}{##1}}}
\expandafter\def\csname PYG@tok@sd\endcsname{\let\PYG@it=\textit\def\PYG@tc##1{\textcolor[rgb]{0.73,0.13,0.13}{##1}}}
\expandafter\def\csname PYG@tok@si\endcsname{\let\PYG@bf=\textbf\def\PYG@tc##1{\textcolor[rgb]{0.73,0.40,0.53}{##1}}}
\expandafter\def\csname PYG@tok@se\endcsname{\let\PYG@bf=\textbf\def\PYG@tc##1{\textcolor[rgb]{0.73,0.40,0.13}{##1}}}
\expandafter\def\csname PYG@tok@sr\endcsname{\def\PYG@tc##1{\textcolor[rgb]{0.73,0.40,0.53}{##1}}}
\expandafter\def\csname PYG@tok@ss\endcsname{\def\PYG@tc##1{\textcolor[rgb]{0.10,0.09,0.49}{##1}}}
\expandafter\def\csname PYG@tok@sx\endcsname{\def\PYG@tc##1{\textcolor[rgb]{0.00,0.50,0.00}{##1}}}
\expandafter\def\csname PYG@tok@m\endcsname{\def\PYG@tc##1{\textcolor[rgb]{0.40,0.40,0.40}{##1}}}
\expandafter\def\csname PYG@tok@gh\endcsname{\let\PYG@bf=\textbf\def\PYG@tc##1{\textcolor[rgb]{0.00,0.00,0.50}{##1}}}
\expandafter\def\csname PYG@tok@gu\endcsname{\let\PYG@bf=\textbf\def\PYG@tc##1{\textcolor[rgb]{0.50,0.00,0.50}{##1}}}
\expandafter\def\csname PYG@tok@gd\endcsname{\def\PYG@tc##1{\textcolor[rgb]{0.63,0.00,0.00}{##1}}}
\expandafter\def\csname PYG@tok@gi\endcsname{\def\PYG@tc##1{\textcolor[rgb]{0.00,0.63,0.00}{##1}}}
\expandafter\def\csname PYG@tok@gr\endcsname{\def\PYG@tc##1{\textcolor[rgb]{1.00,0.00,0.00}{##1}}}
\expandafter\def\csname PYG@tok@ge\endcsname{\let\PYG@it=\textit}
\expandafter\def\csname PYG@tok@gs\endcsname{\let\PYG@bf=\textbf}
\expandafter\def\csname PYG@tok@gp\endcsname{\let\PYG@bf=\textbf\def\PYG@tc##1{\textcolor[rgb]{0.00,0.00,0.50}{##1}}}
\expandafter\def\csname PYG@tok@go\endcsname{\def\PYG@tc##1{\textcolor[rgb]{0.53,0.53,0.53}{##1}}}
\expandafter\def\csname PYG@tok@gt\endcsname{\def\PYG@tc##1{\textcolor[rgb]{0.00,0.27,0.87}{##1}}}
\expandafter\def\csname PYG@tok@err\endcsname{\def\PYG@bc##1{\setlength{\fboxsep}{0pt}\fcolorbox[rgb]{1.00,0.00,0.00}{1,1,1}{\strut ##1}}}
\expandafter\def\csname PYG@tok@kc\endcsname{\let\PYG@bf=\textbf\def\PYG@tc##1{\textcolor[rgb]{0.00,0.50,0.00}{##1}}}
\expandafter\def\csname PYG@tok@kd\endcsname{\let\PYG@bf=\textbf\def\PYG@tc##1{\textcolor[rgb]{0.00,0.50,0.00}{##1}}}
\expandafter\def\csname PYG@tok@kn\endcsname{\let\PYG@bf=\textbf\def\PYG@tc##1{\textcolor[rgb]{0.00,0.50,0.00}{##1}}}
\expandafter\def\csname PYG@tok@kr\endcsname{\let\PYG@bf=\textbf\def\PYG@tc##1{\textcolor[rgb]{0.00,0.50,0.00}{##1}}}
\expandafter\def\csname PYG@tok@bp\endcsname{\def\PYG@tc##1{\textcolor[rgb]{0.00,0.50,0.00}{##1}}}
\expandafter\def\csname PYG@tok@vc\endcsname{\def\PYG@tc##1{\textcolor[rgb]{0.10,0.09,0.49}{##1}}}
\expandafter\def\csname PYG@tok@vg\endcsname{\def\PYG@tc##1{\textcolor[rgb]{0.10,0.09,0.49}{##1}}}
\expandafter\def\csname PYG@tok@vi\endcsname{\def\PYG@tc##1{\textcolor[rgb]{0.10,0.09,0.49}{##1}}}
\expandafter\def\csname PYG@tok@sb\endcsname{\def\PYG@tc##1{\textcolor[rgb]{0.73,0.13,0.13}{##1}}}
\expandafter\def\csname PYG@tok@sc\endcsname{\def\PYG@tc##1{\textcolor[rgb]{0.73,0.13,0.13}{##1}}}
\expandafter\def\csname PYG@tok@s2\endcsname{\def\PYG@tc##1{\textcolor[rgb]{0.73,0.13,0.13}{##1}}}
\expandafter\def\csname PYG@tok@sh\endcsname{\def\PYG@tc##1{\textcolor[rgb]{0.73,0.13,0.13}{##1}}}
\expandafter\def\csname PYG@tok@s1\endcsname{\def\PYG@tc##1{\textcolor[rgb]{0.73,0.13,0.13}{##1}}}
\expandafter\def\csname PYG@tok@mb\endcsname{\def\PYG@tc##1{\textcolor[rgb]{0.40,0.40,0.40}{##1}}}
\expandafter\def\csname PYG@tok@mf\endcsname{\def\PYG@tc##1{\textcolor[rgb]{0.40,0.40,0.40}{##1}}}
\expandafter\def\csname PYG@tok@mh\endcsname{\def\PYG@tc##1{\textcolor[rgb]{0.40,0.40,0.40}{##1}}}
\expandafter\def\csname PYG@tok@mi\endcsname{\def\PYG@tc##1{\textcolor[rgb]{0.40,0.40,0.40}{##1}}}
\expandafter\def\csname PYG@tok@il\endcsname{\def\PYG@tc##1{\textcolor[rgb]{0.40,0.40,0.40}{##1}}}
\expandafter\def\csname PYG@tok@mo\endcsname{\def\PYG@tc##1{\textcolor[rgb]{0.40,0.40,0.40}{##1}}}
\expandafter\def\csname PYG@tok@ch\endcsname{\let\PYG@it=\textit\def\PYG@tc##1{\textcolor[rgb]{0.25,0.50,0.50}{##1}}}
\expandafter\def\csname PYG@tok@cm\endcsname{\let\PYG@it=\textit\def\PYG@tc##1{\textcolor[rgb]{0.25,0.50,0.50}{##1}}}
\expandafter\def\csname PYG@tok@cpf\endcsname{\let\PYG@it=\textit\def\PYG@tc##1{\textcolor[rgb]{0.25,0.50,0.50}{##1}}}
\expandafter\def\csname PYG@tok@c1\endcsname{\let\PYG@it=\textit\def\PYG@tc##1{\textcolor[rgb]{0.25,0.50,0.50}{##1}}}
\expandafter\def\csname PYG@tok@cs\endcsname{\let\PYG@it=\textit\def\PYG@tc##1{\textcolor[rgb]{0.25,0.50,0.50}{##1}}}

\def\PYGZob{\char`\{}
\def\PYGZcb{\char`\}}

\def\PYGZgt{\char`\>}

\def\PYGZhy{\char`\-}

\def\PYGZdq{\char`\"}

\makeatother

\newmintedfile{c}{}

\providecommand{\ket}[1]{\left|#1 \right\rangle}
\usetikzlibrary{decorations.pathreplacing,decorations.pathmorphing}

\newenvironment{bnfsplit}[1][0.7\textwidth]
 {\minipage[t]{#1}$}
 {$\endminipage}

\newcommand*{\code}{\texttt}
\begin{document}

\title{Open Quantum Assembly Language}
\author{Andrew W. Cross, Lev S. Bishop, John A. Smolin, Jay M. Gambetta}
\date{January 10th, 2017}

\maketitle

\section{Background}

Software architectures, compilers, and languages specifically for quantum computing have been studied by the academic community for more than a decade (\cite{selinger04,gay06,svore06,haner16} and references therein). Researchers have implemented software and simulators that can be used in practice to study quantum algorithms at many scales. While we cannot survey this work here, we list a few of these projects, several of which include software that has been made readily available: Liquid \cite{ws14,liquid}, Scaffold \cite{scaffold,scaffcc}, Quipper \cite{valiron15,quipper,green13}, ProjectQ \cite{steiger16,projectq}, QCL \cite{omer03,qcl}, Quiddpro \cite{quiddpro,viamontes05}, Chisel-q \cite{chiselq,chisel}, and Quil \cite{Quil,smith16}.

\begin{figure}[h]
\hspace{-0.2cm}
\includegraphics[width=17cm]{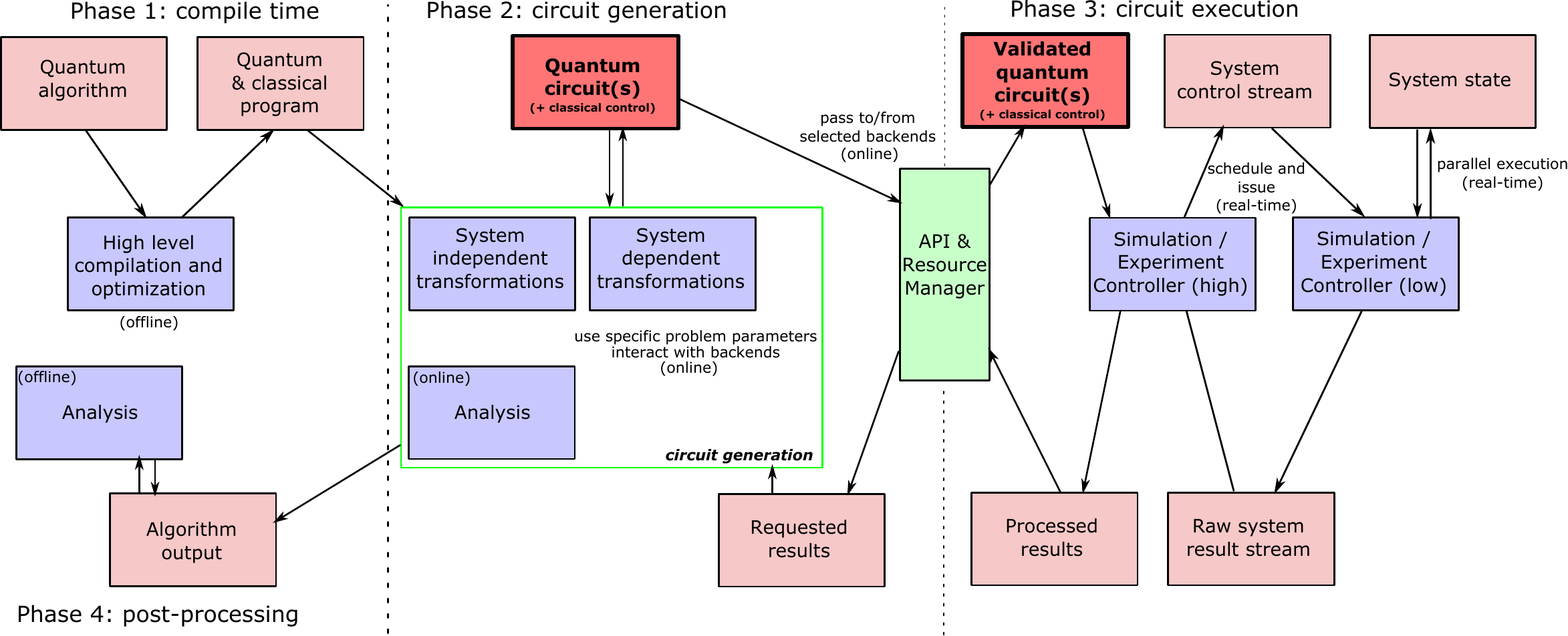}
\caption{Block diagrams of processes (blue) and abstractions (red) to transform and execute a quantum algorithm. The emphasized quantum circuit abstraction is the main focus of this document. The API and Resource Manager (green) represents the gateway to backend processes for circuit execution. Dashed vertical lines separate offline, online, and real-time processes.
\label{fig:diagram}}
\end{figure}

Our goal in this document is to describe an interface language for the Quantum Experience that enables experiments with small depth quantum circuits. The language can be generated by the Composer, hand-written, or targeted by higher level software tools, such as those above. Before we do so, we discuss quantum programs in general to provide context. General quantum programs require coordination of quantum and classical parts of the computation. One way to think about general quantum programs is to identify their distinct phases of execution \cite{green13}. Fig.~\ref{fig:diagram} shows a high-level diagram of the processes and abstractions involved in specifying a quantum algorithm, transforming the algorithm into executable form, running an experiment or simulation, and analyzing the results. A key idea throughout these processes is the use of intermediate representations. An intermediate representation (IR) of a computation is neither its source language description, nor the target machine instructions, but something in between. Compilers may use several IRs during the process of translating and optimizing a program.

{\bf Compilation.} This phase takes place on a classical computer in a setting where specific problem parameters are not yet known and no interaction with the quantum computer is required, i.e. it is offline. The input is source code describing a quantum algorithm and any compile time parameters. The output is a combined quantum/classical program expressed using a high level IR. During this phase, it is possible to compile classical procedures into object code and make initial passes that do not require complete knowledge of the problem parameters.

{\bf Circuit generation.} This takes place on a classical computer in an environment where specific problem parameters are now known, and some interaction with the quantum computer may occur, i.e. this is an online phase. The input is a quantum/classical program expressed using a high level IR, as well as all remaining problem parameters. The output is a collection of quantum circuits, or quantum basic blocks, together with associated classical control instructions and classical object code needed at run-time. A basic block is a straight-line code sequence with no branches (except at the entry and exit points). Since feedback can occur on multiple time scales, the quantum circuits may include instructions for fast feedback. Other classical control instructions outside of the quantum circuit basic block include, for example, run-time parameter computations and measurement-dependent branches. External classical object code could include algorithms to process measurement outcomes into control flow conditions or results, or to generate new basic blocks on the fly. The output of circuit generation is expressed using a quantum circuit IR. Further circuit generation may occur based on processed measurement results.

{\bf Execution.} This takes place on physical quantum computer controllers in a real-time environment, i.e. the quantum computer is active. The input is a collection of quantum circuits and associated run-time control statements expressed using a quantum circuit IR. The input is processed by a high-level controller into a stream of real-time instructions in a low-level format that corresponds to physical operations. These are executed on a low-level controller, and a corresponding results stream provides measurement data back to the high-level controller when needed. In general, the high level controller (or virtual machine) can execute classical control instructions and external object code. The output of circuit execution is a collection of processed measurement results returned from the high-level controller.

{\bf Post-processing.} This takes place on a classical computer after all real-time processing is complete. The input is a collection of processed measurement results, and the output is intermediate results for further circuit generation and/or the final result of the quantum computation.

Our model of program execution on the Quantum Experience does not allow fully general classical computations in the loop with quantum computations, as described above, because qubits remain coherent for a limited time. Quantum programs are broken into distinct circuits whose quantum outputs cannot be carried over into the next circuit. Classical computation is done between quantum circuit executions. Users actively participate in the circuit generation phase and manually implement part of feedback path through the high level controller in Fig.~\ref{fig:diagram}, observing outcomes from the previous quantum circuit and choosing the next quantum circuit to execute. Making use of an API to the execution phase, users can write their own software for compilation and circuit generation that interacts with the hardware over a sequence of quantum circuit executions. After obtaining all of the processed results, users may post-process the data offline.

We specify part of a quantum circuit intermediate representation based on the quantum circuit model, a standard formalism for quantum computation \cite{NC00}. The quantum circuit abstraction is emphasized in Fig.~\ref{fig:diagram}. The IR expresses quantum circuits with fast feedback, such as might constitute the basic blocks of a full-featured IR. A basic block is a straight-line code sequence with no branches (except at the entry and exit points). We have chosen to include statements that are essential for near-term experiments and that we believe will be present in any future IR. The representation will be quite familiar to experts.

The human-readable form of our quantum circuit IR is based on ``quantum assembly language'' \cite{qasm2circ,qasmtools,svore06,quale,dousti16} or QASM (pronounced {\it kazm}). QASM is a simple text language that describes generic quantum circuits. QASM can represent a completely “unrolled” quantum program whose parameters have all been specified. Most QASM variants assume a discrete set of quantum gates, but our IR is designed to control a physical system with a parameterized gate set. While we use the term ``quantum assembly language'', this is merely an analogy and should not be taken too far.

Open QASM represents universal physical circuits, so we propose a built-in gate basis of arbitrary single-qubit gates and a two-qubit entangling gate (CNOT) \cite{barenco95}. We choose a simple language without higher level programming primitives. We define different gate sets using a subroutine-like mechanism that hierarchically specifies new unitary gates in terms of built-in gates and previously defined gate subroutines. In this way, the built-in basis is used to define hardware-supported operations via standard header files. The subroutine mechanism allows limited code reuse by hierarchically defining more complex operations \cite{scaffold,dousti16}. We also add instructions that model a quantum-classical interface, specifically measurement, state reset, and the most elemental classical feedback.

The remaining sections of this document specify Open QASM and provide examples.

\section{Language}\label{sec:spec}

The syntax of the human-readable form of Open QASM has elements of C and assembly languages. The first (non-comment) line of an Open QASM program must be \code{OPENQASM M.m;} indicating a major version~M and minor version~m. Version 2.0 is described in this document. The version keyword cannot occur multiple times in a file. Statements are separated by semicolons. Whitespace is ignored. The language is case sensitive. Comments begin with a pair of forward slashes and end with a new line. The statement \code{include "filename";} continues parsing \code{filename} as if the contents of the file were pasted at the location of the \code{include} statement. The path is specified relative to the current working directory.

The only storage types of Open QASM (version 2.0) are classical and quantum registers, which are one-dimensional arrays of bits and qubits, respectively. The statement \code{qreg name[size];} declares an array of qubits (quantum register) with the given name and size. Identifiers, such as \code{name}, must start with a lowercase letter and can contain alpha-numeric characters and underscores. The label \code{name[j]} refers to a qubit of this register, where $j\in \{0,1,\dots,\mathrm{size}(\mathrm{name})-1\}$. The qubits are initialized to $|0\rangle$. Likewise, \code{creg name[size];} declares an array of bits (register) with the given name and size. The label \code{name[j]} refers to a bit of this register, where $j\in \{0,1,\dots,\mathrm{size}(\mathrm{name})-1\}$. The bits are initialized to $0$.

\begin{figure}
\begin{minipage}{.2\textwidth}
\begin{tikzpicture}[scale=1.000000,x=1pt,y=1pt]
\filldraw[color=white] (0.000000, -7.500000) rectangle (18.000000, 52.500000);
% Drawing wires
% Line 1: q0 W q[0]
\draw[color=black] (0.000000,45.000000) -- (18.000000,45.000000);
\draw[color=black] (0.000000,45.000000) node[left] {$q[0]$};
% Line 2: q1 W q[1]
\draw[color=black] (0.000000,30.000000) -- (18.000000,30.000000);
\draw[color=black] (0.000000,30.000000) node[left] {$q[1]$};
% Line 3: r0 W r[0]
\draw[color=black] (0.000000,15.000000) -- (18.000000,15.000000);
\draw[color=black] (0.000000,15.000000) node[left] {$r[0]$};
% Line 4: r1 W r[1]
\draw[color=black] (0.000000,0.000000) -- (18.000000,0.000000);
\draw[color=black] (0.000000,0.000000) node[left] {$r[1]$};
% Done with wires; drawing gates
% Line 5: q0 +r0		% CX q[0],r[0];
\draw (9.000000, 52.500000) node[text width=144pt,above,text centered] {CX q[0],r[0];};
\draw (9.000000,45.000000) -- (9.000000,15.000000);
\filldraw (9.000000, 45.000000) circle(1.500000pt);
\begin{scope}
\draw[fill=white] (9.000000, 15.000000) circle(3.000000pt);
\clip (9.000000, 15.000000) circle(3.000000pt);
\draw (6.000000, 15.000000) -- (12.000000, 15.000000);
\draw (9.000000, 12.000000) -- (9.000000, 18.000000);
\end{scope}
% Done with gates; drawing ending labels
% Done with ending labels; drawing cut lines and comments
% Done with comments
\end{tikzpicture}
\end{minipage}
\begin{minipage}{.2\textwidth}
\begin{tikzpicture}[scale=1.000000,x=1pt,y=1pt]
\filldraw[color=white] (0.000000, -7.500000) rectangle (24.000000, 52.500000);
% Drawing wires
% Line 1: q0 W q[0]
\draw[color=black] (0.000000,45.000000) -- (24.000000,45.000000);
\draw[color=black] (0.000000,45.000000) node[left] {$q[0]$};
% Line 2: q1 W q[1]
\draw[color=black] (0.000000,30.000000) -- (24.000000,30.000000);
\draw[color=black] (0.000000,30.000000) node[left] {$q[1]$};
% Line 3: r0 W r[0]
\draw[color=black] (0.000000,15.000000) -- (24.000000,15.000000);
\draw[color=black] (0.000000,15.000000) node[left] {$r[0]$};
% Line 4: r1 W r[1]
\draw[color=black] (0.000000,0.000000) -- (24.000000,0.000000);
\draw[color=black] (0.000000,0.000000) node[left] {$r[1]$};
% Done with wires; drawing gates
% Line 5: q0 +r0		% CX q,r;
\draw (9.000000, 52.500000) node[text width=144pt,above,text centered] {CX q,r;};
\draw (9.000000,45.000000) -- (9.000000,15.000000);
\filldraw (9.000000, 45.000000) circle(1.500000pt);
\begin{scope}
\draw[fill=white] (9.000000, 15.000000) circle(3.000000pt);
\clip (9.000000, 15.000000) circle(3.000000pt);
\draw (6.000000, 15.000000) -- (12.000000, 15.000000);
\draw (9.000000, 12.000000) -- (9.000000, 18.000000);
\end{scope}
% Line 6: q1 +r1
\draw (15.000000,30.000000) -- (15.000000,0.000000);
\filldraw (15.000000, 30.000000) circle(1.500000pt);
\begin{scope}
\draw[fill=white] (15.000000, 0.000000) circle(3.000000pt);
\clip (15.000000, 0.000000) circle(3.000000pt);
\draw (12.000000, 0.000000) -- (18.000000, 0.000000);
\draw (15.000000, -3.000000) -- (15.000000, 3.000000);
\end{scope}
% Done with gates; drawing ending labels
% Done with ending labels; drawing cut lines and comments
% Done with comments
\end{tikzpicture}
\end{minipage}
\begin{minipage}{.2\textwidth}
\begin{tikzpicture}[scale=1.000000,x=1pt,y=1pt]
\filldraw[color=white] (0.000000, -7.500000) rectangle (36.000000, 52.500000);
% Drawing wires
% Line 1: q0 W q[0]
\draw[color=black] (0.000000,45.000000) -- (36.000000,45.000000);
\draw[color=black] (0.000000,45.000000) node[left] {$q[0]$};
% Line 2: q1 W q[1]
\draw[color=black] (0.000000,30.000000) -- (36.000000,30.000000);
\draw[color=black] (0.000000,30.000000) node[left] {$q[1]$};
% Line 3: r0 W r[0]
\draw[color=black] (0.000000,15.000000) -- (36.000000,15.000000);
\draw[color=black] (0.000000,15.000000) node[left] {$r[0]$};
% Line 4: r1 W r[1]
\draw[color=black] (0.000000,0.000000) -- (36.000000,0.000000);
\draw[color=black] (0.000000,0.000000) node[left] {$r[1]$};
% Done with wires; drawing gates
% Line 5: q0 +r0		% CX q,r[0];
\draw (9.000000, 52.500000) node[text width=144pt,above,text centered] {CX q,r[0];};
\draw (9.000000,45.000000) -- (9.000000,15.000000);
\filldraw (9.000000, 45.000000) circle(1.500000pt);
\begin{scope}
\draw[fill=white] (9.000000, 15.000000) circle(3.000000pt);
\clip (9.000000, 15.000000) circle(3.000000pt);
\draw (6.000000, 15.000000) -- (12.000000, 15.000000);
\draw (9.000000, 12.000000) -- (9.000000, 18.000000);
\end{scope}
% Line 6: q1 +r0
\draw (27.000000,30.000000) -- (27.000000,15.000000);
\filldraw (27.000000, 30.000000) circle(1.500000pt);
\begin{scope}
\draw[fill=white] (27.000000, 15.000000) circle(3.000000pt);
\clip (27.000000, 15.000000) circle(3.000000pt);
\draw (24.000000, 15.000000) -- (30.000000, 15.000000);
\draw (27.000000, 12.000000) -- (27.000000, 18.000000);
\end{scope}
% Done with gates; drawing ending labels
% Done with ending labels; drawing cut lines and comments
% Done with comments
\end{tikzpicture}
\end{minipage}
\begin{minipage}{.2\textwidth}
\begin{tikzpicture}[scale=1.000000,x=1pt,y=1pt]
\filldraw[color=white] (0.000000, -7.500000) rectangle (36.000000, 52.500000);
% Drawing wires
% Line 1: q0 W q[0]
\draw[color=black] (0.000000,45.000000) -- (36.000000,45.000000);
\draw[color=black] (0.000000,45.000000) node[left] {$q[0]$};
% Line 2: q1 W q[1]
\draw[color=black] (0.000000,30.000000) -- (36.000000,30.000000);
\draw[color=black] (0.000000,30.000000) node[left] {$q[1]$};
% Line 3: r0 W r[0]
\draw[color=black] (0.000000,15.000000) -- (36.000000,15.000000);
\draw[color=black] (0.000000,15.000000) node[left] {$r[0]$};
% Line 4: r1 W r[1]
\draw[color=black] (0.000000,0.000000) -- (36.000000,0.000000);
\draw[color=black] (0.000000,0.000000) node[left] {$r[1]$};
% Done with wires; drawing gates
% Line 5: q0 +r0		% CX q[0],r;
\draw (9.000000, 52.500000) node[text width=144pt,above,text centered] {CX q[0],r;};
\draw (9.000000,45.000000) -- (9.000000,15.000000);
\filldraw (9.000000, 45.000000) circle(1.500000pt);
\begin{scope}
\draw[fill=white] (9.000000, 15.000000) circle(3.000000pt);
\clip (9.000000, 15.000000) circle(3.000000pt);
\draw (6.000000, 15.000000) -- (12.000000, 15.000000);
\draw (9.000000, 12.000000) -- (9.000000, 18.000000);
\end{scope}
% Line 6: q0 +r1
\draw (27.000000,45.000000) -- (27.000000,0.000000);
\filldraw (27.000000, 45.000000) circle(1.500000pt);
\begin{scope}
\draw[fill=white] (27.000000, 0.000000) circle(3.000000pt);
\clip (27.000000, 0.000000) circle(3.000000pt);
\draw (24.000000, 0.000000) -- (30.000000, 0.000000);
\draw (27.000000, -3.000000) -- (27.000000, 3.000000);
\end{scope}
% Done with gates; drawing ending labels
% Done with ending labels; drawing cut lines and comments
% Done with comments
\end{tikzpicture}
\end{minipage}
\caption{The built-in two-qubit entangling gate is the controlled-NOT gate. If \code{a} and \code{b} are qubits, the statement \code{CX a,b;} applies a controlled-NOT (CNOT) gate that flips the target qubit \code{b} iff the control qubit \code{a} is one. If \code{a} and \code{b} are quantum registers, the statement applies CNOT gates between corresponding qubits of each register. There is a similar meaning when \code{a} is a qubit and \code{b} is a quantum register and vice versa.
\label{fig:cnot}}
\end{figure}
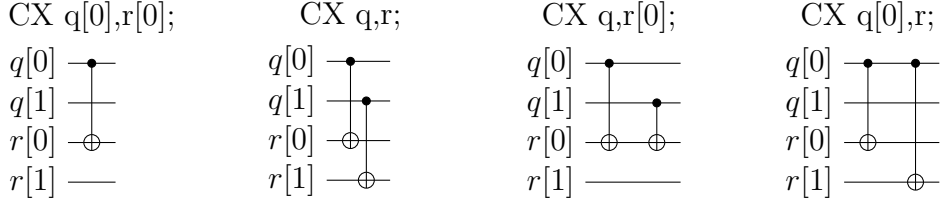

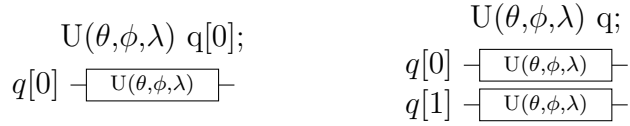
\begin{figure}
\hspace{.2\textwidth}
\begin{minipage}{.2\textwidth}
\begin{tikzpicture}[scale=1.000000,x=1pt,y=1pt]
\filldraw[color=white] (0.000000, -7.500000) rectangle (62.000000, 7.500000);
% Drawing wires
% Line 1: q0 W q[0]
\draw[color=black] (0.000000,0.000000) -- (62.000000,0.000000);
\draw[color=black] (0.000000,0.000000) node[left] {$q[0]$};
% Done with wires; drawing gates
% Line 2: q0 G {U($\theta$,$\phi$,$\lambda$)} width=50 % U($\theta$,$\phi$,$\lambda$) q[0];
\draw (31.000000, 7.500000) node[text width=144pt,above,text centered] {U($\theta$,$\phi$,$\lambda$) q[0];};
\begin{scope}
\draw[fill=white] (31.000000, -0.000000) +(-45.000000:35.355339pt and 8.485281pt) -- +(45.000000:35.355339pt and 8.485281pt) -- +(135.000000:35.355339pt and 8.485281pt) -- +(225.000000:35.355339pt and 8.485281pt) -- cycle;
\clip (31.000000, -0.000000) +(-45.000000:35.355339pt and 8.485281pt) -- +(45.000000:35.355339pt and 8.485281pt) -- +(135.000000:35.355339pt and 8.485281pt) -- +(225.000000:35.355339pt and 8.485281pt) -- cycle;
\draw (31.000000, -0.000000) node {\scriptsize{U($\theta$,$\phi$,$\lambda$)}};
\end{scope}
% Done with gates; drawing ending labels
% Done with ending labels; drawing cut lines and comments
% Done with comments
\end{tikzpicture}
\end{minipage}
\hspace{.1\textwidth}
\begin{minipage}{.2\textwidth}
\begin{tikzpicture}[scale=1.000000,x=1pt,y=1pt]
\filldraw[color=white] (0.000000, -7.500000) rectangle (62.000000, 22.500000);
% Drawing wires
% Line 1: q0 W q[0]
\draw[color=black] (0.000000,15.000000) -- (62.000000,15.000000);
\draw[color=black] (0.000000,15.000000) node[left] {$q[0]$};
% Line 2: q1 W q[1]
\draw[color=black] (0.000000,0.000000) -- (62.000000,0.000000);
\draw[color=black] (0.000000,0.000000) node[left] {$q[1]$};
% Done with wires; drawing gates
% Line 3: q0 G {U($\theta$,$\phi$,$\lambda$)} width=50 % U($\theta$,$\phi$,$\lambda$) q;
\draw (31.000000, 22.500000) node[text width=144pt,above,text centered] {U($\theta$,$\phi$,$\lambda$) q;};
\begin{scope}
\draw[fill=white] (31.000000, 15.000000) +(-45.000000:35.355339pt and 8.485281pt) -- +(45.000000:35.355339pt and 8.485281pt) -- +(135.000000:35.355339pt and 8.485281pt) -- +(225.000000:35.355339pt and 8.485281pt) -- cycle;
\clip (31.000000, 15.000000) +(-45.000000:35.355339pt and 8.485281pt) -- +(45.000000:35.355339pt and 8.485281pt) -- +(135.000000:35.355339pt and 8.485281pt) -- +(225.000000:35.355339pt and 8.485281pt) -- cycle;
\draw (31.000000, 15.000000) node {\scriptsize{U($\theta$,$\phi$,$\lambda$)}};
\end{scope}
% Line 4: q1 G {U($\theta$,$\phi$,$\lambda$)} width=50
\begin{scope}
\draw[fill=white] (31.000000, -0.000000) +(-45.000000:35.355339pt and 8.485281pt) -- +(45.000000:35.355339pt and 8.485281pt) -- +(135.000000:35.355339pt and 8.485281pt) -- +(225.000000:35.355339pt and 8.485281pt) -- cycle;
\clip (31.000000, -0.000000) +(-45.000000:35.355339pt and 8.485281pt) -- +(45.000000:35.355339pt and 8.485281pt) -- +(135.000000:35.355339pt and 8.485281pt) -- +(225.000000:35.355339pt and 8.485281pt) -- cycle;
\draw (31.000000, -0.000000) node {\scriptsize{U($\theta$,$\phi$,$\lambda$)}};
\end{scope}
% Done with gates; drawing ending labels
% Done with ending labels; drawing cut lines and comments
% Done with comments
\end{tikzpicture}
\end{minipage}
\caption{The single-qubit unitary gates are built in. These gates are parameterized by three real parameters $\theta$, $\phi$, and $\lambda$. If the argument \code{q} is a quantum register, the statement applies \code{size(q)} gates in parallel to the qubits of the register.\label{fig:utpl}}
\end{figure}

The built-in universal gate basis is ``CNOT + $U(2)$''. There is one built-in two-qubit gate (Fig.~\ref{fig:cnot})
\begin{equation}
\mathrm{CNOT} := \left(\begin{array}{cccc}
1 & 0 & 0 & 0 \\
0 & 1 & 0 & 0 \\
0 & 0 & 0 & 1 \\
0 & 0 & 1 & 0 \end{array}\right)
\end{equation}
called the controlled-NOT gate. The statement \code{CX a,b;} applies a CNOT gate that flips the target qubit \code{b} if and only if the control qubit \code{a} is one. The arguments cannot refer to the same qubit. Built-in gates have reserved uppercase keywords. If \code{a} and \code{b} are quantum registers {\em with the same size}, the statement means apply \code{CX a[j], b[j];} for each index \code{j} into register \code{a}. If instead, \code{a} is a qubit and \code{b} is a quantum register, the statement means apply \code{CX a, b[j];} for each index \code{j} into register \code{b}. Finally, if \code{a} is a quantum register and \code{b} is a qubit, the statement means apply \code{CX a[j], b;} for each index \code{j} into register \code{a}.

All of the single-qubit unitary gates are also built in (Fig.~\ref{fig:utpl}) and parameterized as
\begin{equation}
U(\theta,\phi,\lambda) := R_z(\phi)R_y(\theta)R_z(\lambda) = \left(\begin{array}{cc} e^{-i(\phi+\lambda)/2}\cos(\theta/2) & -e^{-i(\phi-\lambda)/2}\sin(\theta/2) \\
e^{i(\phi-\lambda)/2}\sin(\theta/2) & e^{i(\phi+\lambda)/2}\cos(\theta/2) \end{array}\right).
\end{equation}
Here $R_y(\theta)=\mathrm{exp}(-i\theta Y/2)$ and $R_z(\phi)=\mathrm{exp}(-i\theta Z/2)$. This specifies any element of $SU(2)$. When \code{a} is a quantum register, the statement \code{U(theta,phi,lambda) a;} means apply \code{U(theta,phi,lambda) a[j];} for each index \code{j} into register \code{a}. The real parameters $\theta\in [0,4\pi)$, $\phi\in [0,4\pi)$, and $\lambda\in [0,4\pi)$ are given by {\it parameter expressions} constructed using in-fix notation. These support scientific calculator features with arbitrary precision real numbers\footnote{Features include scientific notation; real arithmetic; logarithmic, trigonometic, and exponential functions; square roots; and the built-in constant $\pi$. The Quantum Experience uses a double precision floating point type for real numbers.}. For example, \code{U(pi/2,0,pi) q[0];} applies a Hadamard gate to qubit \code{q[0]}. Open QASM (version 2.0) does not provide a mechanism for computing parameters based on measurement outcomes.

New gates can be defined as unitary subroutines using the built-in gates, as shown in Fig.~\ref{fig:gate}. These can be viewed as macros whose expansion we defer until run-time. Gates are defined by statements of the form
\begin{Verbatim}[commandchars=\\\{\}]
\PYG{c+c1}{// comment}
\PYG{n}{gate} \PYG{n+nf}{name}\PYG{p}{(}\PYG{n}{params}\PYG{p}{)} \PYG{n}{qargs}
\PYG{p}{\PYGZob{}}
  \PYG{n}{body}
\PYG{p}{\PYGZcb{}}
\end{Verbatim}
where the optional parameter list \code{params} is a comma-separated list of variable parameter names, and the argument list \code{qargs} is a comma-separated list of qubit arguments. Both the parameter names and qubit arguments are identifiers. If there are no variable parameters, the parentheses are optional. At least one qubit argument is required. The first comment may contain documentation, such as TeX markup, to be associated with the gate. The arguments in \code{qargs} cannot be indexed within the body of the gate definition. 
\begin{Verbatim}[commandchars=\\\{\}]
\PYG{c+c1}{// this is ok:}
\PYG{n}{gate} \PYG{n}{g} \PYG{n}{a}
\PYG{p}{\PYGZob{}}
  \PYG{n}{U}\PYG{p}{(}\PYG{l+m+mi}{0}\PYG{p}{,}\PYG{l+m+mi}{0}\PYG{p}{,}\PYG{l+m+mi}{0}\PYG{p}{)} \PYG{n}{a}\PYG{p}{;}
\PYG{p}{\PYGZcb{}}
\PYG{c+c1}{// this is invalid:}
\PYG{n}{gate} \PYG{n}{g} \PYG{n}{a}
\PYG{p}{\PYGZob{}}
  \PYG{n}{U}\PYG{p}{(}\PYG{l+m+mi}{0}\PYG{p}{,}\PYG{l+m+mi}{0}\PYG{p}{,}\PYG{l+m+mi}{0}\PYG{p}{)} \PYG{n}{a}\PYG{p}{[}\PYG{l+m+mi}{0}\PYG{p}{];}
\PYG{p}{\PYGZcb{}}
\end{Verbatim}
Only built-in gate statements, calls to previously defined gates, and barrier statements can appear in \code{body}. The statements in the body can only refer to the symbols given in the parameter or argument list, and these symbols are scoped only to the subroutine body. An empty body corresponds to the identity gate. Subroutines must be declared before use and cannot call themselves. The statement \code{name(params) qargs;} applies the subroutine, and the variable parameters \code{params} are given as parameter expressions. The gate can be applied to any combination of qubits and quantum registers {\em of the same size} as shown in the following example. The quantum circuit given by
\begin{Verbatim}[commandchars=\\\{\}]
\PYG{n}{gate} \PYG{n}{g} \PYG{n}{qb0}\PYG{p}{,}\PYG{n}{qb1}\PYG{p}{,}\PYG{n}{qb2}\PYG{p}{,}\PYG{n}{qb3}
\PYG{p}{\PYGZob{}}
  \PYG{c+c1}{// body}
\PYG{p}{\PYGZcb{}}
\PYG{n}{qreg} \PYG{n}{qr0}\PYG{p}{[}\PYG{l+m+mi}{1}\PYG{p}{];}
\PYG{n}{qreg} \PYG{n}{qr1}\PYG{p}{[}\PYG{l+m+mi}{2}\PYG{p}{];}
\PYG{n}{qreg} \PYG{n}{qr2}\PYG{p}{[}\PYG{l+m+mi}{3}\PYG{p}{];}
\PYG{n}{qreg} \PYG{n}{qr3}\PYG{p}{[}\PYG{l+m+mi}{2}\PYG{p}{];}
\PYG{n}{g} \PYG{n}{qr0}\PYG{p}{[}\PYG{l+m+mi}{0}\PYG{p}{],}\PYG{n}{qr1}\PYG{p}{,}\PYG{n}{qr2}\PYG{p}{[}\PYG{l+m+mi}{0}\PYG{p}{],}\PYG{n}{qr3}\PYG{p}{;} \PYG{c+c1}{// ok}
\PYG{n}{g} \PYG{n}{qr0}\PYG{p}{[}\PYG{l+m+mi}{0}\PYG{p}{],}\PYG{n}{qr2}\PYG{p}{,}\PYG{n}{qr1}\PYG{p}{[}\PYG{l+m+mi}{0}\PYG{p}{],}\PYG{n}{qr3}\PYG{p}{;} \PYG{c+c1}{// error!}
\end{Verbatim}
has a second-to-last line that means
\begin{algorithmic}[0]
 \For{$j \gets 0, 1$}
 \State \code{g qr0[0],qr1[j],qr2[0],qr3[j];}
 \EndFor
\end{algorithmic}
We provide this so that user-defined gates can be applied in parallel like the built-in gates.

\begin{figure}
\hspace{.1\textwidth}
\begin{minipage}{.4\textwidth}
\begin{tikzpicture}[scale=1.000000,x=1pt,y=1pt]
\filldraw[color=white] (0.000000, -7.500000) rectangle (306.000000, 22.500000);
% Drawing wires
% Line 1: q0 W q[0]
\draw[color=black] (0.000000,15.000000) -- (306.000000,15.000000);
\draw[color=black] (0.000000,15.000000) node[left] {$q[0]$};
% Line 2: q1 W q[1]
\draw[color=black] (0.000000,0.000000) -- (306.000000,0.000000);
\draw[color=black] (0.000000,0.000000) node[left] {$q[1]$};
% Done with wires; drawing gates
% Line 3: q0 q1 G {cu1$\left(\frac{\pi}{2}\right)$} width=35
\draw (23.500000,15.000000) -- (23.500000,0.000000);
\begin{scope}
\draw[fill=white] (23.500000, 7.500000) +(-45.000000:24.748737pt and 19.091883pt) -- +(45.000000:24.748737pt and 19.091883pt) -- +(135.000000:24.748737pt and 19.091883pt) -- +(225.000000:24.748737pt and 19.091883pt) -- cycle;
\clip (23.500000, 7.500000) +(-45.000000:24.748737pt and 19.091883pt) -- +(45.000000:24.748737pt and 19.091883pt) -- +(135.000000:24.748737pt and 19.091883pt) -- +(225.000000:24.748737pt and 19.091883pt) -- cycle;
\draw (23.500000, 7.500000) node {{cu1$\left(\frac{\pi}{2}\right)$}};
\end{scope}
% Line 4: =
\draw[fill=white,color=white] (53.000000, -6.000000) rectangle (68.000000, 21.000000);
\draw (60.500000, 7.500000) node {$=$};
% Line 5: q0 G {U(0,0,$\frac{\pi}{4}$)} width=50
\begin{scope}
\draw[fill=white] (105.000000, 15.000000) +(-45.000000:35.355339pt and 8.485281pt) -- +(45.000000:35.355339pt and 8.485281pt) -- +(135.000000:35.355339pt and 8.485281pt) -- +(225.000000:35.355339pt and 8.485281pt) -- cycle;
\clip (105.000000, 15.000000) +(-45.000000:35.355339pt and 8.485281pt) -- +(45.000000:35.355339pt and 8.485281pt) -- +(135.000000:35.355339pt and 8.485281pt) -- +(225.000000:35.355339pt and 8.485281pt) -- cycle;
\draw (105.000000, 15.000000) node {\scriptsize{U(0,0,$\frac{\pi}{4}$)}};
\end{scope}
% Line 6: q0 +q1
\draw (145.000000,15.000000) -- (145.000000,0.000000);
\filldraw (145.000000, 15.000000) circle(1.500000pt);
\begin{scope}
\draw[fill=white] (145.000000, 0.000000) circle(3.000000pt);
\clip (145.000000, 0.000000) circle(3.000000pt);
\draw (142.000000, 0.000000) -- (148.000000, 0.000000);
\draw (145.000000, -3.000000) -- (145.000000, 3.000000);
\end{scope}
% Line 7: q1 G {U(0,0,$-\frac{\pi}{4}$)} width=60
\begin{scope}
\draw[fill=white] (190.000000, -0.000000) +(-45.000000:42.426407pt and 8.485281pt) -- +(45.000000:42.426407pt and 8.485281pt) -- +(135.000000:42.426407pt and 8.485281pt) -- +(225.000000:42.426407pt and 8.485281pt) -- cycle;
\clip (190.000000, -0.000000) +(-45.000000:42.426407pt and 8.485281pt) -- +(45.000000:42.426407pt and 8.485281pt) -- +(135.000000:42.426407pt and 8.485281pt) -- +(225.000000:42.426407pt and 8.485281pt) -- cycle;
\draw (190.000000, -0.000000) node {\scriptsize{U(0,0,$-\frac{\pi}{4}$)}};
\end{scope}
% Line 8: q0 +q1
\draw (235.000000,15.000000) -- (235.000000,0.000000);
\filldraw (235.000000, 15.000000) circle(1.500000pt);
\begin{scope}
\draw[fill=white] (235.000000, 0.000000) circle(3.000000pt);
\clip (235.000000, 0.000000) circle(3.000000pt);
\draw (232.000000, 0.000000) -- (238.000000, 0.000000);
\draw (235.000000, -3.000000) -- (235.000000, 3.000000);
\end{scope}
% Line 9: q1 G {U(0,0,$\frac{\pi}{4}$)} width=50
\begin{scope}
\draw[fill=white] (275.000000, -0.000000) +(-45.000000:35.355339pt and 8.485281pt) -- +(45.000000:35.355339pt and 8.485281pt) -- +(135.000000:35.355339pt and 8.485281pt) -- +(225.000000:35.355339pt and 8.485281pt) -- cycle;
\clip (275.000000, -0.000000) +(-45.000000:35.355339pt and 8.485281pt) -- +(45.000000:35.355339pt and 8.485281pt) -- +(135.000000:35.355339pt and 8.485281pt) -- +(225.000000:35.355339pt and 8.485281pt) -- cycle;
\draw (275.000000, -0.000000) node {\scriptsize{U(0,0,$\frac{\pi}{4}$)}};
\end{scope}
% Done with gates; drawing ending labels
% Done with ending labels; drawing cut lines and comments
% Done with comments
\end{tikzpicture}
\end{minipage}

\hspace{.3\textwidth}
\begin{minipage}{.3\textwidth}
\begin{Verbatim}[commandchars=\\\{\}]
\PYG{n}{gate} \PYG{n+nf}{cu1}\PYG{p}{(}\PYG{n}{lambda}\PYG{p}{)} \PYG{n}{a}\PYG{p}{,}\PYG{n}{b}
\PYG{p}{\PYGZob{}}
  \PYG{n}{U}\PYG{p}{(}\PYG{l+m+mi}{0}\PYG{p}{,}\PYG{l+m+mi}{0}\PYG{p}{,}\PYG{n}{theta}\PYG{o}{/}\PYG{l+m+mi}{2}\PYG{p}{)} \PYG{n}{a}\PYG{p}{;}
  \PYG{n}{CX} \PYG{n}{a}\PYG{p}{,}\PYG{n}{b}\PYG{p}{;}
  \PYG{n}{U}\PYG{p}{(}\PYG{l+m+mi}{0}\PYG{p}{,}\PYG{l+m+mi}{0}\PYG{p}{,}\PYG{o}{\PYGZhy{}}\PYG{n}{theta}\PYG{o}{/}\PYG{l+m+mi}{2}\PYG{p}{)} \PYG{n}{b}\PYG{p}{;}
  \PYG{n}{CX} \PYG{n}{a}\PYG{p}{,}\PYG{n}{b}\PYG{p}{;}
  \PYG{n}{U}\PYG{p}{(}\PYG{l+m+mi}{0}\PYG{p}{,}\PYG{l+m+mi}{0}\PYG{p}{,}\PYG{n}{theta}\PYG{o}{/}\PYG{l+m+mi}{2}\PYG{p}{)} \PYG{n}{b}\PYG{p}{;}
\PYG{p}{\PYGZcb{}}
\PYG{n}{cu1}\PYG{p}{(}\PYG{n}{pi}\PYG{o}{/}\PYG{l+m+mi}{2}\PYG{p}{)} \PYG{n}{q}\PYG{p}{[}\PYG{l+m+mi}{0}\PYG{p}{],}\PYG{n}{q}\PYG{p}{[}\PYG{l+m+mi}{1}\PYG{p}{];}
\end{Verbatim}
\end{minipage}

\caption{New gates are defined as unitary subroutines. The gates are applied using the statement \code{name(params) qargs;} just like the built-in gates. The parentheses are optional if there are no parameters. The gate $\mathrm{cu1}(\theta)$ corresponds to the unitary matrix $\mathrm{diag}(1,1,1,e^{i\theta})$ up to a global phase.
\label{fig:gate}}
\end{figure}

To support gates whose physical implementation may be possible, but whose definition is unspecified, we provide an ``opaque'' gate declaration. This may be used in practice in several instances. For example, the system may evolve under some fixed but uncharacterized drift Hamiltonian for some fixed amount of time. The system might be subject to an $n$-qubit operator whose parameters are computationally challenging to estimate. The syntax for an opaque gate declaration is the same as a gate declaration but without a body.

Measurement is shown in Fig.~\ref{fig:measure}. The statement \code{measure qubit|qreg -> bit|creg;} measures the qubit(s) in the $Z$-basis and records the measurement outcome(s) by overwriting the bit(s). Measurement corresponds to a projection onto one of the eigenstates of $Z$, and qubit(s) are immediately available for further quantum computation. Both arguments must be register-type, or both must be bit-type. If both arguments are register-type and have the same size, the statement \code{measure a -> b;} means apply \code{measure a[j] -> b[j];} for each index \code{j} into register \code{a}.

The \code{reset qubit|qreg;} statement resets a qubit or quantum register to the state $|0\rangle$. This corresponds to a partial trace over those qubits (i.e. discarding them) before replacing them with $|0\rangle\langle 0|$, as shown in Fig.~\ref{fig:prepare}.

\begin{figure}
\begin{minipage}{.3\textwidth}
\centering\begin{tikzpicture}[scale=1.000000,x=1pt,y=1pt]
\filldraw[color=white] (0.000000, -7.500000) rectangle (24.000000, 22.500000);
% Drawing wires
% Line 2: q0 W q[0]
\draw[color=black] (0.000000,15.000000) -- (12.000000,15.000000);
\draw[color=black] (12.000000,15.00000) -- (24.000000,15.00000);
% \draw[color=black] (12.000000,15.500000) -- (24.000000,15.500000);
\draw[color=black] (0.000000,15.000000) node[left] {$q[0]$};
% Line 3: c0 W c[0]
\draw[color=black] (0.000000,0.000000) -- (24.000000,0.000000);
\draw[color=black] (0.000000,-1.5000000) -- (24.000000,-1.5000000);
\draw[color=black] (0.000000,0.000000) node[left] {$c[0]$};
% Done with wires; drawing gates
% Line 4: q0:cwire +c0 %% or
%\draw (30.000000, 13.00000) node[text width=144pt,below,text centered] {\scriptsize or};
% \draw (11.500000,15.000000) -- (11.500000,0.000000);
 \draw (12.00000,15.000000) -- (12.00000,0.000000);
%\filldraw (12.000000, 15.000000) circle(1.500000pt);
\begin{scope}
%\draw[fill=white] (12.000000, 0.000000) circle(3.000000pt);
 \clip (12.000000, 0.000000) circle(3.000000pt);
\draw (9.000000, 0.000000) -- (15.000000, 0.000000);
% \draw (12.000000, -3.000000) -- (12.000000, 3.000000);
\end{scope}
\draw[fill=white] (6.000000, 9.000000) rectangle (18.000000, 21.000000);
\draw[very thin] (12.000000, 15.600000) arc (90:150:6.000000pt);
\draw[very thin] (12.000000, 15.600000) arc (90:30:6.000000pt);
\draw[->,>=stealth] (12.000000, 9.600000) -- +(80:10.392305pt);
\draw[->,>=stealth] (12.000000, 10.00000)  -- +(270:10.392305pt); % arrowhead
% Done with gates; drawing ending labels
% Done with ending labels; drawing cut lines and comments
% Done with comments
\end{tikzpicture}
\end{minipage}
\begin{minipage}{.3\textwidth}
\centering\begin{tikzpicture}[scale=1.000000,x=1pt,y=1pt]
\filldraw[color=white] (0.000000, -7.500000) rectangle (24.000000, 7.500000);
% Drawing wires
% Line 2: q0 W q[0] c[0]
\draw[color=black] (0.000000,0.000000) -- (12.000000,0.000000);
\draw[color=black] (12.000000,-0.500000) -- (24.000000,-0.500000);
\draw[color=black] (12.000000,0.500000) -- (24.000000,0.500000);
\draw[color=black] (0.000000,0.000000) node[left] {$q[0]$};
% Done with wires; drawing gates
% Line 3: q0 M %% as the \\ final op. \\ on q[0]
%\draw (12.000000, -7.500000) node[text width=120pt,below,text centered] {\scriptsize as the \\ final op. \\ on q[0]};
\draw[fill=white] (6.000000, -6.000000) rectangle (18.000000, 6.000000);
\draw[very thin] (12.000000, 0.600000) arc (90:150:6.000000pt);
\draw[very thin] (12.000000, 0.600000) arc (90:30:6.000000pt);
\draw[->,>=stealth] (12.000000, -5.400000) -- +(80:10.392305pt);
% Done with gates; drawing ending labels
\draw[color=black] (24.000000,0.000000) node[right] {$c[0]$};
%\draw (55.000000, 5.00000) node[text width=144pt,below,text centered] {\scriptsize or};
% Done with ending labels; drawing cut lines and comments
% Done with comments
\end{tikzpicture}
\end{minipage}
\begin{minipage}{.3\textwidth}
\centering\begin{tikzpicture}[scale=1.000000,x=1pt,y=1pt]
\filldraw[color=white] (0.000000, -7.500000) rectangle (44.000000, 22.500000);
% Drawing wires
% Line 2: q0 W q[0]
\draw[color=black] (0.000000,15.000000) -- (32.000000,15.000000);
\draw[color=black] (32.000000,15.00000) -- (44.000000,15.00000);
%\draw[color=black] (32.000000,15.500000) -- (44.000000,15.500000);
\draw[color=black] (0.000000,15.000000) node[left] {$q[0]$};
% Line 3: c W c
\draw[color=black] (0.000000,0.000000) -- (44.000000,0.000000);
\draw[color=black] (0.000000,0.000000) node[left] {$c$};
% Done with wires; drawing gates
% Line 4: c / 2
\draw (6.000000, -6.000000) -- (14.000000, 6.000000);
\draw (10.000000, 3.000000) node[right] {$\scriptstyle{2}$};
% Line 5: q0:cwire +c %% 0
\draw (28.000000, 10.00000) node[text width=144pt,below,text centered] {\scriptsize 0};
\draw (32.00000,15.000000) -- (32.00000,0.000000);
%\draw (32.500000,15.000000) -- (32.500000,0.000000);
\filldraw (32.000000, 15.000000) circle(1.500000pt);
\begin{scope}
% \draw[fill=white] (32.000000, 0.000000) circle(3.000000pt);
\clip (32.000000, 0.000000) circle(3.000000pt);
\draw (29.000000, 0.000000) -- (35.000000, 0.000000);
%\draw (32.000000, -3.000000) -- (32.000000, 3.000000);
\end{scope}
\draw[fill=white] (26.000000, 9.000000) rectangle (38.000000, 21.000000);
\draw[very thin] (32.000000, 15.600000) arc (90:150:6.000000pt);
\draw[very thin] (32.000000, 15.600000) arc (90:30:6.000000pt);
\draw[->,>=stealth] (32.000000, 9.600000) -- +(80:10.392305pt);
\draw[->,>=stealth] (32.000000, 10.00000)  -- +(270:10.392305pt); % arrowhead
\draw[color=black] (0.000000,-1.5000000) -- (44.000000,-1.5000000);
% Done with gates; drawing ending labels
% Done with ending labels; drawing cut lines and comments
% Done with comments
\end{tikzpicture}
\end{minipage}

\begin{minipage}{.3\textwidth}
\centering\begin{tikzpicture}[scale=1.000000,x=1pt,y=1pt]
\filldraw[color=white] (0.000000, -7.500000) rectangle (64.000000, 22.500000);
% Drawing wires
% Line 2: q W q[0]
\draw[color=black] (0.000000,15.000000) -- (32.000000,15.000000);
\draw[color=black] (32.000000,15.00000) -- (64.000000,15.00000);
% \draw[color=black] (32.000000,15.500000) -- (64.000000,15.500000);
\draw[color=black] (0.000000,15.000000) node[left] {$q[0]$};
% Line 3: c W c[0]
\draw[color=black] (0.000000,0.000000) -- (64.000000,0.000000);
\draw[color=black] (0.000000,0.000000) node[left] {$c[0]$};
% Done with wires; drawing gates
% Line 4: q / 2
\draw (6.000000, 9.000000) -- (14.000000, 21.000000);
\draw (12.000000, 18.000000) node[right] {$\scriptstyle{2}$};
% Line 5: c / 2
\draw (6.000000, -6.000000) -- (14.000000, 6.000000);
\draw (12.000000, 3.000000) node[right] {$\scriptstyle{2}$};
% Line 6: q:cwire +c
%\draw (31.500000,15.000000) -- (31.500000,0.000000);
\draw (32.00000,15.000000) -- (32.00000,0.000000);
\filldraw (32.000000, 15.000000) circle(1.500000pt);
\begin{scope}
%\draw[fill=white] (32.000000, 0.000000) circle(3.000000pt);
\clip (32.000000, 0.000000) circle(3.000000pt);
\draw (29.000000, 0.000000) -- (35.000000, 0.000000);
% \draw (32.000000, -3.000000) -- (32.000000, 3.000000);
\end{scope}
\draw[fill=white] (26.000000, 9.000000) rectangle (38.000000, 21.000000);
\draw[very thin] (32.000000, 15.600000) arc (90:150:6.000000pt);
\draw[very thin] (32.000000, 15.600000) arc (90:30:6.000000pt);
\draw[->,>=stealth] (32.000000, 9.600000) -- +(80:10.392305pt);
% Line 7: q / 2
\draw (50.000000, 9.000000) -- (58.000000, 21.000000);
\draw (56.000000, 18.000000) node[right] {$\scriptstyle{2}$};
% Line 8: c / 2
\draw (50.000000, -6.000000) -- (58.000000, 6.000000);
\draw (56.000000, 3.000000) node[right] {$\scriptstyle{2}$};
%\draw (72.000000, 13.00000) node[text width=144pt,below,text centered] {\scriptsize or};
\draw[->,>=stealth] (32.000000, 10.00000)  -- +(270:10.392305pt); % arrowhead
\draw[color=black] (0.000000,-1.5000000) -- (64.000000,-1.5000000);
% Done with gates; drawing ending labels
% Done with ending labels; drawing cut lines and comments
% Done with comments
\end{tikzpicture}
\end{minipage}
\begin{minipage}{.3\textwidth}
\centering\begin{tikzpicture}[scale=1.000000,x=1pt,y=1pt]
\filldraw[color=white] (0.000000, -7.500000) rectangle (36.000000, 52.500000);
% Drawing wires
% Line 1: q0 W q[0]
\draw[color=black] (0.000000,45.000000) -- (12.000000,45.000000);
%\draw[color=black] (12.000000,44.500000) -- (36.000000,44.500000);
\draw[color=black] (12.000000,45.00000) -- (36.000000,45.00000);
\draw[color=black] (0.000000,45.000000) node[left] {$q[0]$};
% Line 2: q1 W q[1]
\draw[color=black] (0.000000,30.000000) -- (24.000000,30.000000);
%\draw[color=black] (24.000000,29.500000) -- (36.000000,29.500000);
\draw[color=black] (24.000000,30.500000) -- (36.000000,30.500000);
\draw[color=black] (0.000000,30.000000) node[left] {$q[1]$};
% Line 3: c0 W c[0]
\draw[color=black] (0.000000,15.000000) -- (36.000000,15.000000);
\draw[color=black] (0.000000,15.000000) node[left] {$c[0]$};
\draw[color=black] (0.000000,13.5000000) -- (36.000000,13.5000000);
% Line 4: c1 W c[1]
\draw[color=black] (0.000000,0.000000) -- (36.000000,0.000000);
\draw[color=black] (0.000000,0.000000) node[left] {$c[1]$};
% Done with wires; drawing gates
% Line 5: q0:cwire +c0
%\draw (11.500000,45.000000) -- (11.500000,15.000000);
\draw (12.00000,45.000000) -- (12.00000,25.000000);
\filldraw (12.000000, 45.000000) circle(1.500000pt);
\begin{scope}
%\draw[fill=white] (12.000000, 15.000000) circle(3.000000pt);
\clip (12.000000, 15.000000) circle(3.000000pt);
\draw (9.000000, 15.000000) -- (15.000000, 15.000000);
%\draw (12.000000, 12.000000) -- (12.000000, 18.000000);
\end{scope}
\draw[fill=white] (6.000000, 39.000000) rectangle (18.000000, 51.000000);
\draw[very thin] (12.000000, 45.600000) arc (90:150:6.000000pt);
\draw[very thin] (12.000000, 45.600000) arc (90:30:6.000000pt);
\draw[->,>=stealth] (12.000000, 39.600000) -- +(80:10.392305pt);
% Line 6: q1:cwire +c1
%\draw (23.500000,30.000000) -- (23.500000,0.000000);
\draw (24.00000,30.000000) -- (24.00000,0.000000);
\filldraw (24.000000, 30.000000) circle(1.500000pt);
\begin{scope}
%\draw[fill=white] (24.000000, 0.000000) circle(3.000000pt);
\clip (24.000000, 0.000000) circle(3.000000pt);
\draw (21.000000, 0.000000) -- (27.000000, 0.000000);
%\draw (24.000000, -3.000000) -- (24.000000, 3.000000);
\end{scope}
\draw[fill=white] (18.000000, 24.000000) rectangle (30.000000, 36.000000);
\draw[very thin] (24.000000, 30.600000) arc (90:150:6.000000pt);
\draw[very thin] (24.000000, 30.600000) arc (90:30:6.000000pt);
\draw[->,>=stealth] (24.000000, 24.600000) -- +(80:10.392305pt);
\draw[->,>=stealth] (24.000000, 10.00000)  -- +(270:10.392305pt); % arrowhead
\draw[->,>=stealth] (12.000000, 25.00000)  -- +(270:10.392305pt); % arrowhead
\draw[color=black] (0.000000,-1.5000000) -- (36.000000,-1.5000000);
%\draw (42.000000, 25.00000) node[text width=144pt,below,text centered] {\scriptsize or};
% Done with gates; drawing ending labels
% Done with ending labels; drawing cut lines and comments
% Done with comments
\end{tikzpicture}
\end{minipage}
\begin{minipage}{.3\textwidth}
\centering\begin{tikzpicture}[scale=1.000000,x=1pt,y=1pt]
\filldraw[color=white] (0.000000, -7.500000) rectangle (68.000000, 37.500000);
% Drawing wires
% Line 1: q0 W q0
\draw[color=black] (0.000000,30.000000) -- (32.000000,30.000000);
%\draw[color=black] (32.000000,29.500000) -- (68.000000,29.500000);
\draw[color=black] (32.000000,30.00000) -- (68.000000,30.00000);
\draw[color=black] (0.000000,30.000000) node[left] {$q[0]$};
% Line 2: q1 W q1
\draw[color=black] (0.000000,15.000000) -- (56.000000,15.000000);
%\draw[color=black] (56.000000,14.500000) -- (68.000000,14.500000);
\draw[color=black] (56.000000,15.00000) -- (68.000000,15.00000);
\draw[color=black] (0.000000,15.000000) node[left] {$q[1]$};
% Line 3: c W c
\draw[color=black] (0.000000,0.000000) -- (68.000000,0.000000);
\draw[color=black] (0.000000,0.000000) node[left] {$c$};
% Done with wires; drawing gates
% Line 4: c / 2
\draw (6.000000, -6.000000) -- (14.000000, 6.000000);
\draw (11.000000, 3.000000) node[right] {$\scriptstyle{2}$};
% Line 5: q0:cwire +c %% 0
\draw (28.000000, 10.00000) node[text width=144pt,below,text centered] {\scriptsize 0};
%\draw (31.500000,30.000000) -- (31.500000,0.000000);
\draw (32.00000,30.000000) -- (32.00000,0.000000);
\filldraw (32.000000, 30.000000) circle(1.500000pt);
\begin{scope}
%\draw[fill=white] (32.000000, 0.000000) circle(3.000000pt);
\clip (32.000000, 0.000000) circle(3.000000pt);
\draw (29.000000, 0.000000) -- (35.000000, 0.000000);
%\draw (32.000000, -3.000000) -- (32.000000, 3.000000);
\end{scope}
\draw[fill=white] (26.000000, 24.000000) rectangle (38.000000, 36.000000);
\draw[very thin] (32.000000, 30.600000) arc (90:150:6.000000pt);
\draw[very thin] (32.000000, 30.600000) arc (90:30:6.000000pt);
\draw[->,>=stealth] (32.000000, 24.600000) -- +(80:10.392305pt);
% Line 6: q1:cwire +c %% 1
\draw (53.000000, 10.00000) node[text width=144pt,below,text centered] {\scriptsize 1};
%\draw (55.500000,15.000000) -- (55.500000,0.000000);
\draw (56.00000,15.000000) -- (56.00000,0.000000);
\filldraw (56.000000, 15.000000) circle(1.500000pt);
\begin{scope}
%\draw[fill=white] (56.000000, 0.000000) circle(3.000000pt);
\clip (56.000000, 0.000000) circle(3.000000pt);
\draw (53.000000, 0.000000) -- (59.000000, 0.000000);
%\draw (56.000000, -3.000000) -- (56.000000, 3.000000);
\end{scope}
\draw[fill=white] (50.000000, 9.000000) rectangle (62.000000, 21.000000);
\draw[very thin] (56.000000, 15.600000) arc (90:150:6.000000pt);
\draw[very thin] (56.000000, 15.600000) arc (90:30:6.000000pt);
\draw[->,>=stealth] (56.000000, 9.600000) -- +(80:10.392305pt);
\draw[color=black] (0.000000,-1.5000000) -- (68.000000,-1.5000000);
\draw[->,>=stealth] (56.000000, 10.00000)  -- +(270:10.392305pt); % arrowhead
\draw[->,>=stealth] (32.000000, 10.00000)  -- +(270:10.392305pt); % arrowhead
% Done with gates; drawing ending labels
% Done with ending labels; drawing cut lines and comments
% Done with comments
\end{tikzpicture}
\end{minipage}
\caption{The \code{measure} statement projectively measures a qubit or each qubit of a quantum register. The measurement projects onto the $Z$-basis and leaves qubits available for further operations. The top row of circuits depicts single-qubit measurement using the statement \code{measure q[0] -> c[0];} while the bottom depicts measurement of an entire register using the statement \code{measure q -> c;}. The center circuit of the top row depicts measurement as the final operation on \code{q[0]}.
\label{fig:measure}}
\end{figure}
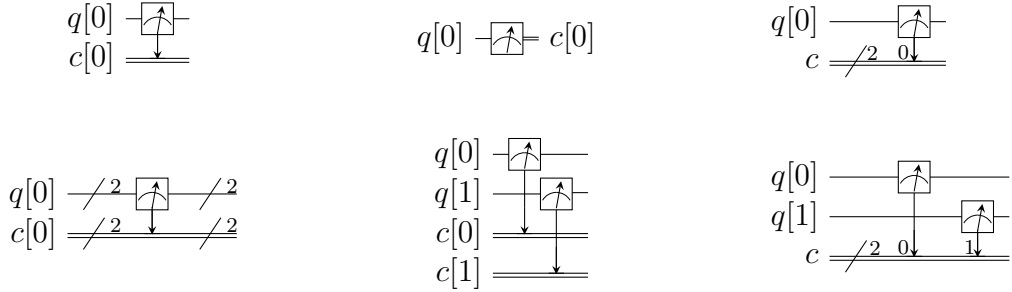

\begin{figure}
\begin{minipage}{.3\textwidth}
\centering
\begin{tikzpicture}[scale=1.000000,x=1pt,y=1pt]
\filldraw[color=white] (0.000000, -7.500000) rectangle (36.000000, 7.500000);
% Drawing wires
% Line 1: q0 W q[0]
\draw[color=black] (0.000000,0.000000) -- (9.000000,0.000000);
\draw[color=black] (34.000000,0.000000) -- (40.000000,0.000000);
\draw[color=black] (0.000000,0.000000) node[left] {$q[0]$};
% Done with wires; drawing gates
\draw (6.000000, 7.000000) node[text width=144pt,above,text centered] {reset q[0];};
% Line 2: q0 OUT 0
\filldraw[color=white] (6.000000, -3.000000) rectangle (12.000000, 3.000000);
\draw (6.000000, -3.000000) -- (6.000000, 3.000000);
%\draw (9.000000, 0.000000) node {$\scriptstyle{0}$};
% Line 3: q0 IN \ket{0}
\filldraw[color=white] (24.000000, -3.000000) rectangle (30.000000, 3.000000);
%\draw (30.000000, -3.000000) -- (30.000000, 3.000000);
\draw (27.000000, 0.000000) node {$\scriptstyle{\ket{0}}$};
% Done with gates; drawing ending labels
% Done with ending labels; drawing cut lines and comments
% Done with comments
\end{tikzpicture}
\end{minipage}
\begin{minipage}{.3\textwidth}
\centering\begin{tikzpicture}[scale=1.000000,x=1pt,y=1pt]
\filldraw[color=white] (0.000000, -7.500000) rectangle (36.000000, 22.500000);
% Drawing wires
% Line 1: q0 W q[0]
\draw[color=black] (0.000000,15.000000) -- (9.000000,15.000000);
\draw[color=black] (34.000000,15.000000) -- (40.000000,15.000000);
\draw[color=black] (0.000000,15.000000) node[left] {$q[0]$};
% Line 2: q1 W q[1]
\draw[color=black] (0.000000,0.000000) -- (9.000000,0.000000);
\draw[color=black] (34.000000,0.000000) -- (40.000000,0.000000);
\draw[color=black] (0.000000,0.000000) node[left] {$q[1]$};
% Done with wires; drawing gates
\draw (6.000000, 25.000000) node[text width=144pt,above,text centered] {reset q;};
% Line 3: q0 OUT 0
\filldraw[color=white] (6.000000, 12.000000) rectangle (12.000000, 18.000000);
\draw (6.000000, 12.000000) -- (6.000000, 18.000000);
%\draw (9.000000, 15.000000) node {$\scriptstyle{0}$};
% Line 4: q1 OUT 0
\filldraw[color=white] (6.000000, -3.000000) rectangle (12.000000, 3.000000);
\draw (6.000000, -3.000000) -- (6.000000, 3.000000);
% \draw (9.000000, 0.000000) node {$\scriptstyle{0}$};
% Line 5: q0 IN \ket{0}
\filldraw[color=white] (24.000000, 12.000000) rectangle (30.000000, 18.000000);
%\draw (30.000000, 12.000000) -- (30.000000, 18.000000);
\draw (27.000000, 15.000000) node {$\scriptstyle{\ket{0}}$};
% Line 6: q1 IN \ket{0}
\filldraw[color=white] (24.000000, -3.000000) rectangle (30.000000, 3.000000);
%\draw (30.000000, -3.000000) -- (30.000000, 3.000000);
\draw (27.000000, 0.000000) node {$\scriptstyle{\ket{0}}$};
% Done with gates; drawing ending labels
% Done with ending labels; drawing cut lines and comments
% Done with comments
\end{tikzpicture}
\end{minipage}
\begin{minipage}{.3\textwidth}
\centering\begin{tikzpicture}[scale=1.000000,x=1pt,y=1pt]
\filldraw[color=white] (0.000000, -7.500000) rectangle (76.000000, 7.500000);
% Drawing wires
% Line 1: q W q
\draw[color=black] (0.000000,0.000000) -- (29.000000,0.000000);
\draw[color=black] (53.000000,0.000000) -- (76.000000,0.000000);
\draw[color=black] (0.000000,0.000000) node[left] {$q$};
% Done with wires; drawing gates
\draw (25.000000, 12.000000) node[text width=144pt,above,text centered] {reset q;};
% Line 2: q / 2
\draw (6.000000, -6.000000) -- (14.000000, 6.000000);
\draw (12.000000, 3.000000) node[right] {$\scriptstyle{2}$};
% Line 3: q OUT 0
\filldraw[color=white] (26.000000, -3.000000) rectangle (32.000000, 3.000000);
\draw (26.000000, -3.000000) -- (26.000000, 3.000000);
% \draw (29.000000, 0.000000) node {$\scriptstyle{0}$};
% Line 4: q IN \ket{0}
\filldraw[color=white] (44.000000, -3.000000) rectangle (50.000000, 3.000000);
% \draw (50.000000, -3.000000) -- (50.000000, 3.000000);
\draw (47.000000, 0.000000) node {$\scriptstyle{\ket{0}}$};
% Line 5: q / 2
\draw (62.000000, -6.000000) -- (70.000000, 6.000000);
\draw (68.000000, 3.000000) node[right] {$\scriptstyle{2}$};
% Done with gates; drawing ending labels
% Done with ending labels; drawing cut lines and comments
% Done with comments
\end{tikzpicture}
\end{minipage}
\caption{The \code{reset} statement prepares a qubit or quantum register in the state $|0\rangle$.
\label{fig:prepare}}
\end{figure}
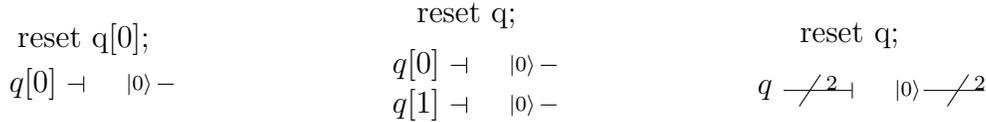

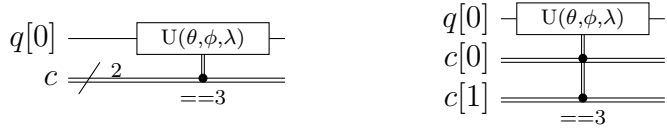
\begin{figure}
\centering
\begin{minipage}{.3\textwidth}
\centering
\begin{tikzpicture}[scale=1.000000,x=1pt,y=1pt]
\filldraw[color=white] (0.000000, -7.500000) rectangle (82.000000, 22.500000);
% Drawing wires
% Line 1: q0 W q[0]
\draw[color=black] (0.000000,15.000000) -- (82.000000,15.000000);
\draw[color=black] (0.000000,15.000000) node[left] {$q[0]$};
% Line 2: c W c
\draw[color=black] (0.000000,0.000000) -- (82.000000,0.000000);
\draw[color=black] (0.000000,0.000000) node[left] {$c$};
% Done with wires; drawing gates
% Line 3: c / 2 %% ==3
\draw (51.000000, -.500000) node[text width=144pt,below,text centered]{\scriptsize ==3};
%\draw (6.000000, -6.000000) -- (14.000000, 6.000000);
\draw (12.000000, 3.000000) node[right] {$\scriptstyle{2}$};
% Line 4: q0 G {u($\theta$,$\phi$,$\lambda$)} c  width=50
\draw (50.5000000,15.000000) -- (50.5000000,0.000000);
\draw (51.5000000,15.000000) -- (51.5000000,0.000000);
\begin{scope}
\draw[fill=white] (51.000000, 15.000000) +(-45.000000:35.355339pt and 8.485281pt) -- +(45.000000:35.355339pt and 8.485281pt) -- +(135.000000:35.355339pt and 8.485281pt) -- +(225.000000:35.355339pt and 8.485281pt) -- cycle;
\clip (51.000000, 15.000000) +(-45.000000:35.355339pt and 8.485281pt) -- +(45.000000:35.355339pt and 8.485281pt) -- +(135.000000:35.355339pt and 8.485281pt) -- +(225.000000:35.355339pt and 8.485281pt) -- cycle;
\draw (51.000000, 15.000000) node {\scriptsize {U($\theta$,$\phi$,$\lambda$)}};
\end{scope}
\filldraw (51.000000, 0.000000) circle(1.500000pt);
\draw[color=black] (0.000000,-1.5000000) -- (82.000000,-1.5000000);
\draw (4.000000, -6.000000) -- (12.000000, 6.000000);
% Done with gates; drawing ending labels
% Done with ending labels; drawing cut lines and comments
% Done with comments
\end{tikzpicture}
\end{minipage}
\begin{minipage}{.3\textwidth}
\centering\begin{tikzpicture}[scale=1.000000,x=1pt,y=1pt]
\filldraw[color=white] (0.000000, -7.500000) rectangle (62.000000, 37.500000);
% Drawing wires
% Line 2: q0 W q[0]
\draw[color=black] (0.000000,30.000000) -- (62.000000,30.000000);
\draw[color=black] (0.000000,30.000000) node[left] {$q[0]$};
% Line 3: c0 W c[0]
\draw[color=black] (0.000000,15.000000) -- (62.000000,15.000000);
\draw[color=black] (0.000000,15.000000) node[left] {$c[0]$};
% Line 4: c1 W c[1]
\draw[color=black] (0.000000,0.000000) -- (62.000000,0.000000);
\draw[color=black] (0.000000,0.000000) node[left] {$c[1]$};
% Done with wires; drawing gates
% Line 5: q0 G {u($\theta$,$\phi$,$\lambda$)} c1  width=50 %% ==3
\draw (31.000000, -.500000) node[text width=144pt,below,text centered] {\scriptsize ==3};
\draw (30.5000000,30.000000) -- (30.5000000,0.000000);
\draw (31.5000000,30.000000) -- (31.5000000,0.000000);
\filldraw (31.000000, 15.000000) circle(1.500000pt); % extra dot
\begin{scope} 
\draw[fill=white] (31.000000, 30.000000) +(-45.000000:35.355339pt and 8.485281pt) -- +(45.000000:35.355339pt and 8.485281pt) -- +(135.000000:35.355339pt and 8.485281pt) -- +(225.000000:35.355339pt and 8.485281pt) -- cycle;
\clip (31.000000, 30.000000) +(-45.000000:35.355339pt and 8.485281pt) -- +(45.000000:35.355339pt and 8.485281pt) -- +(135.000000:35.355339pt and 8.485281pt) -- +(225.000000:35.355339pt and 8.485281pt) -- cycle;
\draw (31.000000, 30.000000) node {\scriptsize{U($\theta$,$\phi$,$\lambda$)}};
\end{scope}
\filldraw (31.000000, 0.000000) circle(1.500000pt);
\draw[color=black] (0.000000,13.5000000) -- (62.000000,13.5000000);
\draw[color=black] (0.000000,-1.5000000) -- (62.000000,-1.5000000);
% Done with gates; drawing ending labels
% Done with ending labels; drawing cut lines and comments
% Done with comments
\end{tikzpicture}
\end{minipage}
\caption{The \code{if} statement applies a quantum operation only if a classical register has the indicated integer value. These circuits depict the statement \code{if(c==3) U(theta,phi,lambda) q[0];}.
\label{fig:if}}
\end{figure}

There is one type of classically-controlled quantum operation: the \code{if} statement shown in Fig.~\ref{fig:if}. The \code{if} statement conditionally executes a quantum operation based on the value of a classical register. This allows measurement outcomes to determine future quantum operations. We choose to have one decision register for simplicity. This register is interpreted as an integer, using the bit at index zero as the low order bit. The quantum operation executes only if the register has the given integer value. Only quantum operations, i.e.\ built-in gates, gate (and opaque) subroutines, preparation, and measurement, can be prefaced by \code{if}. A quantum program with a parameter that depends on values that are known only at run-time can be rewritten using a sequence of \code{if} statements. Specifically, for a single-parameter gate with $n$ bits of precision, we may choose to write $2^n$ statements, only one of which is executed, or we can decompose the parameterized gate into a sequence of $n$ conditional gates.

The \code{barrier} instruction prevents optimizations from reordering gates across its source line. For example,
\begin{Verbatim}[commandchars=\\\{\}]
\PYG{n}{CX} \PYG{n}{r}\PYG{p}{[}\PYG{l+m+mi}{0}\PYG{p}{],}\PYG{n}{r}\PYG{p}{[}\PYG{l+m+mi}{1}\PYG{p}{];}
\PYG{n}{h} \PYG{n}{q}\PYG{p}{[}\PYG{l+m+mi}{0}\PYG{p}{];}
\PYG{n}{h} \PYG{n}{s}\PYG{p}{[}\PYG{l+m+mi}{0}\PYG{p}{];}
\PYG{n}{barrier} \PYG{n}{r}\PYG{p}{,}\PYG{n}{q}\PYG{p}{[}\PYG{l+m+mi}{0}\PYG{p}{];}
\PYG{n}{h} \PYG{n}{s}\PYG{p}{[}\PYG{l+m+mi}{0}\PYG{p}{];}
\PYG{n}{CX} \PYG{n}{r}\PYG{p}{[}\PYG{l+m+mi}{1}\PYG{p}{],}\PYG{n}{r}\PYG{p}{[}\PYG{l+m+mi}{0}\PYG{p}{];}
\PYG{n}{CX} \PYG{n}{r}\PYG{p}{[}\PYG{l+m+mi}{0}\PYG{p}{],}\PYG{n}{r}\PYG{p}{[}\PYG{l+m+mi}{1}\PYG{p}{];}
\end{Verbatim}
will prevent an attempt to combine the CNOT gates but will allow the pair of \code{h s[0];} gates to cancel.

Open QASM statements are summarized in Table~\ref{tab:qasm:new}. The grammar is presented in Appendix~\ref{app:grammar}.

\begin{landscape}
\begin{table}[htbp]
\begin{threeparttable}
\caption{Open QASM language statements (version 2.0)\label{tab:qasm:new}}
\begin{tabular}{@{}lll@{}}
\toprule
Statement & Description & Example \\
\midrule
\code{OPENQASM 2.0;} & Denotes a file in Open QASM format\tnote{a} & \code{OPENQASM 2.0;}\\
\code{qreg name[size];} & Declare a named register of qubits & \code{qreg q[5];} \\
\code{creg name[size];} & Declare a named register of bits & \code{creg c[5];} \\
\code{include "filename";} & Open and parse another source file & \code{include "qelib1.inc";} \\ 
\code{gate name(params) qargs \string{ body \string}} & Declare a unitary gate &  (see text) \\
\code{opaque name(params) qargs;} & Declare an opaque gate & (see text) \\
\code{// comment text} & Comment a line of text & \code{// oops!} \\
\midrule
\code{U(theta,phi,lambda) qubit|qreg;} & Apply built-in single qubit gate(s)\tnote{b} & \code{U(pi/2,2*pi/3,0) q[0];} \\
\code{CX qubit|qreg,qubit|qreg;} & Apply built-in CNOT gate(s) & \code{CX q[0],q[1];} \\
\code{measure qubit|qreg -> bit|creg;} & Make measurement(s) in $Z$ basis & \code{measure q -> c;} \\
\code{reset qubit|qreg;} & Prepare qubit(s) in $|0\rangle$ & \code{reset q[0];} \\ 
\code{gatename(params) qargs;} & Apply a user-defined unitary gate &  \code{crz(pi/2) q[1],q[0];} \\
\code{if(creg==int) qop;} & Conditionally apply quantum operation & \code{if(c==5) CX q[0],q[1];} \\ 
\midrule
\code{barrier qargs;} & Prevent transformations across this source line & \code{barrier q[0],q[1];} \\
\bottomrule
\end{tabular}
\begin{tablenotes}
\item[a] This must appear as the first non-comment line of the file.
\item[b] The parameters \code{theta}, \code{phi}, and \code{lambda} are given by {\em parameter expressions}; see text and Appendix~\ref{app:grammar}.
\end{tablenotes}
\end{threeparttable}
\end{table}
\end{landscape}
\section{Examples}

This section gives several examples of quantum circuits expressed in Open QASM (version 2.0). The circuits use a gate basis defined for the Quantum Experience.

\subsection{Quantum Experience standard header}

The Quantum Experience standard header defines the gates that are implemented by the hardware, gates that appear in the Quantum Experience composer, and a hierarchy of additional user-defined gates. Our approach is to define physical gates that the hardware implements in terms of the abstract gates \code{U} and \code{CX}. The current physical gates supported by the Quantum Experience are a superset of the abstract gates, but this is not true of all physical gate sets and devices. Choosing to use abstract gates merely to define physical gates gives some flexibility to add or change physical gates at a later time without changing Open QASM. We believe this approach is preferable to invisibly compiling abstract gates to physical gates or to changing the underlying set of abstract gates whenever the hardware changes.

The Quantum Experience currently implements the controlled-NOT gate via the cross-resonance interaction and implements three distinct types of single-qubit gates. The one-parameter gate
\begin{equation}
u_1(\lambda) := \mathrm{diag}(1,e^{i\lambda}) \sim U(0,0,\lambda) = R_z(\lambda)
\end{equation}
changes the phase of a carrier without applying any pulses. The symbol ``$\sim$'' denotes equivalence up to a global phase. The gate
\begin{equation}
u_2(\phi,\lambda) := U(\pi/2,\phi,\lambda) = R_z(\phi+\frac{\pi}{2})R_x(\pi/2)R_z(\lambda-\frac{\pi}{2})
\end{equation}
uses a single $\pi/2$-pulse. The most general single-qubit gate
\begin{equation}
u_3(\theta,\phi,\lambda) := U(\theta,\phi,\lambda) = R_z(\phi+3\pi)R_x(\pi/2)R_z(\theta+\pi)R_x(\pi/2)R_z(\lambda)
\end{equation}
uses a pair of $\pi/2$-pulses.

\begin{Verbatim}[commandchars=\\\{\}]
\PYG{c+c1}{// Quantum Experience (QE) Standard Header}
\PYG{c+c1}{// file: qelib1.inc}

\PYG{c+c1}{// \PYGZhy{}\PYGZhy{}\PYGZhy{} QE Hardware primitives \PYGZhy{}\PYGZhy{}\PYGZhy{}}

\PYG{c+c1}{// 3\PYGZhy{}parameter 2\PYGZhy{}pulse single qubit gate}
\PYG{n}{gate} \PYG{n+nf}{u3}\PYG{p}{(}\PYG{n}{theta}\PYG{p}{,}\PYG{n}{phi}\PYG{p}{,}\PYG{n}{lambda}\PYG{p}{)} \PYG{n}{q} \PYG{p}{\PYGZob{}} \PYG{n}{U}\PYG{p}{(}\PYG{n}{theta}\PYG{p}{,}\PYG{n}{phi}\PYG{p}{,}\PYG{n}{lambda}\PYG{p}{)} \PYG{n}{q}\PYG{p}{;} \PYG{p}{\PYGZcb{}}
\PYG{c+c1}{// 2\PYGZhy{}parameter 1\PYGZhy{}pulse single qubit gate}
\PYG{n}{gate} \PYG{n+nf}{u2}\PYG{p}{(}\PYG{n}{phi}\PYG{p}{,}\PYG{n}{lambda}\PYG{p}{)} \PYG{n}{q} \PYG{p}{\PYGZob{}} \PYG{n}{U}\PYG{p}{(}\PYG{n}{pi}\PYG{o}{/}\PYG{l+m+mi}{2}\PYG{p}{,}\PYG{n}{phi}\PYG{p}{,}\PYG{n}{lambda}\PYG{p}{)} \PYG{n}{q}\PYG{p}{;} \PYG{p}{\PYGZcb{}}
\PYG{c+c1}{// 1\PYGZhy{}parameter 0\PYGZhy{}pulse single qubit gate}
\PYG{n}{gate} \PYG{n+nf}{u1}\PYG{p}{(}\PYG{n}{lambda}\PYG{p}{)} \PYG{n}{q} \PYG{p}{\PYGZob{}} \PYG{n}{U}\PYG{p}{(}\PYG{l+m+mi}{0}\PYG{p}{,}\PYG{l+m+mi}{0}\PYG{p}{,}\PYG{n}{lambda}\PYG{p}{)} \PYG{n}{q}\PYG{p}{;} \PYG{p}{\PYGZcb{}}
\PYG{c+c1}{// controlled\PYGZhy{}NOT}
\PYG{n}{gate} \PYG{n}{cx} \PYG{n}{c}\PYG{p}{,}\PYG{n}{t} \PYG{p}{\PYGZob{}} \PYG{n}{CX} \PYG{n}{c}\PYG{p}{,}\PYG{n}{t}\PYG{p}{;} \PYG{p}{\PYGZcb{}}
\PYG{c+c1}{// idle gate (identity)}
\PYG{n}{gate} \PYG{n}{id} \PYG{n}{a} \PYG{p}{\PYGZob{}} \PYG{n}{U}\PYG{p}{(}\PYG{l+m+mi}{0}\PYG{p}{,}\PYG{l+m+mi}{0}\PYG{p}{,}\PYG{l+m+mi}{0}\PYG{p}{)} \PYG{n}{a}\PYG{p}{;} \PYG{p}{\PYGZcb{}}

\PYG{c+c1}{// \PYGZhy{}\PYGZhy{}\PYGZhy{} QE Standard Gates \PYGZhy{}\PYGZhy{}\PYGZhy{}}

\PYG{c+c1}{// Pauli gate: bit\PYGZhy{}flip}
\PYG{n}{gate} \PYG{n}{x} \PYG{n}{a} \PYG{p}{\PYGZob{}} \PYG{n}{u3}\PYG{p}{(}\PYG{n}{pi}\PYG{p}{,}\PYG{l+m+mi}{0}\PYG{p}{,}\PYG{n}{pi}\PYG{p}{)} \PYG{n}{a}\PYG{p}{;} \PYG{p}{\PYGZcb{}}
\PYG{c+c1}{// Pauli gate: bit and phase flip}
\PYG{n}{gate} \PYG{n}{y} \PYG{n}{a} \PYG{p}{\PYGZob{}} \PYG{n}{u3}\PYG{p}{(}\PYG{n}{pi}\PYG{p}{,}\PYG{n}{pi}\PYG{o}{/}\PYG{l+m+mi}{2}\PYG{p}{,}\PYG{n}{pi}\PYG{o}{/}\PYG{l+m+mi}{2}\PYG{p}{)} \PYG{n}{a}\PYG{p}{;} \PYG{p}{\PYGZcb{}}
\PYG{c+c1}{// Pauli gate: phase flip}
\PYG{n}{gate} \PYG{n}{z} \PYG{n}{a} \PYG{p}{\PYGZob{}} \PYG{n}{u1}\PYG{p}{(}\PYG{n}{pi}\PYG{p}{)} \PYG{n}{a}\PYG{p}{;} \PYG{p}{\PYGZcb{}}
\PYG{c+c1}{// Clifford gate: Hadamard}
\PYG{n}{gate} \PYG{n}{h} \PYG{n}{a} \PYG{p}{\PYGZob{}} \PYG{n}{u2}\PYG{p}{(}\PYG{l+m+mi}{0}\PYG{p}{,}\PYG{n}{pi}\PYG{p}{)} \PYG{n}{a}\PYG{p}{;} \PYG{p}{\PYGZcb{}}
\PYG{c+c1}{// Clifford gate: sqrt(Z) phase gate}
\PYG{n}{gate} \PYG{n}{s} \PYG{n}{a} \PYG{p}{\PYGZob{}} \PYG{n}{u1}\PYG{p}{(}\PYG{n}{pi}\PYG{o}{/}\PYG{l+m+mi}{2}\PYG{p}{)} \PYG{n}{a}\PYG{p}{;} \PYG{p}{\PYGZcb{}}
\PYG{c+c1}{// Clifford gate: conjugate of sqrt(Z)}
\PYG{n}{gate} \PYG{n}{sdg} \PYG{n}{a} \PYG{p}{\PYGZob{}} \PYG{n}{u1}\PYG{p}{(}\PYG{o}{\PYGZhy{}}\PYG{n}{pi}\PYG{o}{/}\PYG{l+m+mi}{2}\PYG{p}{)} \PYG{n}{a}\PYG{p}{;} \PYG{p}{\PYGZcb{}}
\PYG{c+c1}{// C3 gate: sqrt(S) phase gate}
\PYG{n}{gate} \PYG{n}{t} \PYG{n}{a} \PYG{p}{\PYGZob{}} \PYG{n}{u1}\PYG{p}{(}\PYG{n}{pi}\PYG{o}{/}\PYG{l+m+mi}{4}\PYG{p}{)} \PYG{n}{a}\PYG{p}{;} \PYG{p}{\PYGZcb{}}
\PYG{c+c1}{// C3 gate: conjugate of sqrt(S)}
\PYG{n}{gate} \PYG{n}{tdg} \PYG{n}{a} \PYG{p}{\PYGZob{}} \PYG{n}{u1}\PYG{p}{(}\PYG{o}{\PYGZhy{}}\PYG{n}{pi}\PYG{o}{/}\PYG{l+m+mi}{4}\PYG{p}{)} \PYG{n}{a}\PYG{p}{;} \PYG{p}{\PYGZcb{}}

\PYG{c+c1}{// \PYGZhy{}\PYGZhy{}\PYGZhy{} Standard rotations \PYGZhy{}\PYGZhy{}\PYGZhy{}}
\PYG{c+c1}{// Rotation around X\PYGZhy{}axis}
\PYG{n}{gate} \PYG{n}{rx}\PYG{p}{(}\PYG{n}{theta}\PYG{p}{)} \PYG{n}{a} \PYG{p}{\PYGZob{}} \PYG{n}{u3}\PYG{p}{(}\PYG{n}{theta}\PYG{p}{,}\PYG{o}{\PYGZhy{}}\PYG{n}{pi}\PYG{o}{/}\PYG{l+m+mi}{2}\PYG{p}{,}\PYG{n}{pi}\PYG{o}{/}\PYG{l+m+mi}{2}\PYG{p}{)} \PYG{n}{a}\PYG{p}{;} \PYG{p}{\PYGZcb{}}
\PYG{c+c1}{// rotation around Y\PYGZhy{}axis}
\PYG{n}{gate} \PYG{n}{ry}\PYG{p}{(}\PYG{n}{theta}\PYG{p}{)} \PYG{n}{a} \PYG{p}{\PYGZob{}} \PYG{n}{u3}\PYG{p}{(}\PYG{n}{theta}\PYG{p}{,}\PYG{l+m+mi}{0}\PYG{p}{,}\PYG{l+m+mi}{0}\PYG{p}{)} \PYG{n}{a}\PYG{p}{;} \PYG{p}{\PYGZcb{}}
\PYG{c+c1}{// rotation around Z axis}
\PYG{n}{gate} \PYG{n}{rz}\PYG{p}{(}\PYG{n}{phi}\PYG{p}{)} \PYG{n}{a} \PYG{p}{\PYGZob{}} \PYG{n}{u1}\PYG{p}{(}\PYG{n}{phi}\PYG{p}{)} \PYG{n}{a}\PYG{p}{;} \PYG{p}{\PYGZcb{}}

\PYG{c+c1}{// \PYGZhy{}\PYGZhy{}\PYGZhy{} QE Standard User\PYGZhy{}Defined Gates  \PYGZhy{}\PYGZhy{}\PYGZhy{}}

\PYG{c+c1}{// controlled\PYGZhy{}Phase}
\PYG{n}{gate} \PYG{n}{cz} \PYG{n}{a}\PYG{p}{,}\PYG{n}{b} \PYG{p}{\PYGZob{}} \PYG{n}{h} \PYG{n}{b}\PYG{p}{;} \PYG{n}{cx} \PYG{n}{a}\PYG{p}{,}\PYG{n}{b}\PYG{p}{;} \PYG{n}{h} \PYG{n}{b}\PYG{p}{;} \PYG{p}{\PYGZcb{}}
\PYG{c+c1}{// controlled\PYGZhy{}Y}
\PYG{n}{gate} \PYG{n}{cy} \PYG{n}{a}\PYG{p}{,}\PYG{n}{b} \PYG{p}{\PYGZob{}} \PYG{n}{sdg} \PYG{n}{b}\PYG{p}{;} \PYG{n}{cx} \PYG{n}{a}\PYG{p}{,}\PYG{n}{b}\PYG{p}{;} \PYG{n}{s} \PYG{n}{b}\PYG{p}{;} \PYG{p}{\PYGZcb{}}
\PYG{c+c1}{// controlled\PYGZhy{}H}
\PYG{n}{gate} \PYG{n}{ch} \PYG{n}{a}\PYG{p}{,}\PYG{n}{b} \PYG{p}{\PYGZob{}}
\PYG{n}{h} \PYG{n}{b}\PYG{p}{;} \PYG{n}{sdg} \PYG{n}{b}\PYG{p}{;}
\PYG{n}{cx} \PYG{n}{a}\PYG{p}{,}\PYG{n}{b}\PYG{p}{;}
\PYG{n}{h} \PYG{n}{b}\PYG{p}{;} \PYG{n}{t} \PYG{n}{b}\PYG{p}{;}
\PYG{n}{cx} \PYG{n}{a}\PYG{p}{,}\PYG{n}{b}\PYG{p}{;}
\PYG{n}{t} \PYG{n}{b}\PYG{p}{;} \PYG{n}{h} \PYG{n}{b}\PYG{p}{;} \PYG{n}{s} \PYG{n}{b}\PYG{p}{;} \PYG{n}{x} \PYG{n}{b}\PYG{p}{;} \PYG{n}{s} \PYG{n}{a}\PYG{p}{;}
\PYG{p}{\PYGZcb{}}
\PYG{c+c1}{// C3 gate: Toffoli}
\PYG{n}{gate} \PYG{n}{ccx} \PYG{n}{a}\PYG{p}{,}\PYG{n}{b}\PYG{p}{,}\PYG{n}{c}
\PYG{p}{\PYGZob{}}
  \PYG{n}{h} \PYG{n}{c}\PYG{p}{;}
  \PYG{n}{cx} \PYG{n}{b}\PYG{p}{,}\PYG{n}{c}\PYG{p}{;} \PYG{n}{tdg} \PYG{n}{c}\PYG{p}{;}
  \PYG{n}{cx} \PYG{n}{a}\PYG{p}{,}\PYG{n}{c}\PYG{p}{;} \PYG{n}{t} \PYG{n}{c}\PYG{p}{;}
  \PYG{n}{cx} \PYG{n}{b}\PYG{p}{,}\PYG{n}{c}\PYG{p}{;} \PYG{n}{tdg} \PYG{n}{c}\PYG{p}{;}
  \PYG{n}{cx} \PYG{n}{a}\PYG{p}{,}\PYG{n}{c}\PYG{p}{;} \PYG{n}{t} \PYG{n}{b}\PYG{p}{;} \PYG{n}{t} \PYG{n}{c}\PYG{p}{;} \PYG{n}{h} \PYG{n}{c}\PYG{p}{;}
  \PYG{n}{cx} \PYG{n}{a}\PYG{p}{,}\PYG{n}{b}\PYG{p}{;} \PYG{n}{t} \PYG{n}{a}\PYG{p}{;} \PYG{n}{tdg} \PYG{n}{b}\PYG{p}{;}
  \PYG{n}{cx} \PYG{n}{a}\PYG{p}{,}\PYG{n}{b}\PYG{p}{;}
\PYG{p}{\PYGZcb{}}
\PYG{c+c1}{// controlled rz rotation}
\PYG{n}{gate} \PYG{n}{crz}\PYG{p}{(}\PYG{n}{lambda}\PYG{p}{)} \PYG{n}{a}\PYG{p}{,}\PYG{n}{b}
\PYG{p}{\PYGZob{}}
  \PYG{n}{u1}\PYG{p}{(}\PYG{n}{lambda}\PYG{o}{/}\PYG{l+m+mi}{2}\PYG{p}{)} \PYG{n}{b}\PYG{p}{;}
  \PYG{n}{cx} \PYG{n}{a}\PYG{p}{,}\PYG{n}{b}\PYG{p}{;}
  \PYG{n}{u1}\PYG{p}{(}\PYG{o}{\PYGZhy{}}\PYG{n}{lambda}\PYG{o}{/}\PYG{l+m+mi}{2}\PYG{p}{)} \PYG{n}{b}\PYG{p}{;}
  \PYG{n}{cx} \PYG{n}{a}\PYG{p}{,}\PYG{n}{b}\PYG{p}{;}
\PYG{p}{\PYGZcb{}}
\PYG{c+c1}{// controlled phase rotation}
\PYG{n}{gate} \PYG{n}{cu1}\PYG{p}{(}\PYG{n}{lambda}\PYG{p}{)} \PYG{n}{a}\PYG{p}{,}\PYG{n}{b}
\PYG{p}{\PYGZob{}}
  \PYG{n}{u1}\PYG{p}{(}\PYG{n}{lambda}\PYG{o}{/}\PYG{l+m+mi}{2}\PYG{p}{)} \PYG{n}{a}\PYG{p}{;}
  \PYG{n}{cx} \PYG{n}{a}\PYG{p}{,}\PYG{n}{b}\PYG{p}{;}
  \PYG{n}{u1}\PYG{p}{(}\PYG{o}{\PYGZhy{}}\PYG{n}{lambda}\PYG{o}{/}\PYG{l+m+mi}{2}\PYG{p}{)} \PYG{n}{b}\PYG{p}{;}
  \PYG{n}{cx} \PYG{n}{a}\PYG{p}{,}\PYG{n}{b}\PYG{p}{;}
  \PYG{n}{u1}\PYG{p}{(}\PYG{n}{lambda}\PYG{o}{/}\PYG{l+m+mi}{2}\PYG{p}{)} \PYG{n}{b}\PYG{p}{;}
\PYG{p}{\PYGZcb{}}
\PYG{c+c1}{// controlled\PYGZhy{}U}
\PYG{n}{gate} \PYG{n}{cu3}\PYG{p}{(}\PYG{n}{theta}\PYG{p}{,}\PYG{n}{phi}\PYG{p}{,}\PYG{n}{lambda}\PYG{p}{)} \PYG{n}{c}\PYG{p}{,} \PYG{n}{t}
\PYG{p}{\PYGZob{}}
  \PYG{c+c1}{// implements controlled\PYGZhy{}U(theta,phi,lambda) with  target t and control c}
  \PYG{n}{u1}\PYG{p}{((}\PYG{n}{lambda}\PYG{o}{\PYGZhy{}}\PYG{n}{phi}\PYG{p}{)}\PYG{o}{/}\PYG{l+m+mi}{2}\PYG{p}{)} \PYG{n}{t}\PYG{p}{;}
  \PYG{n}{cx} \PYG{n}{c}\PYG{p}{,}\PYG{n}{t}\PYG{p}{;}
  \PYG{n}{u3}\PYG{p}{(}\PYG{o}{\PYGZhy{}}\PYG{n}{theta}\PYG{o}{/}\PYG{l+m+mi}{2}\PYG{p}{,}\PYG{l+m+mi}{0}\PYG{p}{,}\PYG{o}{\PYGZhy{}}\PYG{p}{(}\PYG{n}{phi}\PYG{o}{+}\PYG{n}{lambda}\PYG{p}{)}\PYG{o}{/}\PYG{l+m+mi}{2}\PYG{p}{)} \PYG{n}{t}\PYG{p}{;}
  \PYG{n}{cx} \PYG{n}{c}\PYG{p}{,}\PYG{n}{t}\PYG{p}{;}
  \PYG{n}{u3}\PYG{p}{(}\PYG{n}{theta}\PYG{o}{/}\PYG{l+m+mi}{2}\PYG{p}{,}\PYG{n}{phi}\PYG{p}{,}\PYG{l+m+mi}{0}\PYG{p}{)} \PYG{n}{t}\PYG{p}{;}
\PYG{p}{\PYGZcb{}}
\end{Verbatim}

\subsection{Quantum teleportation}

\begin{figure}
\centering
%! \usetikzlibrary{decorations.pathreplacing,decorations.pathmorphing}
\begin{tikzpicture}[scale=1.000000,x=1pt,y=1pt]
\filldraw[color=white] (0.000000, -7.500000) rectangle (320.000000, 82.500000);
% Drawing wires
% Line 1: q0 W q[0]\ket{0}
\draw[color=black] (0.000000,75.000000) -- (164.000000,75.000000);
%\draw[color=black] (164.000000,74.500000) -- (311.000000,74.500000);
\draw[color=black] (164.000000,75.00000) -- (311.000000,75.00000);
\draw[color=black] (0.000000,75.000000) node[left] {$q[0]\ket{0}$};
% Line 2: q1 W q[1]\ket{0}
\draw[color=black] (0.000000,60.000000) -- (176.000000,60.000000);
%\draw[color=black] (176.000000,59.500000) -- (311.000000,59.500000);
\draw[color=black] (176.000000,60.00000) -- (311.000000,60.00000);
\draw[color=black] (0.000000,60.000000) node[left] {$q[1]\ket{0}$};
% Line 3: q2 W q[2]\ket{0}
\draw[color=black] (0.000000,45.000000) -- (290.000000,45.000000);
%\draw[color=black] (290.000000,44.500000) -- (311.000000,44.500000);
\draw[color=black] (290.000000,45.00000) -- (311.000000,45.00000);
\draw[color=black] (0.000000,45.000000) node[left] {$q[2]\ket{0}$};
% Line 4: c00 W c0[0]\ 0
\draw[color=black] (0.000000,30.000000) -- (311.000000,30.000000);
\draw[color=black] (0.000000,29.000000) -- (311.000000,29.000000);
\draw[color=black] (0.000000,30.000000) node[left] {$c0[0]\ 0$};
% Line 5: c10 W c1[0]\ 0
\draw[color=black] (0.000000,15.000000) -- (311.000000,15.000000);
\draw[color=black] (0.000000,14.000000) -- (311.000000,14.000000);
\draw[color=black] (0.000000,15.000000) node[left] {$c1[0]~0$};
% Line 6: c20 W c2[0]\ 0
\draw[color=black] (0.000000,0.000000) -- (311.000000,0.000000);
\draw[color=black] (0.000000,-1.000000) -- (311.000000,-1.000000);
\draw[color=black] (0.000000,0.000000) node[left] {$c2[0]\ 0$};
% Done with wires; drawing gates
% Line 7: q0 G {u$_{3}$($0.3,0.2,0.1$)} width=80
\begin{scope}
\draw[fill=white] (46.000000, 75.000000) +(-45.000000:56.568542pt and 8.485281pt) -- +(45.000000:56.568542pt and 8.485281pt) -- +(135.000000:56.568542pt and 8.485281pt) -- +(225.000000:56.568542pt and 8.485281pt) -- cycle;
\clip (46.000000, 75.000000) +(-45.000000:56.568542pt and 8.485281pt) -- +(45.000000:56.568542pt and 8.485281pt) -- +(135.000000:56.568542pt and 8.485281pt) -- +(225.000000:56.568542pt and 8.485281pt) -- cycle;
\draw (46.000000, 75.000000) node {{u$_{3}$($0.3,0.2,0.1$)}};
\end{scope}
% Line 8: q1 H
\begin{scope}
\draw[fill=white] (46.000000, 60.000000) +(-45.000000:8.485281pt and 8.485281pt) -- +(45.000000:8.485281pt and 8.485281pt) -- +(135.000000:8.485281pt and 8.485281pt) -- +(225.000000:8.485281pt and 8.485281pt) -- cycle;
\clip (46.000000, 60.000000) +(-45.000000:8.485281pt and 8.485281pt) -- +(45.000000:8.485281pt and 8.485281pt) -- +(135.000000:8.485281pt and 8.485281pt) -- +(225.000000:8.485281pt and 8.485281pt) -- cycle;
\draw (46.000000, 60.000000) node {$H$};
\end{scope}
% Line 9: +q2 q1
\draw (101.000000,60.000000) -- (101.000000,45.000000);
\begin{scope}
\draw[fill=white] (101.000000, 45.000000) circle(3.000000pt);
\clip (101.000000, 45.000000) circle(3.000000pt);
\draw (98.000000, 45.000000) -- (104.000000, 45.000000);
\draw (101.000000, 42.000000) -- (101.000000, 48.000000);
\end{scope}
\filldraw (101.000000, 60.000000) circle(1.500000pt);
% Line 10: q0 BARRIER
\draw[decorate,decoration={zigzag,amplitude=1pt,segment length=4}] (110.000000,67.500000) -- (110.000000,82.500000);
% Line 11: q1 BARRIER
\draw[decorate,decoration={zigzag,amplitude=1pt,segment length=4}] (110.000000,52.500000) -- (110.000000,67.500000);
% Line 12: q2 BARRIER
\draw[decorate,decoration={zigzag,amplitude=1pt,segment length=4}] (110.000000,37.500000) -- (110.000000,52.500000);
% Line 13: +q1 q0
\draw (119.000000,75.000000) -- (119.000000,60.000000);
\begin{scope}
\draw[fill=white] (119.000000, 60.000000) circle(3.000000pt);
\clip (119.000000, 60.000000) circle(3.000000pt);
\draw (116.000000, 60.000000) -- (122.000000, 60.000000);
\draw (119.000000, 57.000000) -- (119.000000, 63.000000);
\end{scope}
\filldraw (119.000000, 75.000000) circle(1.500000pt);
% Line 14: q0 H
\begin{scope}
\draw[fill=white] (140.000000, 75.000000) +(-45.000000:8.485281pt and 8.485281pt) -- +(45.000000:8.485281pt and 8.485281pt) -- +(135.000000:8.485281pt and 8.485281pt) -- +(225.000000:8.485281pt and 8.485281pt) -- cycle;
\clip (140.000000, 75.000000) +(-45.000000:8.485281pt and 8.485281pt) -- +(45.000000:8.485281pt and 8.485281pt) -- +(135.000000:8.485281pt and 8.485281pt) -- +(225.000000:8.485281pt and 8.485281pt) -- cycle;
\draw (140.000000, 75.000000) node {$H$};
\end{scope}
% Line 15: TOUCH
% Line 16: q0:cwire +c00
%\draw (163.500000,75.000000) -- (163.500000,30.000000);
\draw (164.00000,75.000000) -- (164.00000,40.000000);
\filldraw (164.000000, 75.000000) circle(1.500000pt);
\begin{scope}
%\draw[fill=white] (164.000000, 30.000000) circle(3.000000pt);
\clip (164.000000, 30.000000) circle(3.000000pt);
\draw (161.000000, 30.000000) -- (167.000000, 30.000000);
%\draw (164.000000, 27.000000) -- (164.000000, 33.000000);
\end{scope}
\draw[fill=white] (158.000000, 69.000000) rectangle (170.000000, 81.000000);
\draw[very thin] (164.000000, 75.600000) arc (90:150:6.000000pt);
\draw[very thin] (164.000000, 75.600000) arc (90:30:6.000000pt);
\draw[->,>=stealth] (164.000000, 69.600000) -- +(80:10.392305pt);
% Line 17: q1:cwire +c10
%\draw (175.500000,60.000000) -- (175.500000,15.000000);
\draw (176.00000,60.000000) -- (176.00000,25.000000);
\filldraw (176.000000, 60.000000) circle(1.500000pt);
\begin{scope}
%\draw[fill=white] (176.000000, 15.000000) circle(3.000000pt);
\clip (176.000000, 15.000000) circle(3.000000pt);
\draw (173.000000, 15.000000) -- (179.000000, 15.000000);
%\draw (176.000000, 12.000000) -- (176.000000, 18.000000);
\end{scope}
\draw[fill=white] (170.000000, 54.000000) rectangle (182.000000, 66.000000);
\draw[very thin] (176.000000, 60.600000) arc (90:150:6.000000pt);
\draw[very thin] (176.000000, 60.600000) arc (90:30:6.000000pt);
\draw[->,>=stealth] (176.000000, 54.600000) -- +(80:10.392305pt);
% Line 18: q2 Z c00 %% ==1
\draw (224.000000, 15.00000) node[text width=144pt,below,text centered] {\scriptsize ==1};
\draw (199.000000,45.000000) -- (199.000000,30.000000);
\draw (201.000000,45.000000) -- (201.000000,30.000000);
\begin{scope}
\draw[fill=white] (200.000000, 45.000000) +(-45.000000:8.485281pt and 8.485281pt) -- +(45.000000:8.485281pt and 8.485281pt) -- +(135.000000:8.485281pt and 8.485281pt) -- +(225.000000:8.485281pt and 8.485281pt) -- cycle;
\clip (200.000000, 45.000000) +(-45.000000:8.485281pt and 8.485281pt) -- +(45.000000:8.485281pt and 8.485281pt) -- +(135.000000:8.485281pt and 8.485281pt) -- +(225.000000:8.485281pt and 8.485281pt) -- cycle;
\draw (200.000000, 45.000000) node {$Z$};
\end{scope}
\filldraw (200.000000, 30.000000) circle(1.500000pt);
% Line 19: q2 X c10 %% ==1
\draw (200.000000, 30.00000) node[text width=144pt,below,text centered] {\scriptsize ==1};
\draw (223.000000,45.000000) -- (223.000000,15.000000);
\draw (225.000000,45.000000) -- (225.000000,15.000000);
\begin{scope}
\draw[fill=white] (224.000000, 45.000000) +(-45.000000:8.485281pt and 8.485281pt) -- +(45.000000:8.485281pt and 8.485281pt) -- +(135.000000:8.485281pt and 8.485281pt) -- +(225.000000:8.485281pt and 8.485281pt) -- cycle;
\clip (224.000000, 45.000000) +(-45.000000:8.485281pt and 8.485281pt) -- +(45.000000:8.485281pt and 8.485281pt) -- +(135.000000:8.485281pt and 8.485281pt) -- +(225.000000:8.485281pt and 8.485281pt) -- cycle;
\draw (224.000000, 45.000000) node {$X$};
\end{scope}
\filldraw (224.000000, 15.000000) circle(1.500000pt);
% Line 20: q2 G {post} width=30
\begin{scope}
\draw[fill=white] (257.000000, 45.000000) +(-45.000000:21.213203pt and 8.485281pt) -- +(45.000000:21.213203pt and 8.485281pt) -- +(135.000000:21.213203pt and 8.485281pt) -- +(225.000000:21.213203pt and 8.485281pt) -- cycle;
\clip (257.000000, 45.000000) +(-45.000000:21.213203pt and 8.485281pt) -- +(45.000000:21.213203pt and 8.485281pt) -- +(135.000000:21.213203pt and 8.485281pt) -- +(225.000000:21.213203pt and 8.485281pt) -- cycle;
\draw (257.000000, 45.000000) node {{post}};
\end{scope}
% Line 21: q2:cwire +c20
%\draw (289.500000,45.000000) -- (289.500000,0.000000);
\draw (290.00000,45.000000) -- (290.00000,0.000000);
\filldraw (290.000000, 45.000000) circle(1.500000pt);
\begin{scope}
%\draw[fill=white] (290.000000, 0.000000) circle(3.000000pt);
\clip (290.000000, 0.000000) circle(3.000000pt);
\draw (287.000000, 0.000000) -- (293.000000, 0.000000);
%\draw (290.000000, -3.000000) -- (290.000000, 3.000000);
\end{scope}
\draw[fill=white] (284.000000, 39.000000) rectangle (296.000000, 51.000000);
\draw[very thin] (290.000000, 45.600000) arc (90:150:6.000000pt);
\draw[very thin] (290.000000, 45.600000) arc (90:30:6.000000pt);
\draw[->,>=stealth] (290.000000, 39.600000) -- +(80:10.392305pt);
% Line 22: TOUCH
% Line 23: q0 OUT 0
\filldraw[color=white] (308.000000, 72.000000) rectangle (314.000000, 78.000000);
\draw (308.000000, 72.000000) -- (308.000000, 78.000000);
%\draw (311.000000, 75.000000) node {$\scriptstyle{0}$};
% Line 24: q1 OUT 0
\filldraw[color=white] (308.000000, 57.000000) rectangle (314.000000, 63.000000);
\draw (308.000000, 57.000000) -- (308.000000, 63.000000);
%\draw (311.000000, 60.000000) node {$\scriptstyle{0}$};
% Line 25: q2 OUT 0
\filldraw[color=white] (308.000000, 42.000000) rectangle (314.000000, 48.000000);
\draw (308.000000, 42.000000) -- (308.000000, 48.000000);
\draw[->,>=stealth] (164.000000, 40.00000)  -- +(270:10.392305pt); % arrowhead
\draw[->,>=stealth] (176.000000, 25.00000)  -- +(270:10.392305pt); % arrowhead
\draw[->,>=stealth] (290.000000, 10.00000)  -- +(270:10.392305pt); % arrowhead
%\draw (311.000000, 45.000000) node {$\scriptstyle{0}$};
% Done with gates; drawing ending labels
% Done with ending labels; drawing cut lines and comments
% Done with comments
\end{tikzpicture}
\caption{Example of quantum teleportation. Qubit \code{q[0]} is prepared by \code{U(0.3,0.2,0.1) q[0];} and teleported to \code{q[2]}.
\label{fig:example:teleport}}
\end{figure}
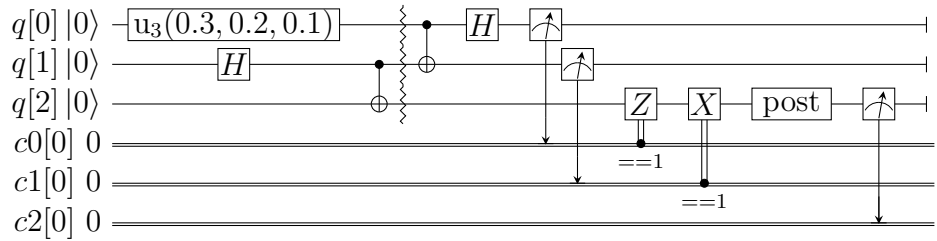

Quantum teleportation (Fig.~\ref{fig:example:teleport}) demonstrates conditional application of future gates based on prior measurement outcomes.

\begin{Verbatim}[commandchars=\\\{\}]
\PYG{c+c1}{// quantum teleportation example}
\PYG{n}{OPENQASM} \PYG{l+m+mf}{2.0}\PYG{p}{;}
\PYG{n}{include} \PYG{l+s}{\PYGZdq{}qelib1.inc\PYGZdq{}}\PYG{p}{;}
\PYG{n}{qreg} \PYG{n}{q}\PYG{p}{[}\PYG{l+m+mi}{3}\PYG{p}{];}
\PYG{n}{creg} \PYG{n}{c0}\PYG{p}{[}\PYG{l+m+mi}{1}\PYG{p}{];}
\PYG{n}{creg} \PYG{n}{c1}\PYG{p}{[}\PYG{l+m+mi}{1}\PYG{p}{];}
\PYG{n}{creg} \PYG{n}{c2}\PYG{p}{[}\PYG{l+m+mi}{1}\PYG{p}{];}
\PYG{c+c1}{// optional post\PYGZhy{}rotation for state tomography}
\PYG{n}{gate} \PYG{n}{post} \PYG{n}{q} \PYG{p}{\PYGZob{}} \PYG{p}{\PYGZcb{}}
\PYG{n}{u3}\PYG{p}{(}\PYG{l+m+mf}{0.3}\PYG{p}{,}\PYG{l+m+mf}{0.2}\PYG{p}{,}\PYG{l+m+mf}{0.1}\PYG{p}{)} \PYG{n}{q}\PYG{p}{[}\PYG{l+m+mi}{0}\PYG{p}{];}
\PYG{n}{h} \PYG{n}{q}\PYG{p}{[}\PYG{l+m+mi}{1}\PYG{p}{];}
\PYG{n}{cx} \PYG{n}{q}\PYG{p}{[}\PYG{l+m+mi}{1}\PYG{p}{],}\PYG{n}{q}\PYG{p}{[}\PYG{l+m+mi}{2}\PYG{p}{];}
\PYG{n}{barrier} \PYG{n}{q}\PYG{p}{;}
\PYG{n}{cx} \PYG{n}{q}\PYG{p}{[}\PYG{l+m+mi}{0}\PYG{p}{],}\PYG{n}{q}\PYG{p}{[}\PYG{l+m+mi}{1}\PYG{p}{];}
\PYG{n}{h} \PYG{n}{q}\PYG{p}{[}\PYG{l+m+mi}{0}\PYG{p}{];}
\PYG{n}{measure} \PYG{n}{q}\PYG{p}{[}\PYG{l+m+mi}{0}\PYG{p}{]} \PYG{o}{\PYGZhy{}\PYGZgt{}} \PYG{n}{c0}\PYG{p}{[}\PYG{l+m+mi}{0}\PYG{p}{];}
\PYG{n}{measure} \PYG{n}{q}\PYG{p}{[}\PYG{l+m+mi}{1}\PYG{p}{]} \PYG{o}{\PYGZhy{}\PYGZgt{}} \PYG{n}{c1}\PYG{p}{[}\PYG{l+m+mi}{0}\PYG{p}{];}
\PYG{k}{if}\PYG{p}{(}\PYG{n}{c0}\PYG{o}{==}\PYG{l+m+mi}{1}\PYG{p}{)} \PYG{n}{z} \PYG{n}{q}\PYG{p}{[}\PYG{l+m+mi}{2}\PYG{p}{];}
\PYG{k}{if}\PYG{p}{(}\PYG{n}{c1}\PYG{o}{==}\PYG{l+m+mi}{1}\PYG{p}{)} \PYG{n}{x} \PYG{n}{q}\PYG{p}{[}\PYG{l+m+mi}{2}\PYG{p}{];}
\PYG{n}{post} \PYG{n}{q}\PYG{p}{[}\PYG{l+m+mi}{2}\PYG{p}{];}
\PYG{n}{measure} \PYG{n}{q}\PYG{p}{[}\PYG{l+m+mi}{2}\PYG{p}{]} \PYG{o}{\PYGZhy{}\PYGZgt{}} \PYG{n}{c2}\PYG{p}{[}\PYG{l+m+mi}{0}\PYG{p}{];}
\end{Verbatim}

\subsection{Quantum Fourier transform}
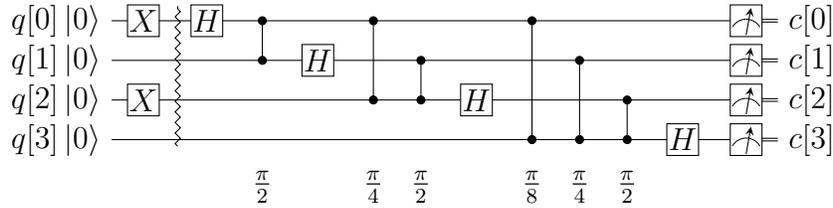
\begin{figure}
\centering
\begin{tikzpicture}[scale=1.000000,x=1pt,y=1pt]
\filldraw[color=white] (0.000000, -7.500000) rectangle (252.000000, 52.500000);
% Drawing wires
% Line 2: q0 W q[0]\ket{0} c[0]
\draw[color=black] (0.000000,45.000000) -- (240.000000,45.000000);
\draw[color=black] (240.000000,44.500000) -- (252.000000,44.500000);
\draw[color=black] (240.000000,45.500000) -- (252.000000,45.500000);
\draw[color=black] (0.000000,45.000000) node[left] {$q[0]\ket{0}$};
% Line 3: q1 W q[1]\ket{0} c[1]
\draw[color=black] (0.000000,30.000000) -- (240.000000,30.000000);
\draw[color=black] (240.000000,29.500000) -- (252.000000,29.500000);
\draw[color=black] (240.000000,30.500000) -- (252.000000,30.500000);
\draw[color=black] (0.000000,30.000000) node[left] {$q[1]\ket{0}$};
% Line 4: q2 W q[2]\ket{0} c[2]
\draw[color=black] (0.000000,15.000000) -- (240.000000,15.000000);
\draw[color=black] (240.000000,14.500000) -- (252.000000,14.500000);
\draw[color=black] (240.000000,15.500000) -- (252.000000,15.500000);
\draw[color=black] (0.000000,15.000000) node[left] {$q[2]\ket{0}$};
% Line 5: q3 W q[3]\ket{0} c[3]
\draw[color=black] (0.000000,0.000000) -- (240.000000,0.000000);
\draw[color=black] (240.000000,-0.500000) -- (252.000000,-0.500000);
\draw[color=black] (240.000000,0.500000) -- (252.000000,0.500000);
\draw[color=black] (0.000000,0.000000) node[left] {$q[3]\ket{0}$};
% Done with wires; drawing gates
% Line 6: q0 X
\begin{scope}
\draw[fill=white] (12.000000, 45.000000) +(-45.000000:8.485281pt and 8.485281pt) -- +(45.000000:8.485281pt and 8.485281pt) -- +(135.000000:8.485281pt and 8.485281pt) -- +(225.000000:8.485281pt and 8.485281pt) -- cycle;
\clip (12.000000, 45.000000) +(-45.000000:8.485281pt and 8.485281pt) -- +(45.000000:8.485281pt and 8.485281pt) -- +(135.000000:8.485281pt and 8.485281pt) -- +(225.000000:8.485281pt and 8.485281pt) -- cycle;
\draw (12.000000, 45.000000) node {$X$};
\end{scope}
% Line 7: q2 X
\begin{scope}
\draw[fill=white] (12.000000, 15.000000) +(-45.000000:8.485281pt and 8.485281pt) -- +(45.000000:8.485281pt and 8.485281pt) -- +(135.000000:8.485281pt and 8.485281pt) -- +(225.000000:8.485281pt and 8.485281pt) -- cycle;
\clip (12.000000, 15.000000) +(-45.000000:8.485281pt and 8.485281pt) -- +(45.000000:8.485281pt and 8.485281pt) -- +(135.000000:8.485281pt and 8.485281pt) -- +(225.000000:8.485281pt and 8.485281pt) -- cycle;
\draw (12.000000, 15.000000) node {$X$};
\end{scope}
\draw[decorate,decoration={zigzag,amplitude=1pt,segment length=4}] (25.000000,-2.500000) -- (25.000000,50.500000);
% Line 8: q0 H
\begin{scope}
\draw[fill=white] (36.000000, 45.000000) +(-45.000000:8.485281pt and 8.485281pt) -- +(45.000000:8.485281pt and 8.485281pt) -- +(135.000000:8.485281pt and 8.485281pt) -- +(225.000000:8.485281pt and 8.485281pt) -- cycle;
\clip (36.000000, 45.000000) +(-45.000000:8.485281pt and 8.485281pt) -- +(45.000000:8.485281pt and 8.485281pt) -- +(135.000000:8.485281pt and 8.485281pt) -- +(225.000000:8.485281pt and 8.485281pt) -- cycle;
\draw (36.000000, 45.000000) node {$H$};
\end{scope}
% Line 9: q0 q1 %% $\frac{\pi}{2}$
\draw (57.000000, -7.500000) node[text width=144pt,below,text centered] {$\frac{\pi}{2}$};
\draw (57.000000,45.000000) -- (57.000000,30.000000);
\filldraw (57.000000, 45.000000) circle(1.500000pt);
\filldraw (57.000000, 30.000000) circle(1.500000pt);
% Line 10: q1 H
\begin{scope}
\draw[fill=white] (78.000000, 30.000000) +(-45.000000:8.485281pt and 8.485281pt) -- +(45.000000:8.485281pt and 8.485281pt) -- +(135.000000:8.485281pt and 8.485281pt) -- +(225.000000:8.485281pt and 8.485281pt) -- cycle;
\clip (78.000000, 30.000000) +(-45.000000:8.485281pt and 8.485281pt) -- +(45.000000:8.485281pt and 8.485281pt) -- +(135.000000:8.485281pt and 8.485281pt) -- +(225.000000:8.485281pt and 8.485281pt) -- cycle;
\draw (78.000000, 30.000000) node {$H$};
\end{scope}
% Line 11: TOUCH
% Line 12: q0 q2 %% $\frac{\pi}{4}$
\draw (99.000000, -7.500000) node[text width=144pt,below,text centered] {$\frac{\pi}{4}$};
\draw (99.000000,45.000000) -- (99.000000,15.000000);
\filldraw (99.000000, 45.000000) circle(1.500000pt);
\filldraw (99.000000, 15.000000) circle(1.500000pt);
% Line 13: q1 q2 %% $\frac{\pi}{2}$
\draw (117.000000, -7.500000) node[text width=144pt,below,text centered] {$\frac{\pi}{2}$};
\draw (117.000000,30.000000) -- (117.000000,15.000000);
\filldraw (117.000000, 30.000000) circle(1.500000pt);
\filldraw (117.000000, 15.000000) circle(1.500000pt);
% Line 14: q2 H
\begin{scope}
\draw[fill=white] (138.000000, 15.000000) +(-45.000000:8.485281pt and 8.485281pt) -- +(45.000000:8.485281pt and 8.485281pt) -- +(135.000000:8.485281pt and 8.485281pt) -- +(225.000000:8.485281pt and 8.485281pt) -- cycle;
\clip (138.000000, 15.000000) +(-45.000000:8.485281pt and 8.485281pt) -- +(45.000000:8.485281pt and 8.485281pt) -- +(135.000000:8.485281pt and 8.485281pt) -- +(225.000000:8.485281pt and 8.485281pt) -- cycle;
\draw (138.000000, 15.000000) node {$H$};
\end{scope}
% Line 15: TOUCH
% Line 16: q0 q3 %% $\frac{\pi}{8}$
\draw (159.000000, -7.500000) node[text width=144pt,below,text centered] {$\frac{\pi}{8}$};
\draw (159.000000,45.000000) -- (159.000000,0.000000);
\filldraw (159.000000, 45.000000) circle(1.500000pt);
\filldraw (159.000000, 0.000000) circle(1.500000pt);
% Line 17: q1 q3 %% $\frac{\pi}{4}$
\draw (177.000000, -7.500000) node[text width=144pt,below,text centered] {$\frac{\pi}{4}$};
\draw (177.000000,30.000000) -- (177.000000,0.000000);
\filldraw (177.000000, 30.000000) circle(1.500000pt);
\filldraw (177.000000, 0.000000) circle(1.500000pt);
% Line 18: q2 q3 %% $\frac{\pi}{2}$
\draw (195.000000, -7.500000) node[text width=144pt,below,text centered] {$\frac{\pi}{2}$};
\draw (195.000000,15.000000) -- (195.000000,0.000000);
\filldraw (195.000000, 15.000000) circle(1.500000pt);
\filldraw (195.000000, 0.000000) circle(1.500000pt);
% Line 19: q3 H
\begin{scope}
\draw[fill=white] (216.000000, -0.000000) +(-45.000000:8.485281pt and 8.485281pt) -- +(45.000000:8.485281pt and 8.485281pt) -- +(135.000000:8.485281pt and 8.485281pt) -- +(225.000000:8.485281pt and 8.485281pt) -- cycle;
\clip (216.000000, -0.000000) +(-45.000000:8.485281pt and 8.485281pt) -- +(45.000000:8.485281pt and 8.485281pt) -- +(135.000000:8.485281pt and 8.485281pt) -- +(225.000000:8.485281pt and 8.485281pt) -- cycle;
\draw (216.000000, -0.000000) node {$H$};
\end{scope}
% Line 20: TOUCH
% Line 21: q0 M
\draw[fill=white] (234.000000, 39.000000) rectangle (246.000000, 51.000000);
\draw[very thin] (240.000000, 45.600000) arc (90:150:6.000000pt);
\draw[very thin] (240.000000, 45.600000) arc (90:30:6.000000pt);
\draw[->,>=stealth] (240.000000, 39.600000) -- +(80:10.392305pt);
% Line 22: q1 M
\draw[fill=white] (234.000000, 24.000000) rectangle (246.000000, 36.000000);
\draw[very thin] (240.000000, 30.600000) arc (90:150:6.000000pt);
\draw[very thin] (240.000000, 30.600000) arc (90:30:6.000000pt);
\draw[->,>=stealth] (240.000000, 24.600000) -- +(80:10.392305pt);
% Line 23: q2 M
\draw[fill=white] (234.000000, 9.000000) rectangle (246.000000, 21.000000);
\draw[very thin] (240.000000, 15.600000) arc (90:150:6.000000pt);
\draw[very thin] (240.000000, 15.600000) arc (90:30:6.000000pt);
\draw[->,>=stealth] (240.000000, 9.600000) -- +(80:10.392305pt);
% Line 24: q3 M
\draw[fill=white] (234.000000, -6.000000) rectangle (246.000000, 6.000000);
\draw[very thin] (240.000000, 0.600000) arc (90:150:6.000000pt);
\draw[very thin] (240.000000, 0.600000) arc (90:30:6.000000pt);
\draw[->,>=stealth] (240.000000, -5.400000) -- +(80:10.392305pt);
% Done with gates; drawing ending labels
\draw[color=black] (252.000000,45.000000) node[right] {$c[0]$};
\draw[color=black] (252.000000,30.000000) node[right] {$c[1]$};
\draw[color=black] (252.000000,15.000000) node[right] {$c[2]$};
\draw[color=black] (252.000000,0.000000) node[right] {$c[3]$};
% Done with ending labels; drawing cut lines and comments
% Done with comments
\end{tikzpicture}
\caption{Example of a $4$-qubit quantum Fourier transform. The circuit applies the QFT to $|1010\rangle$ and measures in the computational basis. The output is read in reverse order \code{c[3], c[2], c[1], c[0]}. \label{fig:example:qft}}
\end{figure}

The quantum Fourier transform (QFT, Fig.~\ref{fig:example:qft}) demonstrates parameter passing to gate subroutines. This circuit applies the QFT to the state $|q_0 q_1 q_2 q_3\rangle=|1010\rangle$ and measures in the computational basis.

\begin{Verbatim}[commandchars=\\\{\}]
\PYG{c+c1}{// quantum Fourier transform}
\PYG{n}{OPENQASM} \PYG{l+m+mf}{2.0}\PYG{p}{;}
\PYG{n}{include} \PYG{l+s}{\PYGZdq{}qelib1.inc\PYGZdq{}}\PYG{p}{;}
\PYG{n}{qreg} \PYG{n}{q}\PYG{p}{[}\PYG{l+m+mi}{4}\PYG{p}{];}
\PYG{n}{creg} \PYG{n}{c}\PYG{p}{[}\PYG{l+m+mi}{4}\PYG{p}{];}
\PYG{n}{x} \PYG{n}{q}\PYG{p}{[}\PYG{l+m+mi}{0}\PYG{p}{];} 
\PYG{n}{x} \PYG{n}{q}\PYG{p}{[}\PYG{l+m+mi}{2}\PYG{p}{];}
\PYG{n}{barrier} \PYG{n}{q}\PYG{p}{;}
\PYG{n}{h} \PYG{n}{q}\PYG{p}{[}\PYG{l+m+mi}{0}\PYG{p}{];}
\PYG{n}{cu1}\PYG{p}{(}\PYG{n}{pi}\PYG{o}{/}\PYG{l+m+mi}{2}\PYG{p}{)} \PYG{n}{q}\PYG{p}{[}\PYG{l+m+mi}{1}\PYG{p}{],}\PYG{n}{q}\PYG{p}{[}\PYG{l+m+mi}{0}\PYG{p}{];}
\PYG{n}{h} \PYG{n}{q}\PYG{p}{[}\PYG{l+m+mi}{1}\PYG{p}{];}
\PYG{n}{cu1}\PYG{p}{(}\PYG{n}{pi}\PYG{o}{/}\PYG{l+m+mi}{4}\PYG{p}{)} \PYG{n}{q}\PYG{p}{[}\PYG{l+m+mi}{2}\PYG{p}{],}\PYG{n}{q}\PYG{p}{[}\PYG{l+m+mi}{0}\PYG{p}{];}
\PYG{n}{cu1}\PYG{p}{(}\PYG{n}{pi}\PYG{o}{/}\PYG{l+m+mi}{2}\PYG{p}{)} \PYG{n}{q}\PYG{p}{[}\PYG{l+m+mi}{2}\PYG{p}{],}\PYG{n}{q}\PYG{p}{[}\PYG{l+m+mi}{1}\PYG{p}{];}
\PYG{n}{h} \PYG{n}{q}\PYG{p}{[}\PYG{l+m+mi}{2}\PYG{p}{];}
\PYG{n}{cu1}\PYG{p}{(}\PYG{n}{pi}\PYG{o}{/}\PYG{l+m+mi}{8}\PYG{p}{)} \PYG{n}{q}\PYG{p}{[}\PYG{l+m+mi}{3}\PYG{p}{],}\PYG{n}{q}\PYG{p}{[}\PYG{l+m+mi}{0}\PYG{p}{];}
\PYG{n}{cu1}\PYG{p}{(}\PYG{n}{pi}\PYG{o}{/}\PYG{l+m+mi}{4}\PYG{p}{)} \PYG{n}{q}\PYG{p}{[}\PYG{l+m+mi}{3}\PYG{p}{],}\PYG{n}{q}\PYG{p}{[}\PYG{l+m+mi}{1}\PYG{p}{];}
\PYG{n}{cu1}\PYG{p}{(}\PYG{n}{pi}\PYG{o}{/}\PYG{l+m+mi}{2}\PYG{p}{)} \PYG{n}{q}\PYG{p}{[}\PYG{l+m+mi}{3}\PYG{p}{],}\PYG{n}{q}\PYG{p}{[}\PYG{l+m+mi}{2}\PYG{p}{];}
\PYG{n}{h} \PYG{n}{q}\PYG{p}{[}\PYG{l+m+mi}{3}\PYG{p}{];}
\PYG{n}{measure} \PYG{n}{q} \PYG{o}{\PYGZhy{}\PYGZgt{}} \PYG{n}{c}\PYG{p}{;}
\end{Verbatim}

\subsection{Inverse QFT followed by measurement}

If the qubits are all measured after the inverse QFT, the measurement commutes with the controls of the \code{cu1} gates, and those gates can be replaced by classically-controlled single qubit rotations (see for example Figure 3.3 in \cite{mermin}). The example demonstrates how to implement this classical control using conditional gates.

\begin{figure}
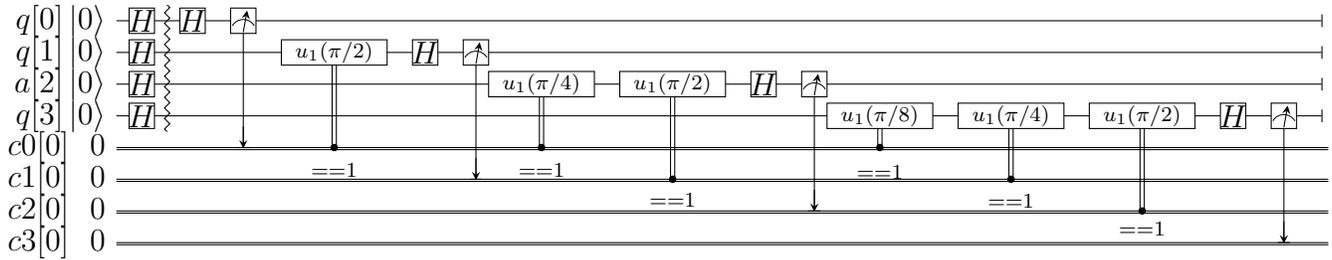

\hspace{-0.8cm}
\begin{minipage}{.8\textwidth}
\include{qpics/c13a}
\end{minipage}
\caption{Example of a $4$-qubit inverse quantum Fourier transform followed by measurement. In this case, the measurement commutes with the controls of the \code{cu1} gates and can be rewritten as shown (see Figure 3.3 in \cite{mermin}). The circuit applies the inverse QFT to the uniform superposition and measures in the computational basis. 
\label{fig:example:iqft}}
\end{figure}
\begin{Verbatim}[commandchars=\\\{\}]
\PYG{c+c1}{// QFT and measure, version 1}
\PYG{n}{OPENQASM} \PYG{l+m+mf}{2.0}\PYG{p}{;}
\PYG{n}{include} \PYG{l+s}{\PYGZdq{}qelib1.inc\PYGZdq{}}\PYG{p}{;}
\PYG{n}{qreg} \PYG{n}{q}\PYG{p}{[}\PYG{l+m+mi}{4}\PYG{p}{];}
\PYG{n}{creg} \PYG{n}{c}\PYG{p}{[}\PYG{l+m+mi}{4}\PYG{p}{];}
\PYG{n}{h} \PYG{n}{q}\PYG{p}{;}
\PYG{n}{barrier} \PYG{n}{q}\PYG{p}{;}
\PYG{n}{h} \PYG{n}{q}\PYG{p}{[}\PYG{l+m+mi}{0}\PYG{p}{];}
\PYG{n}{measure} \PYG{n}{q}\PYG{p}{[}\PYG{l+m+mi}{0}\PYG{p}{]} \PYG{o}{\PYGZhy{}\PYGZgt{}} \PYG{n}{c}\PYG{p}{[}\PYG{l+m+mi}{0}\PYG{p}{];}
\PYG{k}{if}\PYG{p}{(}\PYG{n}{c}\PYG{o}{==}\PYG{l+m+mi}{1}\PYG{p}{)} \PYG{n}{u1}\PYG{p}{(}\PYG{n}{pi}\PYG{o}{/}\PYG{l+m+mi}{2}\PYG{p}{)} \PYG{n}{q}\PYG{p}{[}\PYG{l+m+mi}{1}\PYG{p}{];}
\PYG{n}{h} \PYG{n}{q}\PYG{p}{[}\PYG{l+m+mi}{1}\PYG{p}{];}
\PYG{n}{measure} \PYG{n}{q}\PYG{p}{[}\PYG{l+m+mi}{1}\PYG{p}{]} \PYG{o}{\PYGZhy{}\PYGZgt{}} \PYG{n}{c}\PYG{p}{[}\PYG{l+m+mi}{1}\PYG{p}{];}
\PYG{k}{if}\PYG{p}{(}\PYG{n}{c}\PYG{o}{==}\PYG{l+m+mi}{1}\PYG{p}{)} \PYG{n}{u1}\PYG{p}{(}\PYG{n}{pi}\PYG{o}{/}\PYG{l+m+mi}{4}\PYG{p}{)} \PYG{n}{q}\PYG{p}{[}\PYG{l+m+mi}{2}\PYG{p}{];}
\PYG{k}{if}\PYG{p}{(}\PYG{n}{c}\PYG{o}{==}\PYG{l+m+mi}{2}\PYG{p}{)} \PYG{n}{u1}\PYG{p}{(}\PYG{n}{pi}\PYG{o}{/}\PYG{l+m+mi}{2}\PYG{p}{)} \PYG{n}{q}\PYG{p}{[}\PYG{l+m+mi}{2}\PYG{p}{];}
\PYG{k}{if}\PYG{p}{(}\PYG{n}{c}\PYG{o}{==}\PYG{l+m+mi}{3}\PYG{p}{)} \PYG{n}{u1}\PYG{p}{(}\PYG{n}{pi}\PYG{o}{/}\PYG{l+m+mi}{2}\PYG{o}{+}\PYG{n}{pi}\PYG{o}{/}\PYG{l+m+mi}{4}\PYG{p}{)} \PYG{n}{q}\PYG{p}{[}\PYG{l+m+mi}{2}\PYG{p}{];}
\PYG{n}{h} \PYG{n}{q}\PYG{p}{[}\PYG{l+m+mi}{2}\PYG{p}{];}
\PYG{n}{measure} \PYG{n}{q}\PYG{p}{[}\PYG{l+m+mi}{2}\PYG{p}{]} \PYG{o}{\PYGZhy{}\PYGZgt{}} \PYG{n}{c}\PYG{p}{[}\PYG{l+m+mi}{2}\PYG{p}{];}
\PYG{k}{if}\PYG{p}{(}\PYG{n}{c}\PYG{o}{==}\PYG{l+m+mi}{1}\PYG{p}{)} \PYG{n}{u1}\PYG{p}{(}\PYG{n}{pi}\PYG{o}{/}\PYG{l+m+mi}{8}\PYG{p}{)} \PYG{n}{q}\PYG{p}{[}\PYG{l+m+mi}{3}\PYG{p}{];}
\PYG{k}{if}\PYG{p}{(}\PYG{n}{c}\PYG{o}{==}\PYG{l+m+mi}{2}\PYG{p}{)} \PYG{n}{u1}\PYG{p}{(}\PYG{n}{pi}\PYG{o}{/}\PYG{l+m+mi}{4}\PYG{p}{)} \PYG{n}{q}\PYG{p}{[}\PYG{l+m+mi}{3}\PYG{p}{];}
\PYG{k}{if}\PYG{p}{(}\PYG{n}{c}\PYG{o}{==}\PYG{l+m+mi}{3}\PYG{p}{)} \PYG{n}{u1}\PYG{p}{(}\PYG{n}{pi}\PYG{o}{/}\PYG{l+m+mi}{4}\PYG{o}{+}\PYG{n}{pi}\PYG{o}{/}\PYG{l+m+mi}{8}\PYG{p}{)} \PYG{n}{q}\PYG{p}{[}\PYG{l+m+mi}{3}\PYG{p}{];}
\PYG{k}{if}\PYG{p}{(}\PYG{n}{c}\PYG{o}{==}\PYG{l+m+mi}{4}\PYG{p}{)} \PYG{n}{u1}\PYG{p}{(}\PYG{n}{pi}\PYG{o}{/}\PYG{l+m+mi}{2}\PYG{p}{)} \PYG{n}{q}\PYG{p}{[}\PYG{l+m+mi}{3}\PYG{p}{];}
\PYG{k}{if}\PYG{p}{(}\PYG{n}{c}\PYG{o}{==}\PYG{l+m+mi}{5}\PYG{p}{)} \PYG{n}{u1}\PYG{p}{(}\PYG{n}{pi}\PYG{o}{/}\PYG{l+m+mi}{2}\PYG{o}{+}\PYG{n}{pi}\PYG{o}{/}\PYG{l+m+mi}{8}\PYG{p}{)} \PYG{n}{q}\PYG{p}{[}\PYG{l+m+mi}{3}\PYG{p}{];}
\PYG{k}{if}\PYG{p}{(}\PYG{n}{c}\PYG{o}{==}\PYG{l+m+mi}{6}\PYG{p}{)} \PYG{n}{u1}\PYG{p}{(}\PYG{n}{pi}\PYG{o}{/}\PYG{l+m+mi}{2}\PYG{o}{+}\PYG{n}{pi}\PYG{o}{/}\PYG{l+m+mi}{4}\PYG{p}{)} \PYG{n}{q}\PYG{p}{[}\PYG{l+m+mi}{3}\PYG{p}{];}
\PYG{k}{if}\PYG{p}{(}\PYG{n}{c}\PYG{o}{==}\PYG{l+m+mi}{7}\PYG{p}{)} \PYG{n}{u1}\PYG{p}{(}\PYG{n}{pi}\PYG{o}{/}\PYG{l+m+mi}{2}\PYG{o}{+}\PYG{n}{pi}\PYG{o}{/}\PYG{l+m+mi}{4}\PYG{o}{+}\PYG{n}{pi}\PYG{o}{/}\PYG{l+m+mi}{8}\PYG{p}{)} \PYG{n}{q}\PYG{p}{[}\PYG{l+m+mi}{3}\PYG{p}{];}
\PYG{n}{h} \PYG{n}{q}\PYG{p}{[}\PYG{l+m+mi}{3}\PYG{p}{];}
\PYG{n}{measure} \PYG{n}{q}\PYG{p}{[}\PYG{l+m+mi}{3}\PYG{p}{]} \PYG{o}{\PYGZhy{}\PYGZgt{}} \PYG{n}{c}\PYG{p}{[}\PYG{l+m+mi}{3}\PYG{p}{];}
\end{Verbatim}

Alternatively, we can decompose the rotations and apply them using fewer statements but more quantum gates. The corresponding circuit for this example is shown in Fig.~\ref{fig:example:iqft}.
\begin{Verbatim}[commandchars=\\\{\}]
\PYG{c+c1}{// QFT and measure, version 2}
\PYG{n}{OPENQASM} \PYG{l+m+mf}{2.0}\PYG{p}{;}
\PYG{n}{include} \PYG{l+s}{\PYGZdq{}qelib1.inc\PYGZdq{}}\PYG{p}{;}
\PYG{n}{qreg} \PYG{n}{q}\PYG{p}{[}\PYG{l+m+mi}{4}\PYG{p}{];}
\PYG{n}{creg} \PYG{n}{c0}\PYG{p}{[}\PYG{l+m+mi}{1}\PYG{p}{];}
\PYG{n}{creg} \PYG{n}{c1}\PYG{p}{[}\PYG{l+m+mi}{1}\PYG{p}{];}
\PYG{n}{creg} \PYG{n}{c2}\PYG{p}{[}\PYG{l+m+mi}{1}\PYG{p}{];}
\PYG{n}{creg} \PYG{n}{c3}\PYG{p}{[}\PYG{l+m+mi}{1}\PYG{p}{];}
\PYG{n}{h} \PYG{n}{q}\PYG{p}{;}
\PYG{n}{barrier} \PYG{n}{q}\PYG{p}{;}
\PYG{n}{h} \PYG{n}{q}\PYG{p}{[}\PYG{l+m+mi}{0}\PYG{p}{];}
\PYG{n}{measure} \PYG{n}{q}\PYG{p}{[}\PYG{l+m+mi}{0}\PYG{p}{]} \PYG{o}{\PYGZhy{}\PYGZgt{}} \PYG{n}{c0}\PYG{p}{[}\PYG{l+m+mi}{0}\PYG{p}{];}
\PYG{k}{if}\PYG{p}{(}\PYG{n}{c0}\PYG{o}{==}\PYG{l+m+mi}{1}\PYG{p}{)} \PYG{n}{u1}\PYG{p}{(}\PYG{n}{pi}\PYG{o}{/}\PYG{l+m+mi}{2}\PYG{p}{)} \PYG{n}{q}\PYG{p}{[}\PYG{l+m+mi}{1}\PYG{p}{];}
\PYG{n}{h} \PYG{n}{q}\PYG{p}{[}\PYG{l+m+mi}{1}\PYG{p}{];}
\PYG{n}{measure} \PYG{n}{q}\PYG{p}{[}\PYG{l+m+mi}{1}\PYG{p}{]} \PYG{o}{\PYGZhy{}\PYGZgt{}} \PYG{n}{c1}\PYG{p}{[}\PYG{l+m+mi}{0}\PYG{p}{];}
\PYG{k}{if}\PYG{p}{(}\PYG{n}{c0}\PYG{o}{==}\PYG{l+m+mi}{1}\PYG{p}{)} \PYG{n}{u1}\PYG{p}{(}\PYG{n}{pi}\PYG{o}{/}\PYG{l+m+mi}{4}\PYG{p}{)} \PYG{n}{q}\PYG{p}{[}\PYG{l+m+mi}{2}\PYG{p}{];}
\PYG{k}{if}\PYG{p}{(}\PYG{n}{c1}\PYG{o}{==}\PYG{l+m+mi}{1}\PYG{p}{)} \PYG{n}{u1}\PYG{p}{(}\PYG{n}{pi}\PYG{o}{/}\PYG{l+m+mi}{2}\PYG{p}{)} \PYG{n}{q}\PYG{p}{[}\PYG{l+m+mi}{2}\PYG{p}{];}
\PYG{n}{h} \PYG{n}{q}\PYG{p}{[}\PYG{l+m+mi}{2}\PYG{p}{];}
\PYG{n}{measure} \PYG{n}{q}\PYG{p}{[}\PYG{l+m+mi}{2}\PYG{p}{]} \PYG{o}{\PYGZhy{}\PYGZgt{}} \PYG{n}{c2}\PYG{p}{[}\PYG{l+m+mi}{0}\PYG{p}{];}
\PYG{k}{if}\PYG{p}{(}\PYG{n}{c0}\PYG{o}{==}\PYG{l+m+mi}{1}\PYG{p}{)} \PYG{n}{u1}\PYG{p}{(}\PYG{n}{pi}\PYG{o}{/}\PYG{l+m+mi}{8}\PYG{p}{)} \PYG{n}{q}\PYG{p}{[}\PYG{l+m+mi}{3}\PYG{p}{];}
\PYG{k}{if}\PYG{p}{(}\PYG{n}{c1}\PYG{o}{==}\PYG{l+m+mi}{1}\PYG{p}{)} \PYG{n}{u1}\PYG{p}{(}\PYG{n}{pi}\PYG{o}{/}\PYG{l+m+mi}{4}\PYG{p}{)} \PYG{n}{q}\PYG{p}{[}\PYG{l+m+mi}{3}\PYG{p}{];}
\PYG{k}{if}\PYG{p}{(}\PYG{n}{c2}\PYG{o}{==}\PYG{l+m+mi}{1}\PYG{p}{)} \PYG{n}{u1}\PYG{p}{(}\PYG{n}{pi}\PYG{o}{/}\PYG{l+m+mi}{2}\PYG{p}{)} \PYG{n}{q}\PYG{p}{[}\PYG{l+m+mi}{3}\PYG{p}{];}
\PYG{n}{h} \PYG{n}{q}\PYG{p}{[}\PYG{l+m+mi}{3}\PYG{p}{];}
\PYG{n}{measure} \PYG{n}{q}\PYG{p}{[}\PYG{l+m+mi}{3}\PYG{p}{]} \PYG{o}{\PYGZhy{}\PYGZgt{}} \PYG{n}{c3}\PYG{p}{[}\PYG{l+m+mi}{0}\PYG{p}{];}
\end{Verbatim}

\subsection{Ripple-carry adder}

\begin{figure}
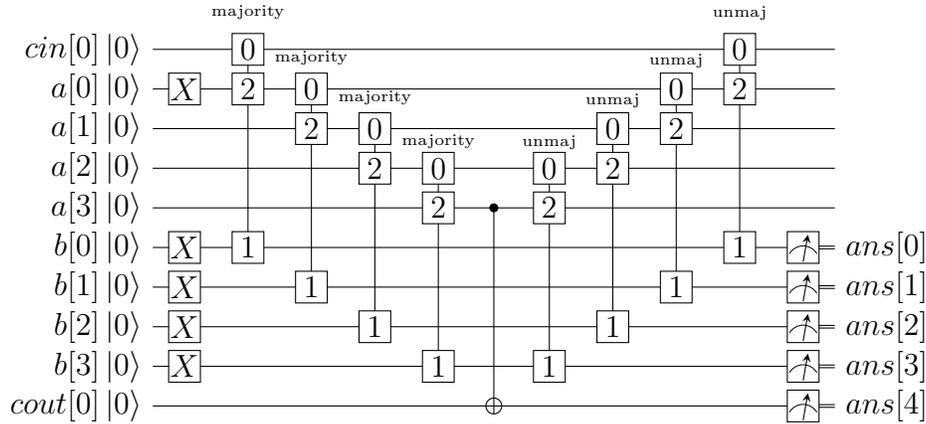

\centering
\include{qpics/c14}
\caption{Example of a quantum ripple-carry adder from \cite{cuccaro04}. This circuit prepares $a=1$, $b=15$ and computes the sum into \code{b} with an output carry \code{cout[0]}.\label{fig:example:add}}
\end{figure}

The ripple-carry adder \cite{cuccaro04} shown in Fig.~\ref{fig:example:add} exhibits hierarchical use of gate subroutines.

\begin{Verbatim}[commandchars=\\\{\}]
\PYG{c+c1}{// quantum ripple\PYGZhy{}carry adder from Cuccaro et al, quant\PYGZhy{}ph/0410184}
\PYG{n}{OPENQASM} \PYG{l+m+mf}{2.0}\PYG{p}{;}
\PYG{n}{include} \PYG{l+s}{\PYGZdq{}qelib1.inc\PYGZdq{}}\PYG{p}{;}
\PYG{n}{gate} \PYG{n}{majority} \PYG{n}{a}\PYG{p}{,}\PYG{n}{b}\PYG{p}{,}\PYG{n}{c} 
\PYG{p}{\PYGZob{}} 
  \PYG{n}{cx} \PYG{n}{c}\PYG{p}{,}\PYG{n}{b}\PYG{p}{;} 
  \PYG{n}{cx} \PYG{n}{c}\PYG{p}{,}\PYG{n}{a}\PYG{p}{;} 
  \PYG{n}{ccx} \PYG{n}{a}\PYG{p}{,}\PYG{n}{b}\PYG{p}{,}\PYG{n}{c}\PYG{p}{;} 
\PYG{p}{\PYGZcb{}}
\PYG{n}{gate} \PYG{n}{unmaj} \PYG{n}{a}\PYG{p}{,}\PYG{n}{b}\PYG{p}{,}\PYG{n}{c} 
\PYG{p}{\PYGZob{}} 
  \PYG{n}{ccx} \PYG{n}{a}\PYG{p}{,}\PYG{n}{b}\PYG{p}{,}\PYG{n}{c}\PYG{p}{;} 
  \PYG{n}{cx} \PYG{n}{c}\PYG{p}{,}\PYG{n}{a}\PYG{p}{;} 
  \PYG{n}{cx} \PYG{n}{a}\PYG{p}{,}\PYG{n}{b}\PYG{p}{;} 
\PYG{p}{\PYGZcb{}}
\PYG{n}{qreg} \PYG{n}{cin}\PYG{p}{[}\PYG{l+m+mi}{1}\PYG{p}{];}
\PYG{n}{qreg} \PYG{n}{a}\PYG{p}{[}\PYG{l+m+mi}{4}\PYG{p}{];}
\PYG{n}{qreg} \PYG{n}{b}\PYG{p}{[}\PYG{l+m+mi}{4}\PYG{p}{];}
\PYG{n}{qreg} \PYG{n}{cout}\PYG{p}{[}\PYG{l+m+mi}{1}\PYG{p}{];}
\PYG{n}{creg} \PYG{n}{ans}\PYG{p}{[}\PYG{l+m+mi}{5}\PYG{p}{];}
\PYG{c+c1}{// set input states}
\PYG{n}{x} \PYG{n}{a}\PYG{p}{[}\PYG{l+m+mi}{0}\PYG{p}{];} \PYG{c+c1}{// a = 0001}
\PYG{n}{x} \PYG{n}{b}\PYG{p}{;}    \PYG{c+c1}{// b = 1111}
\PYG{c+c1}{// add a to b, storing result in b}
\PYG{n}{majority} \PYG{n}{cin}\PYG{p}{[}\PYG{l+m+mi}{0}\PYG{p}{],}\PYG{n}{b}\PYG{p}{[}\PYG{l+m+mi}{0}\PYG{p}{],}\PYG{n}{a}\PYG{p}{[}\PYG{l+m+mi}{0}\PYG{p}{];}
\PYG{n}{majority} \PYG{n}{a}\PYG{p}{[}\PYG{l+m+mi}{0}\PYG{p}{],}\PYG{n}{b}\PYG{p}{[}\PYG{l+m+mi}{1}\PYG{p}{],}\PYG{n}{a}\PYG{p}{[}\PYG{l+m+mi}{1}\PYG{p}{];}
\PYG{n}{majority} \PYG{n}{a}\PYG{p}{[}\PYG{l+m+mi}{1}\PYG{p}{],}\PYG{n}{b}\PYG{p}{[}\PYG{l+m+mi}{2}\PYG{p}{],}\PYG{n}{a}\PYG{p}{[}\PYG{l+m+mi}{2}\PYG{p}{];}
\PYG{n}{majority} \PYG{n}{a}\PYG{p}{[}\PYG{l+m+mi}{2}\PYG{p}{],}\PYG{n}{b}\PYG{p}{[}\PYG{l+m+mi}{3}\PYG{p}{],}\PYG{n}{a}\PYG{p}{[}\PYG{l+m+mi}{3}\PYG{p}{];}
\PYG{n}{cx} \PYG{n}{a}\PYG{p}{[}\PYG{l+m+mi}{3}\PYG{p}{],}\PYG{n}{cout}\PYG{p}{[}\PYG{l+m+mi}{0}\PYG{p}{];}
\PYG{n}{unmaj} \PYG{n}{a}\PYG{p}{[}\PYG{l+m+mi}{2}\PYG{p}{],}\PYG{n}{b}\PYG{p}{[}\PYG{l+m+mi}{3}\PYG{p}{],}\PYG{n}{a}\PYG{p}{[}\PYG{l+m+mi}{3}\PYG{p}{];}
\PYG{n}{unmaj} \PYG{n}{a}\PYG{p}{[}\PYG{l+m+mi}{1}\PYG{p}{],}\PYG{n}{b}\PYG{p}{[}\PYG{l+m+mi}{2}\PYG{p}{],}\PYG{n}{a}\PYG{p}{[}\PYG{l+m+mi}{2}\PYG{p}{];}
\PYG{n}{unmaj} \PYG{n}{a}\PYG{p}{[}\PYG{l+m+mi}{0}\PYG{p}{],}\PYG{n}{b}\PYG{p}{[}\PYG{l+m+mi}{1}\PYG{p}{],}\PYG{n}{a}\PYG{p}{[}\PYG{l+m+mi}{1}\PYG{p}{];}
\PYG{n}{unmaj} \PYG{n}{cin}\PYG{p}{[}\PYG{l+m+mi}{0}\PYG{p}{],}\PYG{n}{b}\PYG{p}{[}\PYG{l+m+mi}{0}\PYG{p}{],}\PYG{n}{a}\PYG{p}{[}\PYG{l+m+mi}{0}\PYG{p}{];}
\PYG{n}{measure} \PYG{n}{b}\PYG{p}{[}\PYG{l+m+mi}{0}\PYG{p}{]} \PYG{o}{\PYGZhy{}\PYGZgt{}} \PYG{n}{ans}\PYG{p}{[}\PYG{l+m+mi}{0}\PYG{p}{];}
\PYG{n}{measure} \PYG{n}{b}\PYG{p}{[}\PYG{l+m+mi}{1}\PYG{p}{]} \PYG{o}{\PYGZhy{}\PYGZgt{}} \PYG{n}{ans}\PYG{p}{[}\PYG{l+m+mi}{1}\PYG{p}{];}
\PYG{n}{measure} \PYG{n}{b}\PYG{p}{[}\PYG{l+m+mi}{2}\PYG{p}{]} \PYG{o}{\PYGZhy{}\PYGZgt{}} \PYG{n}{ans}\PYG{p}{[}\PYG{l+m+mi}{2}\PYG{p}{];}
\PYG{n}{measure} \PYG{n}{b}\PYG{p}{[}\PYG{l+m+mi}{3}\PYG{p}{]} \PYG{o}{\PYGZhy{}\PYGZgt{}} \PYG{n}{ans}\PYG{p}{[}\PYG{l+m+mi}{3}\PYG{p}{];}
\PYG{n}{measure} \PYG{n}{cout}\PYG{p}{[}\PYG{l+m+mi}{0}\PYG{p}{]} \PYG{o}{\PYGZhy{}\PYGZgt{}} \PYG{n}{ans}\PYG{p}{[}\PYG{l+m+mi}{4}\PYG{p}{];}
\end{Verbatim}

\subsection{Randomized benchmarking}

A complete randomized benchmarking experiment could be described by a high level program. After passing through the upper phases of compilation, the program consists of many quantum circuits and associated classical control. Benchmarking is a particularly simple example because there is no data dependence between these quantum circuits. 

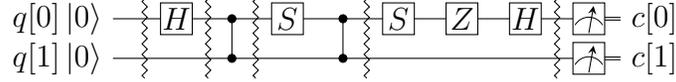
\begin{figure}
\centering
%! \usetikzlibrary{decorations.pathreplacing,decorations.pathmorphing}
\begin{tikzpicture}[scale=1.000000,x=1pt,y=1pt]
\filldraw[color=white] (0.000000, -7.500000) rectangle (192.000000, 22.500000);
% Drawing wires
% Line 1: q0 W q[0]\ket{0} c[0]
\draw[color=black] (0.000000,15.000000) -- (180.000000,15.000000);
\draw[color=black] (180.000000,14.500000) -- (192.000000,14.500000);
\draw[color=black] (180.000000,15.500000) -- (192.000000,15.500000);
\draw[color=black] (0.000000,15.000000) node[left] {$q[0]\ket{0}$};
% Line 2: q1 W q[1]\ket{0} c[1]
\draw[color=black] (0.000000,0.000000) -- (180.000000,0.000000);
\draw[color=black] (180.000000,-0.500000) -- (192.000000,-0.500000);
\draw[color=black] (180.000000,0.500000) -- (192.000000,0.500000);
\draw[color=black] (0.000000,0.000000) node[left] {$q[1]\ket{0}$};
% Done with wires; drawing gates
% Line 3: BARRIER
\draw[decorate,decoration={zigzag,amplitude=1pt,segment length=4}] (12.000000,-7.500000) -- (12.000000,22.500000);
% Line 4: q0 H
\begin{scope}
\draw[fill=white] (24.000000, 15.000000) +(-45.000000:8.485281pt and 8.485281pt) -- +(45.000000:8.485281pt and 8.485281pt) -- +(135.000000:8.485281pt and 8.485281pt) -- +(225.000000:8.485281pt and 8.485281pt) -- cycle;
\clip (24.000000, 15.000000) +(-45.000000:8.485281pt and 8.485281pt) -- +(45.000000:8.485281pt and 8.485281pt) -- +(135.000000:8.485281pt and 8.485281pt) -- +(225.000000:8.485281pt and 8.485281pt) -- cycle;
\draw (24.000000, 15.000000) node {$H$};
\end{scope}
% Line 5: BARRIER
\draw[decorate,decoration={zigzag,amplitude=1pt,segment length=4}] (36.000000,-7.500000) -- (36.000000,22.500000);
% Line 6: q0 q1
\draw (45.000000,15.000000) -- (45.000000,0.000000);
\filldraw (45.000000, 15.000000) circle(1.500000pt);
\filldraw (45.000000, 0.000000) circle(1.500000pt);
% Line 7: BARRIER
\draw[decorate,decoration={zigzag,amplitude=1pt,segment length=4}] (54.000000,-7.500000) -- (54.000000,22.500000);
% Line 8: q0 G $S$
\begin{scope}
\draw[fill=white] (66.000000, 15.000000) +(-45.000000:8.485281pt and 8.485281pt) -- +(45.000000:8.485281pt and 8.485281pt) -- +(135.000000:8.485281pt and 8.485281pt) -- +(225.000000:8.485281pt and 8.485281pt) -- cycle;
\clip (66.000000, 15.000000) +(-45.000000:8.485281pt and 8.485281pt) -- +(45.000000:8.485281pt and 8.485281pt) -- +(135.000000:8.485281pt and 8.485281pt) -- +(225.000000:8.485281pt and 8.485281pt) -- cycle;
\draw (66.000000, 15.000000) node {$S$};
\end{scope}
% Line 9: q0 q1
\draw (87.000000,15.000000) -- (87.000000,0.000000);
\filldraw (87.000000, 15.000000) circle(1.500000pt);
\filldraw (87.000000, 0.000000) circle(1.500000pt);
% Line 10: BARRIER
\draw[decorate,decoration={zigzag,amplitude=1pt,segment length=4}] (96.000000,-7.500000) -- (96.000000,22.500000);
% Line 11: q0 G $S$
\begin{scope}
\draw[fill=white] (108.000000, 15.000000) +(-45.000000:8.485281pt and 8.485281pt) -- +(45.000000:8.485281pt and 8.485281pt) -- +(135.000000:8.485281pt and 8.485281pt) -- +(225.000000:8.485281pt and 8.485281pt) -- cycle;
\clip (108.000000, 15.000000) +(-45.000000:8.485281pt and 8.485281pt) -- +(45.000000:8.485281pt and 8.485281pt) -- +(135.000000:8.485281pt and 8.485281pt) -- +(225.000000:8.485281pt and 8.485281pt) -- cycle;
\draw (108.000000, 15.000000) node {$S$};
\end{scope}
% Line 12: q0 Z
\begin{scope}
\draw[fill=white] (132.000000, 15.000000) +(-45.000000:8.485281pt and 8.485281pt) -- +(45.000000:8.485281pt and 8.485281pt) -- +(135.000000:8.485281pt and 8.485281pt) -- +(225.000000:8.485281pt and 8.485281pt) -- cycle;
\clip (132.000000, 15.000000) +(-45.000000:8.485281pt and 8.485281pt) -- +(45.000000:8.485281pt and 8.485281pt) -- +(135.000000:8.485281pt and 8.485281pt) -- +(225.000000:8.485281pt and 8.485281pt) -- cycle;
\draw (132.000000, 15.000000) node {$Z$};
\end{scope}
% Line 13: q0 H
\begin{scope}
\draw[fill=white] (156.000000, 15.000000) +(-45.000000:8.485281pt and 8.485281pt) -- +(45.000000:8.485281pt and 8.485281pt) -- +(135.000000:8.485281pt and 8.485281pt) -- +(225.000000:8.485281pt and 8.485281pt) -- cycle;
\clip (156.000000, 15.000000) +(-45.000000:8.485281pt and 8.485281pt) -- +(45.000000:8.485281pt and 8.485281pt) -- +(135.000000:8.485281pt and 8.485281pt) -- +(225.000000:8.485281pt and 8.485281pt) -- cycle;
\draw (156.000000, 15.000000) node {$H$};
\end{scope}
% Line 14: BARRIER
\draw[decorate,decoration={zigzag,amplitude=1pt,segment length=4}] (168.000000,-7.500000) -- (168.000000,22.500000);
% Line 15: q0 M
\draw[fill=white] (174.000000, 9.000000) rectangle (186.000000, 21.000000);
\draw[very thin] (180.000000, 15.600000) arc (90:150:6.000000pt);
\draw[very thin] (180.000000, 15.600000) arc (90:30:6.000000pt);
\draw[->,>=stealth] (180.000000, 9.600000) -- +(80:10.392305pt);
% Line 16: q1 M
\draw[fill=white] (174.000000, -6.000000) rectangle (186.000000, 6.000000);
\draw[very thin] (180.000000, 0.600000) arc (90:150:6.000000pt);
\draw[very thin] (180.000000, 0.600000) arc (90:30:6.000000pt);
\draw[->,>=stealth] (180.000000, -5.400000) -- +(80:10.392305pt);
% Done with gates; drawing ending labels
\draw[color=black] (192.000000,15.000000) node[right] {$c[0]$};
\draw[color=black] (192.000000,0.000000) node[right] {$c[1]$};
% Done with ending labels; drawing cut lines and comments
% Done with comments
\end{tikzpicture}
\caption{Example of a two-qubit randomized benchmarking (RB) sequence over the basis $\langle H, S, CZ, X, Y, Z\rangle$. Barriers separate the implementations of each Clifford gate. An RB experiment consists of many sequences. Each sequence runs some number of times (``shots'').\label{fig:example:rb}}
\end{figure}

Each circuit is a sequence of random Clifford gates composed from a set of basic gates (Fig.~\ref{fig:example:rb} uses the gate set \code{h}, \code{s}, \code{cz}, and Paulis). If the gate set differs from the built-in gate set, new gates can be defined using the \code{gate} statement. Each of the randomly-chosen Clifford gates is separated from prior and future gates by barrier instructions to prevent the sequence from simplifying to the identity as a result of subsequent transformations.

\begin{Verbatim}[commandchars=\\\{\}]
\PYG{c+c1}{// One randomized benchmarking sequence}
\PYG{n}{OPENQASM} \PYG{l+m+mf}{2.0}\PYG{p}{;}
\PYG{n}{include} \PYG{l+s}{\PYGZdq{}qelib1.inc\PYGZdq{}}\PYG{p}{;}
\PYG{n}{qreg} \PYG{n}{q}\PYG{p}{[}\PYG{l+m+mi}{2}\PYG{p}{];}
\PYG{n}{creg} \PYG{n}{c}\PYG{p}{[}\PYG{l+m+mi}{2}\PYG{p}{];}
\PYG{n}{h} \PYG{n}{q}\PYG{p}{[}\PYG{l+m+mi}{0}\PYG{p}{];}
\PYG{n}{barrier} \PYG{n}{q}\PYG{p}{;}
\PYG{n}{cz} \PYG{n}{q}\PYG{p}{[}\PYG{l+m+mi}{0}\PYG{p}{],}\PYG{n}{q}\PYG{p}{[}\PYG{l+m+mi}{1}\PYG{p}{];}
\PYG{n}{barrier} \PYG{n}{q}\PYG{p}{;}
\PYG{n}{s} \PYG{n}{q}\PYG{p}{[}\PYG{l+m+mi}{0}\PYG{p}{];}
\PYG{n}{cz} \PYG{n}{q}\PYG{p}{[}\PYG{l+m+mi}{0}\PYG{p}{],}\PYG{n}{q}\PYG{p}{[}\PYG{l+m+mi}{1}\PYG{p}{];}
\PYG{n}{barrier} \PYG{n}{q}\PYG{p}{;}
\PYG{n}{s} \PYG{n}{q}\PYG{p}{[}\PYG{l+m+mi}{0}\PYG{p}{];}
\PYG{n}{z} \PYG{n}{q}\PYG{p}{[}\PYG{l+m+mi}{0}\PYG{p}{];}
\PYG{n}{h} \PYG{n}{q}\PYG{p}{[}\PYG{l+m+mi}{0}\PYG{p}{];}
\PYG{n}{barrier} \PYG{n}{q}\PYG{p}{;}
\PYG{n}{measure} \PYG{n}{q} \PYG{o}{\PYGZhy{}\PYGZgt{}} \PYG{n}{c}\PYG{p}{;}
\end{Verbatim}

\subsection{Quantum process tomography}

\begin{figure}
\centering
%! \usetikzlibrary{decorations.pathreplacing,decorations.pathmorphing}
\begin{tikzpicture}[scale=1.000000,x=1pt,y=1pt]
\filldraw[color=white] (0.000000, -7.500000) rectangle (142.000000, 7.500000);
% Drawing wires
% Line 2: q0 W q[0]\ket{0} c[0]
\draw[color=black] (0.000000,0.000000) -- (130.000000,0.000000);
\draw[color=black] (130.000000,-0.500000) -- (142.000000,-0.500000);
\draw[color=black] (130.000000,0.500000) -- (142.000000,0.500000);
\draw[color=black] (0.000000,0.000000) node[left] {$q[0]\ket{0}$};
% Done with wires; drawing gates
% Line 3: q0 G:width=30 $pre$
\begin{scope}
\draw[fill=white] (21.000000, -0.000000) +(-45.000000:21.213203pt and 8.485281pt) -- +(45.000000:21.213203pt and 8.485281pt) -- +(135.000000:21.213203pt and 8.485281pt) -- +(225.000000:21.213203pt and 8.485281pt) -- cycle;
\clip (21.000000, -0.000000) +(-45.000000:21.213203pt and 8.485281pt) -- +(45.000000:21.213203pt and 8.485281pt) -- +(135.000000:21.213203pt and 8.485281pt) -- +(225.000000:21.213203pt and 8.485281pt) -- cycle;
\draw (21.000000, -0.000000) node {\scriptsize$pre$};
\end{scope}
% Line 4: BARRIER
\draw[decorate,decoration={zigzag,amplitude=1pt,segment length=4}] (42.000000,-7.500000) -- (42.000000,7.500000);
% Line 5: q0 H
\begin{scope}
\draw[fill=white] (54.000000, -0.000000) +(-45.000000:8.485281pt and 8.485281pt) -- +(45.000000:8.485281pt and 8.485281pt) -- +(135.000000:8.485281pt and 8.485281pt) -- +(225.000000:8.485281pt and 8.485281pt) -- cycle;
\clip (54.000000, -0.000000) +(-45.000000:8.485281pt and 8.485281pt) -- +(45.000000:8.485281pt and 8.485281pt) -- +(135.000000:8.485281pt and 8.485281pt) -- +(225.000000:8.485281pt and 8.485281pt) -- cycle;
\draw (54.000000, -0.000000) node {$H$};
\end{scope}
% Line 6: BARRIER
\draw[decorate,decoration={zigzag,amplitude=1pt,segment length=4}] (66.000000,-7.500000) -- (66.000000,7.500000);
% Line 7: q0 G:width=40 $post$
\begin{scope}
\draw[fill=white] (92.000000, -0.000000) +(-45.000000:28.284271pt and 8.485281pt) -- +(45.000000:28.284271pt and 8.485281pt) -- +(135.000000:28.284271pt and 8.485281pt) -- +(225.000000:28.284271pt and 8.485281pt) -- cycle;
\clip (92.000000, -0.000000) +(-45.000000:28.284271pt and 8.485281pt) -- +(45.000000:28.284271pt and 8.485281pt) -- +(135.000000:28.284271pt and 8.485281pt) -- +(225.000000:28.284271pt and 8.485281pt) -- cycle;
\draw (92.000000, -0.000000) node {\scriptsize$post$};
\end{scope}
% Line 8: q0 M
\draw[fill=white] (124.000000, -6.000000) rectangle (136.000000, 6.000000);
\draw[very thin] (130.000000, 0.600000) arc (90:150:6.000000pt);
\draw[very thin] (130.000000, 0.600000) arc (90:30:6.000000pt);
\draw[->,>=stealth] (130.000000, -5.400000) -- +(80:10.392305pt);
% Done with gates; drawing ending labels
\draw[color=black] (142.000000,0.000000) node[right] {$c[0]$};
% Done with ending labels; drawing cut lines and comments
% Done with comments
\end{tikzpicture}
\caption{Example of a single-qubit quantum process tomography circuit. The \code{pre} and \code{post} gates are described by a higher-level program that generates intermediate code containing several independent circuits. Each circuit is executed some number of times (``shots'') to compute statistics from which the \code{h} gate process is reconstructed. Barriers separate the process under study from the pre- and post- gates. \label{fig:example:qpt}}
\end{figure}
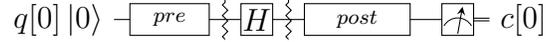

As in randomized benchmarking, a high-level program describes a quantum process tomography (QPT) experiment. Each program compiles to intermediate code with several independent quantum circuits that can each be described using Open QASM (version 2.0). Fig.~\ref{fig:example:qpt} shows QPT of a Hadamard gate. Each circuit is identical except for the definitions of the \code{pre} and \code{post} gates. The empty definitions in the current example are placeholders that define identity gates. For textbook QPT, the \code{pre} and \code{post} gates are both taken from the set $\{I,H,SH\}$ to prepare $|0\rangle$, $|+\rangle$, and $|+i\rangle$ and measure in the $Z$, $X$, and $Y$ basis.

\begin{Verbatim}[commandchars=\\\{\}]
\PYG{n}{OPENQASM} \PYG{l+m+mf}{2.0}\PYG{p}{;}
\PYG{n}{include} \PYG{l+s}{\PYGZdq{}qelib1.inc\PYGZdq{}}\PYG{p}{;}
\PYG{n}{gate} \PYG{n}{pre} \PYG{n}{q} \PYG{p}{\PYGZob{}} \PYG{p}{\PYGZcb{}}   \PYG{c+c1}{// pre\PYGZhy{}rotation}
\PYG{n}{gate} \PYG{n}{post} \PYG{n}{q} \PYG{p}{\PYGZob{}} \PYG{p}{\PYGZcb{}}  \PYG{c+c1}{// post\PYGZhy{}rotation}
\PYG{n}{qreg} \PYG{n}{q}\PYG{p}{[}\PYG{l+m+mi}{1}\PYG{p}{];}
\PYG{n}{creg} \PYG{n}{c}\PYG{p}{[}\PYG{l+m+mi}{1}\PYG{p}{];}
\PYG{n}{pre} \PYG{n}{q}\PYG{p}{[}\PYG{l+m+mi}{0}\PYG{p}{];}
\PYG{n}{barrier} \PYG{n}{q}\PYG{p}{;}
\PYG{n}{h} \PYG{n}{q}\PYG{p}{[}\PYG{l+m+mi}{0}\PYG{p}{];}
\PYG{n}{barrier} \PYG{n}{q}\PYG{p}{;}
\PYG{n}{post} \PYG{n}{q}\PYG{p}{[}\PYG{l+m+mi}{0}\PYG{p}{];}
\PYG{n}{measure} \PYG{n}{q}\PYG{p}{[}\PYG{l+m+mi}{0}\PYG{p}{]} \PYG{o}{\PYGZhy{}\PYGZgt{}} \PYG{n}{c}\PYG{p}{[}\PYG{l+m+mi}{0}\PYG{p}{];}
\end{Verbatim}

\subsection{Quantum error-correction}

\begin{figure}
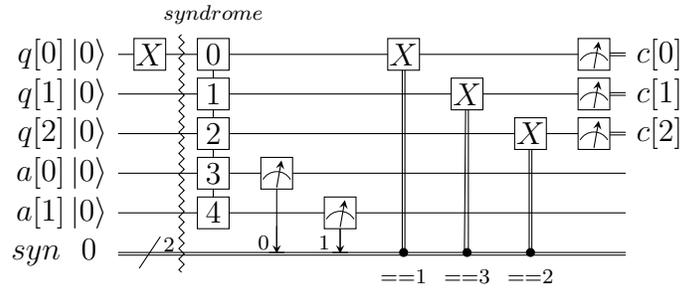

\centering
\include{qpics/c17}
\caption{Example of a quantum bit-flip repetition code. The circuit begins with the (classical) encoded state $|000\rangle$, applies an error on \code{q[0]}, and uses feedback on a syndrome measurement to correct the error.
\label{fig:example:qec3}}
\end{figure}

This example of the 3-bit quantum repetition code (Fig.~\ref{fig:example:qec3}) demonstrates how Open QASM (version 2.0) can express simple quantum error-correction circuits.

\begin{Verbatim}[commandchars=\\\{\}]
\PYG{c+c1}{// Repetition code syndrome measurement}
\PYG{n}{OPENQASM} \PYG{l+m+mf}{2.0}\PYG{p}{;}
\PYG{n}{include} \PYG{l+s}{\PYGZdq{}qelib1.inc\PYGZdq{}}\PYG{p}{;}
\PYG{n}{qreg} \PYG{n}{q}\PYG{p}{[}\PYG{l+m+mi}{3}\PYG{p}{];}
\PYG{n}{qreg} \PYG{n}{a}\PYG{p}{[}\PYG{l+m+mi}{2}\PYG{p}{];}
\PYG{n}{creg} \PYG{n}{c}\PYG{p}{[}\PYG{l+m+mi}{3}\PYG{p}{];}
\PYG{n}{creg} \PYG{n}{syn}\PYG{p}{[}\PYG{l+m+mi}{2}\PYG{p}{];}
\PYG{n}{gate} \PYG{n}{syndrome} \PYG{n}{d1}\PYG{p}{,}\PYG{n}{d2}\PYG{p}{,}\PYG{n}{d3}\PYG{p}{,}\PYG{n}{a1}\PYG{p}{,}\PYG{n}{a2} 
\PYG{p}{\PYGZob{}} 
  \PYG{n}{cx} \PYG{n}{d1}\PYG{p}{,}\PYG{n}{a1}\PYG{p}{;} \PYG{n}{cx} \PYG{n}{d2}\PYG{p}{,}\PYG{n}{a1}\PYG{p}{;} 
  \PYG{n}{cx} \PYG{n}{d2}\PYG{p}{,}\PYG{n}{a2}\PYG{p}{;} \PYG{n}{cx} \PYG{n}{d3}\PYG{p}{,}\PYG{n}{a2}\PYG{p}{;} 
\PYG{p}{\PYGZcb{}}
\PYG{n}{x} \PYG{n}{q}\PYG{p}{[}\PYG{l+m+mi}{0}\PYG{p}{];} \PYG{c+c1}{// error}
\PYG{n}{barrier} \PYG{n}{q}\PYG{p}{;}
\PYG{n}{syndrome} \PYG{n}{q}\PYG{p}{[}\PYG{l+m+mi}{0}\PYG{p}{],}\PYG{n}{q}\PYG{p}{[}\PYG{l+m+mi}{1}\PYG{p}{],}\PYG{n}{q}\PYG{p}{[}\PYG{l+m+mi}{2}\PYG{p}{],}\PYG{n}{a}\PYG{p}{[}\PYG{l+m+mi}{0}\PYG{p}{],}\PYG{n}{a}\PYG{p}{[}\PYG{l+m+mi}{1}\PYG{p}{];}
\PYG{n}{measure} \PYG{n}{a} \PYG{o}{\PYGZhy{}\PYGZgt{}} \PYG{n}{syn}\PYG{p}{;}
\PYG{k}{if}\PYG{p}{(}\PYG{n}{syn}\PYG{o}{==}\PYG{l+m+mi}{1}\PYG{p}{)} \PYG{n}{x} \PYG{n}{q}\PYG{p}{[}\PYG{l+m+mi}{0}\PYG{p}{];}
\PYG{k}{if}\PYG{p}{(}\PYG{n}{syn}\PYG{o}{==}\PYG{l+m+mi}{2}\PYG{p}{)} \PYG{n}{x} \PYG{n}{q}\PYG{p}{[}\PYG{l+m+mi}{2}\PYG{p}{];}
\PYG{k}{if}\PYG{p}{(}\PYG{n}{syn}\PYG{o}{==}\PYG{l+m+mi}{3}\PYG{p}{)} \PYG{n}{x} \PYG{n}{q}\PYG{p}{[}\PYG{l+m+mi}{1}\PYG{p}{];}
\PYG{n}{measure} \PYG{n}{q} \PYG{o}{\PYGZhy{}\PYGZgt{}} \PYG{n}{c}\PYG{p}{;}
\end{Verbatim}

\section{Acknowledgements}

This document represents ideas and contributions from the IBM Quantum Computing group as a whole. We acknowledge suggestions and discussions with the IBM Quantum Experience community \cite{qe}. We thank Abigail Cross for typesetting the figures and proof-reading the document. We thank Tom Draper and Sandy Kutin for the $\langle\mathrm{q}|\mathrm{pic}\rangle$ package \cite{qpic}, which was used for initial typesetting of the quantum circuits. We acknowledge partial support from the IBM Research Frontiers Institute.

\appendix

\section{Open QASM Grammar}\label{app:grammar}

\begin{bnf*}
\bnfprod{mainprogram}
{\begin{bnfsplit}\bnfts{OPENQASM} \bnfsp \bnfpn{real} \bnfsp \bnfts{;} \bnfsp \bnfpn{program}
\end{bnfsplit}
}\\
\bnfprod{program}
{\begin{bnfsplit}
\bnfpn{statement}
\bnfor \bnfpn{program} \bnfsp \bnfpn{statement}
\end{bnfsplit}
}\\
\bnfprod{statement}
{\begin{bnfsplit}\bnfpn{decl} \\
\bnfor \bnfpn{gatedecl} \bnfsp \bnfpn{goplist} \bnfsp \bnfts{\}} \\
\bnfor \bnfpn{gatedecl} \bnfsp \bnfts{\}} \\ 
\bnfor \bnfts{opaque} \bnfsp \bnfpn{id} \bnfsp \bnfpn{idlist} \bnfsp \bnfts{;} \\
\bnfor \bnfts{opaque} \bnfsp \bnfpn{id} \bnfsp \bnfts{(} \bnfsp \bnfts{)} \bnfsp \bnfpn{idlist} \bnfsp \bnfts{;}
\bnfor \bnfts{opaque} \bnfsp \bnfpn{id} \bnfsp \bnfts{(} \bnfsp \bnfpn{idlist} \bnfsp \bnfts{)} \bnfsp \bnfpn{idlist} \bnfsp \bnfts{;} \\
\bnfor \bnfpn{qop} \\
\bnfor \bnfts{if} \bnfsp \bnfts{(} \bnfsp \bnfpn{id} \bnfsp \bnfts{==} \bnfsp \bnfpn{nninteger} \bnfsp \bnfts{)} \bnfsp \bnfpn{qop} \\
\bnfor \bnfts{barrier} \bnfsp \bnfpn{anylist} \bnfsp \bnfts{;}
\end{bnfsplit}
}\\
\bnfprod{decl}
{\begin{bnfsplit}\bnfts{qreg} \bnfsp \bnfpn{id} \bnfsp \bnfts{[} \bnfsp \bnfpn{nninteger} \bnfsp \bnfts{]} \bnfsp \bnfts{;}
\bnfor \bnfts{creg} \bnfsp \bnfpn{id} \bnfsp \bnfts{[} \bnfsp \bnfpn{nninteger} \bnfsp \bnfts{]} \bnfsp \bnfts{;}
\end{bnfsplit}
}\\
\bnfprod{gatedecl}
{\begin{bnfsplit}\bnfts{gate} \bnfsp \bnfpn{id} \bnfsp \bnfpn{idlist} \bnfsp \bnfts{\{} \\
\bnfor \bnfts{gate} \bnfsp \bnfpn{id} \bnfsp \bnfts{(} \bnfsp \bnfts{)} \bnfsp \bnfpn{idlist} \bnfsp \bnfts{\{} \\
\bnfor \bnfts{gate} \bnfsp \bnfpn{id} \bnfsp \bnfts{(} \bnfsp \bnfpn{idlist} \bnfsp \bnfts{)} \bnfpn{idlist} \bnfsp \bnfts{\{}
\end{bnfsplit}
}\\
\bnfprod{goplist}
{\begin{bnfsplit}\bnfpn{uop} \\
\bnfor \bnfts{barrier} \bnfsp \bnfpn{idlist} \bnfsp \bnfts{;} \\
\bnfor \bnfpn{goplist} \bnfsp \bnfpn{uop} \\
\bnfor \bnfpn{goplist} \bnfsp \bnfts{barrier} \bnfsp \bnfpn{idlist} \bnfsp \bnfts{;}
\end{bnfsplit}
}\\
\bnfprod{qop}
{\begin{bnfsplit}\bnfpn{uop} \\
\bnfor \bnfts{measure} \bnfsp \bnfpn{argument} \bnfsp \bnfts{-} \bnfsp \bnfts{>} \bnfsp \bnfpn{argument} \bnfsp \bnfts{;} \\
\bnfor \bnfts{reset} \bnfsp \bnfpn{argument} \bnfsp \bnfts{;}
\end{bnfsplit}
}\\
\bnfprod{uop}
{\begin{bnfsplit}\bnfts{U} \bnfsp \bnfts{(} \bnfsp \bnfpn{explist} \bnfsp \bnfts{)} \bnfsp \bnfpn{argument} \bnfsp \bnfts{;} \\
\bnfor \bnfts{CX} \bnfsp \bnfpn{argument} \bnfsp \bnfts{,} \bnfsp \bnfpn{argument} \bnfsp \bnfts{;} \\
\bnfor \bnfpn{id} \bnfsp \bnfpn{anylist} \bnfsp \bnfts{;}
\bnfor \bnfpn{id} \bnfsp \bnfts{(} \bnfsp \bnfts{)} \bnfsp \bnfpn{anylist} \bnfsp \bnfts{;} \\
\bnfor \bnfpn{id} \bnfsp \bnfts{(} \bnfsp \bnfpn{explist} \bnfsp \bnfts{)} \bnfsp \bnfpn{anylist} \bnfsp \bnfts{;}
\end{bnfsplit}
}\\
\bnfprod{anylist}
{\begin{bnfsplit}\bnfpn{idlist}
\bnfor \bnfpn{mixedlist}
\end{bnfsplit}
}\\
\bnfprod{idlist}
{\begin{bnfsplit}\bnfpn{id}
\bnfor \bnfpn{idlist} \bnfsp \bnfts{,} \bnfsp \bnfpn{id}
\end{bnfsplit}
}\\
\bnfprod{mixedlist}
{\begin{bnfsplit}\bnfpn{id} \bnfsp \bnfts{[} \bnfsp \bnfpn{nninteger} \bnfsp \bnfts{]}
\bnfor \bnfpn{mixedlist} \bnfsp \bnfts{,} \bnfsp \bnfpn{id} \\
\bnfor \bnfpn{mixedlist} \bnfsp \bnfts{,} \bnfsp \bnfpn{id} \bnfsp \bnfts{[} \bnfsp \bnfpn{nninteger} \bnfsp \bnfts{]} \\
\bnfor \bnfpn{idlist} \bnfsp \bnfts{,} \bnfsp \bnfpn{id} \bnfts{[} \bnfsp \bnfpn{nninteger} \bnfsp \bnfts{]} \\
\end{bnfsplit}
}\\
\bnfprod{argument}
{\begin{bnfsplit}\bnfpn{id}
\bnfor \bnfpn{id} \bnfsp \bnfts{[} \bnfsp \bnfpn{nninteger} \bnfsp \bnfts{]}
\end{bnfsplit}
}\\
\bnfprod{explist}
{\begin{bnfsplit}\bnfpn{exp}
\bnfor \bnfpn{explist} \bnfsp \bnfts{,} \bnfsp \bnfpn{exp}
\end{bnfsplit}
}\\
\bnfprod{exp}
{\begin{bnfsplit} \bnfpn{real}
\bnfor \bnfpn{nninteger}
\bnfor \bnfts{pi}
\bnfor \bnfpn{id} \\
\bnfor \bnfpn{exp} \bnfsp \bnfts{+} \bnfsp \bnfpn{exp}
\bnfor \bnfpn{exp} \bnfsp \bnfts{-} \bnfsp \bnfpn{exp}
\bnfor \bnfpn{exp} \bnfsp \bnfts{*} \bnfsp \bnfpn{exp} \\
\bnfor \bnfpn{exp} \bnfsp \bnfts{/} \bnfsp \bnfpn{exp}
\bnfor \bnfts{-} \bnfsp \bnfpn{exp}
\bnfor \bnfpn{exp} \bnfsp \bnfts{\^} \bnfsp \bnfpn{exp} \\
\bnfor \bnfts{(} \bnfsp \bnfpn{exp} \bnfsp \bnfts{)}
\bnfor \bnfpn{unaryop} \bnfsp \bnfts{(} \bnfsp \bnfpn{exp} \bnfsp \bnfts{)} \\
\end{bnfsplit}
}\\
\bnfprod{unaryop}
{\begin{bnfsplit}\bnfts{sin}
\bnfor \bnfts{cos}
\bnfor \bnfts{tan}
\bnfor \bnfts{exp}
\bnfor \bnfts{ln}
\bnfor \bnfts{sqrt}
\end{bnfsplit}
}\\
\end{bnf*}

This is a simplified grammar for Open QASM presented in Backus-Naur form. The unlisted productions $\langle\mathrm{id}\rangle$, $\langle\mathrm{real}\rangle$ and $\langle\mathrm{nninteger}\rangle$ are defined by the regular expressions:
\begin{verbatim}
id        := [a-z][A-Za-z0-9_]*
real      := ([0-9]+\.[0-9]*|[0-9]*\.[0-9]+)([eE][-+]?[0-9]+)?
nninteger := [1-9]+[0-9]*|0
\end{verbatim}
Not all programs produced using this grammar are valid Open QASM circuits. As explained in Section~\ref{sec:spec}, there are additional rules concerning valid arguments, parameters, declarations, and identifiers, as well as the standard operator precedence rules in the parameter expressions.

\end{document}